\documentclass[12pt,english,aps, final]{revtex4}
\usepackage[T1]{fontenc}
\usepackage[latin9]{inputenc}
\usepackage{array}
\usepackage{multirow}
\usepackage{amsmath}
\usepackage{amssymb}
\usepackage{cancel}
\usepackage{graphicx}
\usepackage{esint}

\usepackage[dvips]{epsfig}
\usepackage{graphics}
\usepackage{wrapfig}
\usepackage{floatflt}

\newcommand{\Li}[2]{{\mbox{Li}}_{#1}\left(#2\right)}
\newcommand{\Cl}[2]{{\mbox{Cl}}_{#1}\left(#2\right)}
\newcommand{\Ls}[2]{{\mbox{Ls}}_{#1}\left(#2\right)}
\newcommand{\sfrac}[2]{\frac{{}_{#1}}{{}^{#2}}}

\makeatletter

\providecommand{\tabularnewline}{\\}

\@ifundefined{textcolor}{}
{%
 \definecolor{BLACK}{gray}{0}
 \definecolor{WHITE}{gray}{1}
 \definecolor{RED}{rgb}{1,0,0}
 \definecolor{GREEN}{rgb}{0,1,0}
 \definecolor{BLUE}{rgb}{0,0,1}
 \definecolor{CYAN}{cmyk}{1,0,0,0}
 \definecolor{MAGENTA}{cmyk}{0,1,0,0}
 \definecolor{YELLOW}{cmyk}{0,0,1,0}
}

\makeatother

\usepackage{babel}
\begin{document}
\title{Recurrence Relations and Dispersive Techniques for Precision Multi-Loop Calculations}
\author{A. Aleksejevs, S. Barkanova and A. I. Davydychev}
\affiliation{Grenfell Campus of Memorial University,  Corner Brook, Newfoundland and Labrador, Canada}
\begin{abstract}

\textit{Ab initio} predictions of two-loop electroweak contributions to observables are increasingly essential for precision collider experiments, yet their evaluation remains very challenging. We connect recurrence techniques and dispersive methods in order to evaluate complex multi-loop Feynman diagrams. By expressing multi-point Passarino-Veltman functions in a two-point basis and using shifted space-time dimensions with recurrence relations, we minimize the number of required dispersive integrals. This approach reduces computation time and enables a precise and efficient analysis of one- and two-loop diagrams.
\end{abstract}
\maketitle

\section{Introduction}

Modern collider and low-energy precision programs are driving sub-percent uncertainties, demanding \textit{ab initio}, two-loop electroweak predictions for multi-scale, multi-leg processes. To meet this precision frontier, theoretical methods must evolve in parallel with experiments. Flagship measurements include MOLLER \citep{MOLLER} at Jefferson Lab (JLab) (low-$Q^2$ determination of $\sin{\theta}_W^2$ from parity-violating M{\o}ller scattering), P2 \citep{P2} at MESA (proton weak charge), and Belle II at KEK SuperKEKB (precision flavour and \textit{CP}-violation studies). Upcoming programs such as SoLID-PVDIS \citep{SOLID} at JLab and the Electron-Ion Collider (EIC) \citep{EIC} at  Brookhaven National Laboratory (BNL) will further require comprehensive higher-order theory predictions across multiple scattering channels.
The accurate theoretical description of electroweak processes has long been a cornerstone of precision tests of the Standard Model (SM). Over the past four decades, the field has progressed from the first analytical one-loop formalisms to modern semi-analytical and numerical two-loop frameworks capable of supporting sub-percent experimental precision. This evolution reflects both major theoretical advances in multi-loop quantum-field-theory techniques and the rising experimental demands of collider and low-energy programs.
The systematic treatment of loop corrections in the electroweak theory was established in the late 1970s and 1980s, culminating in the Passarino-Veltman (PV) tensor-reduction method. The PV algorithm provided a general prescription for expressing one-loop tensor integrals in terms of scalar functions, establishing the algebraic foundation for all subsequent higher-order calculations \citep{PV1979}.

During the following decade, Refs.~\citep{Hollik1990,  Denner1993} produced comprehensive reviews and explicit one-loop calculations relevant to CERN's Large Electron-Positron (LEP) collider precision physics, codifying gauge-invariant renormalization schemes and parameter definitions. These works became standard references for both theoretical and computational approaches to radiative corrections.
The extension to two-loop order demanded a deeper algebraic understanding of Feynman integrals.  References ~\citep{Tarasov96, Tarasov97} introduced dimension-recurrence and propagator-power-reduction identities that relate integrals in $D$ and $D\pm2$ dimensions. These relations enable systematic reduction of higher-rank and higher-dimensional integrals to a minimal set of master integrals.   This formalism remains foundational for modern two-loop reduction algorithms and underpins most symbolic-manipulation packages used today.
Concurrently,  in \citep{PLB91}, a complementary algebraic framework was developed for reducing tensor Feynman integrals to scalar ones using dimension shifting recurrence relations (see also \citep{GG, BGHPS, EGZ} and references therein).

Analytic evaluation of increasingly complex diagrams soon became infeasible due to the proliferation of mass scales, external invariants, and threshold singularities. This challenge prompted the development of semi-analytical and numerical approaches such as sector-decomposition methods, differential-equation systems for master integrals, and dispersion-relation techniques \citep{Lee, SmrBook, Kotikov, BSch, BSSch}.
A major step was the differential-equation approach for master integrals, in which systems of linear equations in kinematic invariants are solved either analytically in canonical ($\varepsilon$-form) basis or numerically using series expansions.  In \citep{Wasser, Henn}, this framework was refined by introducing canonical-basis and uniform-weight formulations, improving both analytic transparency and numerical stability.  

A comprehensive two-loop framework was developed in \citep{AA-1, AA-2, AA-3, AA-4, AA-5},  specifically for polarized M{\o}ller scattering: an essential channel for future parity-violation experiments. The group systematically advanced from reducible two-loop and quadratic one-loop contributions to two-loop irreducible self-energies, vertex and box calculations. These results quantified higher-order electroweak effects in parity-violating asymmetries, establishing reliable theoretical uncertainties at the sub-percent level,  which are critical for the forthcoming MOLLER experiment.

Beyond differential-equation and sector-decomposition strategies, a distinct line of development has been the dispersive approach introduced  in \citep{AA1}-\citep{AB19a}. This methodology expresses multi-point Passarino-Veltman functions in a two-point-function basis, replacing sub-loop insertions with effective propagators represented by dispersion integrals. The result is a semi-analytical bridge between purely analytic amplitude reductions and purely numerical integration techniques.

The dispersive framework preserves key analytic properties --- threshold behavior, unitarity cuts, and gauge invariance --- while reducing computational demands.  It is especially attractive for low-energy observables where delicate cancellations between diagrams require high numerical precision. Furthermore, real-experiment implementation often involves acceptance and energy-threshold cuts that significantly affect radiative corrections; the dispersive formulation allows these to be incorporated naturally at the numerical-integration stage.  An example of an application of the dispersive approach to multi-loop ``banana" integrals can be found in Ref.~\cite{ChYCh}.  Another example is the precision studies of the Higgs boson at future colliders, which requires the inclusion of next-to-next-to-leading-order (NNLO) electroweak corrections.  In \citep{FR1a},  the planar and non-planar double-box topologies with multiple massive propagators in the loops are computed numerically by transforming one of the sub-loops using Feynman parametrization and a dispersion representation.  This was followed by \citep{FR1b},  which provides a complete calculation of the NNLO electroweak corrections involving closed fermion loops. 

Parallel developments in the phenomenological sector culminated in partial two-loop electroweak predictions for key observables. 
In Refs.~\citep{Dubovyk1} and \citep{Freitas1a}, the full set of fermionic and bosonic two-loop corrections to Z-boson observables was produced, while \citep{Erler} addressed hadronic effects in M{\o}ller scattering at NNLO.  More recently,  \citep{Schw} achieved the analytic evaluation of electroweak double-box integrals relevant for M{\o}ller processes. Other important developments can be found in Refs.  \citep{Freitas1}-\citep{FeyC3}.

These advances collectively establish a robust infrastructure for high-precision electroweak phenomenology.  The comprehensive review on updated measurements and higher-order theoretical corrections are available in \citep{Erler1} and \citep{PDG2024}.
Despite impressive progress, complete NNLO electroweak results remain available only for selected processes due to the technical difficulty of two-loop calculations. Closed-fermion-loop NNLO corrections have been achieved for several key observables, but the general problem of fully automated two-loop amplitude generation, reduction, and evaluation remains open. Semi-numerical strategies, such as the dispersive and differential-equation methods, have proven especially effective, balancing analytical control with numerical tractability.
The growing complexity of precision calculations arises from the need to handle diagrams with both more external legs and additional loop orders.  Each loop introduces new mass and momentum scales, overlapping ultraviolet and infrared divergences, and complicated threshold structures, dramatically increasing algebraic and numerical challenges. At one loop, PV-style analytic reductions allow tensor integrals to be expressed through a small set of scalar functions, often in closed form. At two loops and beyond, however, the explosion of topologies and kinematic invariants typically renders complete analytic evaluation impractical.
To address these challenges, the dispersive method reformulates a multi-loop integral as a sequence of nested one-loop integrals via spectral representations. This effectively transforms a two-loop problem into integrals over well-behaved spectral densities, isolating singular behavior in analytically controlled functions that can be integrated numerically with high stability. By combining dimensional shifting, recurrence relations, and dispersion theory, the approach achieves algebraic reduction to a minimal set of independent integrals while maintaining analytic transparency.

The present study extends our previous dispersive framework by incorporating shifted-dimension tensor decomposition and dimension-lowering recurrence relations directly into the dispersive representation. These relations, originally formulated within Tarasov dimension-recursion approach, are adapted here to operate on spectral integrals, leading to a minimal set of independent dispersive building blocks. This algebraic reduction significantly decreases computational time while maintaining high numerical precision.
We demonstrate the applicability of this formalism to one-loop self-energy, triangle, and box diagrams, as well as a two-loop example in which one-loop sub-block is represented through corresponding dispersive integral. The results confirm that the dispersive-recurrence combination provides a  stable and robust framework for the precision electroweak calculations.
This methodology therefore represents a major step toward an automated, high-precision, \textit{ab initio} framework for two-loop quantum-field-theory calculations---one capable of supporting the next generation of parity-violation and flavour-physics experiments, including MOLLER, P2, Belle II, and the future programs at EIC. 

The paper is organized as follows. In Sec.  II, we describe the methods involved in our study, including the tensor decomposition, recurrence relations, the momentum expansion and the dispersion approach. In Sec.  III, we present a numerical comparison of the one-loop three- and four-point functions calculated using the described techniques against results from the Collier numerical library. In Sec.  IV we discuss the application of the presented techniques to two-loop calculations and compare our numerical results with those obtained using other available approaches. Finally, in Sec.  V, we summarize the main ideas and results.

\section{Methodology}

\subsection{Tensor decomposition}

\begin{figure}
\includegraphics[width=0.6\textwidth]{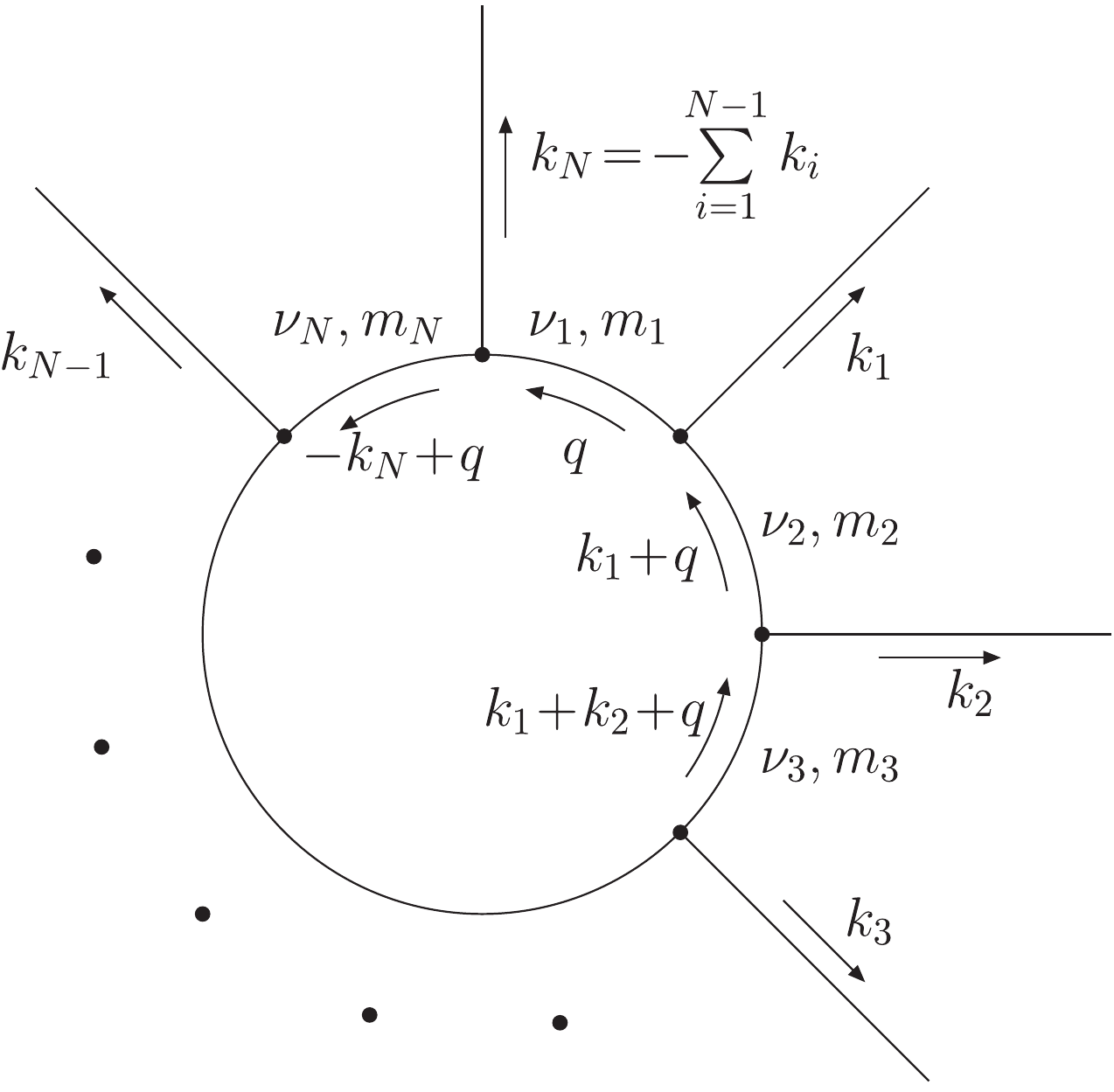}
\caption{The one-loop $N$-point diagram in the notation
corresponding to tensors $T^{(N)}_{\mu_1\ldots\mu_M}$}
\label{fig1k}
\end{figure}

The tensor $N$-point function of an arbitrary rank $M$ is defined as (see Fig.~\ref{fig1k})
\begin{eqnarray}
\label{tensor_T_1_N}
T^{(N)}_{\mu_1\ldots\mu_M} &\!\!=\!\!& 
\frac{\mu^{4-D}\; e^{\gamma_E (4-D)/2}}{{\rm i}\pi^{D/2}}
\int{\rm d}^Dq\; 
\nonumber \\ && \times
\frac{q_{\mu_1}\ldots q_{\mu_M}}
{\left[q^2-m_1^2\right] \left[(k_1\!+\!q)^2-m_2^2\right] 
\left[(k_1\!+\!k_2\!+\!q)^2-m_3^2\right] \ldots
\left[(k_1\!+\ldots+\!k_{N-1}\!+\!q)^2-m_N^2\right]
} , \hspace*{5mm}
\end{eqnarray}
where $D=4-2\varepsilon$ is the space-time dimension ($\varepsilon$ is the standard dimensional regularization parameter) and $\mu$ is the mass scale. $T^{(N)}_{\mu_1\ldots\mu_M}$ can be decomposed in terms of the scalar Passarino-Veltman functions 
$Z_{0\ldots 0 1\ldots 1 2\ldots 2\ldots (N-1)\ldots(N-1)}$ 
as (see, e.g., in Ref.~\cite{Hahn})
\begin{eqnarray}
\label{tensor_T_2_N}
T^{(N)}_{\mu_1\ldots\mu_M} &=&
\mathop{\sum_{l, n_1, n_2}}_{2l+n_1+n_2+\ldots+n_{N-1}=M}
\left\{ [g]^l [k_1]^{n_1} [k_1+k_2]^{n_2}
\ldots [k_1+k_2+\ldots+k_{N-1}]^{n_{N-1}}
\right\}_{\mu_1\ldots\mu_M}\;
\nonumber \\ && \hspace*{35mm} \times
Z_{\underset{2l}{\underbrace{\scriptstyle{0...0}}}\;
\underset{n_1}{\underbrace{\scriptstyle{1...1}}}\;
\underset{n_2}{\underbrace{\scriptstyle{2...2}}}\;
\ldots
\underset{n_{N-1}}{\underbrace{\scriptstyle{(N-1)...(N-1)}}}\;
}\; ,
\end{eqnarray}
where 
$\left\{ [g]^l [k_1]^{n_1} [k_1+k_2]^{n_2}
\ldots [k_1+k_2+\ldots+k_{N-1}]^{n_{N-1}}
\right\}_{\mu_1\ldots\mu_M}$ 
is the symmetrized tensor structure containing $l$ metric tensors $g$, $n_1$ vectors $k_1$, $n_2$ vectors $k_1+k_2$, $\ldots$, and $n_{N-1}$ vectors $k_1+\ldots+k_{N-1}$ ($2l+n_1+n_2+\ldots+n_{N-1}=M$).

\begin{figure}
\includegraphics[width=0.65\textwidth]{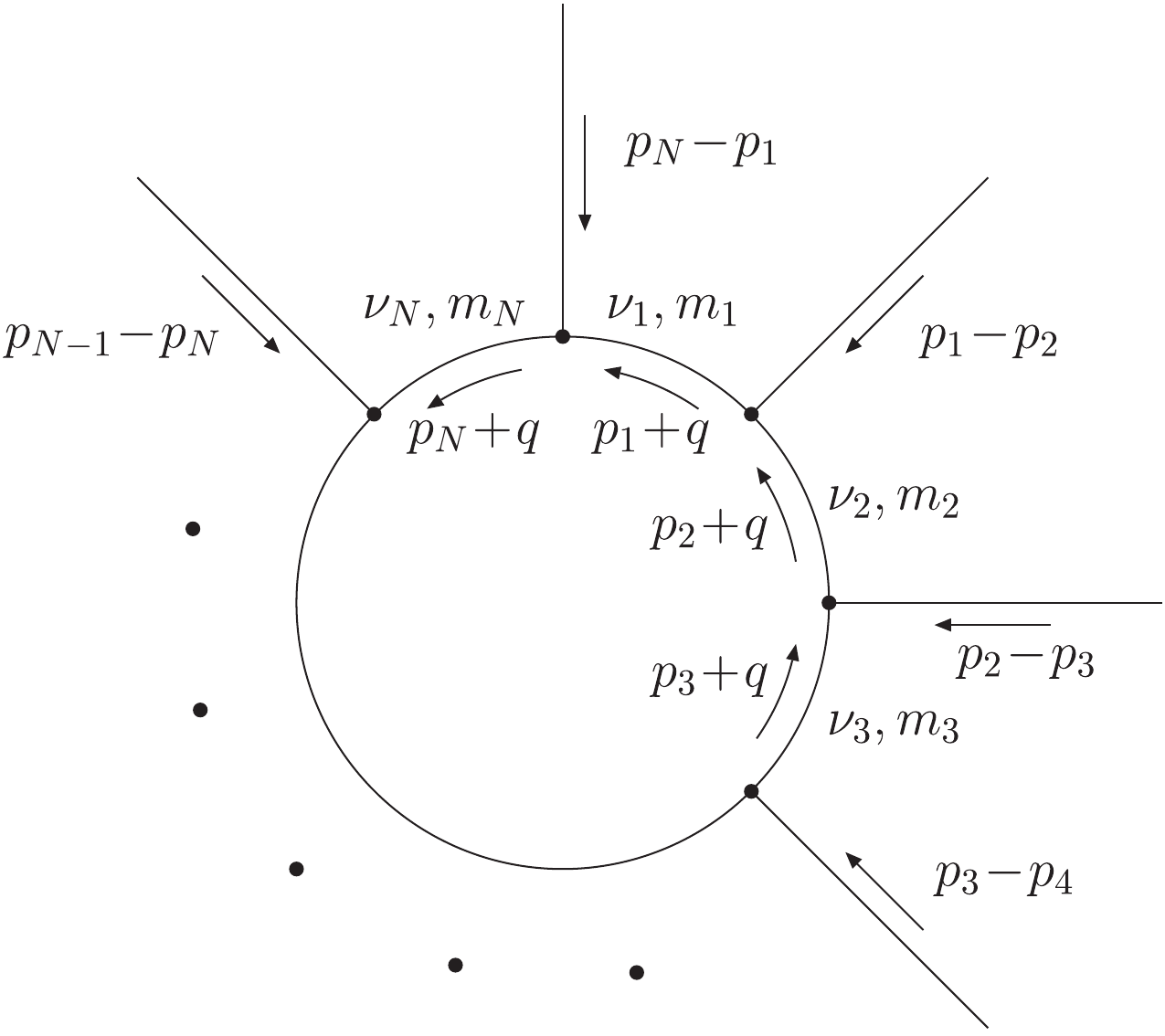}
\caption{The one-loop $N$-point diagram in the notation
corresponding to tensors $J^{(N)}_{\mu_1\ldots\mu_M}$}
\label{fig2p}
\end{figure}

To be able to employ the tensor decomposition approach developed in  Ref.~\cite{PLB91}, let us define the scalar $N$-point integral as (see Fig.~\ref{fig2p})
\begin{equation}
\label{JN_scalar_def}
J^{(N)}(D;\nu_1,\nu_2,\ldots,\nu_N) = 
\int \frac{{\rm d}^D q}
{\left[(p_1+q)^2-m_1^2\right]^{\nu_1} 
\left[(p_2+q)^2-m_2^2\right]^{\nu_2} \ldots
\left[(p_N+q)^2-m_N^2\right]^{\nu_N}} \;
\end{equation}
(it is clear that this scalar integral depends only on the squared momenta $(p_i-p_j)^2$ with $i<j<N$).  Similarly, the $N$-point tensor integral of an arbitrary rank $M$ is defined as
\begin{equation}
\label{JN_tensor_def}
J^{(N)}_{\mu_1\ldots\mu_M}(D;\nu_1,\nu_2,\ldots, \nu_N) = 
\int{\rm d}^Dq\; \frac{q_{\mu_1}\ldots q_{\mu_M}}
{\left[(p_1+q)^2-m_1^2\right]^{\nu_1} 
\left[(p_2+q)^2-m_2^2\right]^{\nu_2} \ldots
\left[(p_N+q)^2-m_N^2\right]^{\nu_3}} \; .
\end{equation}
Then the general tensor decomposition formula (see Eq.~(11) of~\cite{PLB91}) yields
\begin{eqnarray}
\label{tensor_J_1_N}
J^{(N)}_{\mu_1\ldots\mu_M}(D;\nu_1,\nu_2,\ldots \nu_N) 
&\!\!\!\!=\!\!\!\!&
\mathop{\sum_{\lambda, \kappa_1, \ldots, \kappa_N}}_{2\lambda+\kappa_1+\ldots+\kappa_N=M} \!\!\!\!\!
\left(-\frac{1}{2}\right)^{\lambda} \;
%(\nu_1)_{\kappa_1} (\nu_2)_{\kappa_2}\ldots (\nu_N)_{\kappa_N}
\left(\prod\limits_{i=1}^N (\nu_i)_{\kappa_i}\right)
\left\{ [g]^\lambda [p_1]^{\kappa_1} [p_2]^{\kappa_2}\ldots [p_N]^{\kappa_N}\right\}_{\mu_1\ldots\mu_M}\;
\hspace*{-5mm}
\nonumber \\
&& \times 
\pi^{\lambda-M}\; J^{(N)}\left(D+2(M-\lambda); \nu_1+\kappa_1, \nu_2+\kappa_2, \ldots, \nu_N+\kappa_N\right) \; , 
\hspace*{5mm}
\end{eqnarray}
where $(\nu)_{\kappa}\equiv\Gamma(\nu+\kappa)/\Gamma(\nu)$ is the Pochhammer symbol, and the scalar integrals $J^{(N)}$ occurring on the rhs have shifted space-time dimension value $D+2(M-\lambda)$. 

Comparing the definitions of $T^{(N)}_{\mu_1\ldots\mu_M}$ (given in Eq.~(\ref{tensor_T_1_N})) and $J^{(N)}_{\mu_1\ldots\mu_M}$ (given in Eq.~(\ref{JN_tensor_def})) we get the following connection formula:
\begin{equation}
T^{(N)}_{\mu_1\ldots\mu_M} = 
\frac{\mu^{4-D} e^{\gamma_E (4-D)/2}}{{\rm i}\pi^{D/2}}\; 
J^{(N)}_{\mu_1\ldots\mu_M}(D;1,\ldots,1)\Bigr|_{p_1=0,\; p_2=k_1,\ldots , p_N=k_1+\ldots + k_{N-1}}\; .
\end{equation}
Considering Eq.~(\ref{tensor_J_1_N}) in the case $p_1=0,\; p_2=k_1,\ldots , p_N=k_1+\ldots+k_{N-1}$ we see that only the term with $\kappa_1=0$ contributes,
\begin{eqnarray}
\label{tensor_J_2_N}
&&J^{(N)}_{\mu_1\ldots\mu_M}(D;\nu_1,\nu_2,\ldots \nu_N)\Bigr|_{p_1=0,\; p_2=k_1,\ldots , p_N=k_1+\ldots+k_{N-1}} 
\nonumber \\
&& \hspace*{5mm} =
\mathop{\sum_{\lambda, \kappa_2, \ldots, \kappa_N}}_{2\lambda+\kappa_2+\ldots+\kappa_N=M} \!\!\!\!
\left(-\frac{1}{2}\right)^{\lambda}\!
\left(\prod\limits_{i=2}^N (\nu_i)_{\kappa_i}\!\right)
\left\{ [g]^\lambda [k_1]^{\kappa_2} [k_1\!+\!k_2]^{\kappa_3}\ldots 
[k_1\!+\ldots +\!k_{N-1}]^{\kappa_N}\right\}_{\mu_1\ldots\mu_M}\;
\nonumber \\ 
&& \hspace*{10mm}\times 
\pi^{\lambda-M}\; J^{(N)}\left(D+2(M\!-\!\lambda); 
\nu_1, \nu_2+\kappa_2, \ldots, \nu_N+\kappa_N\right)\Bigr|_{p_1=0,\; p_2=k_1,\ldots , p_N=k_1+\ldots+k_{N-1}} . 
\hspace*{5mm}
\end{eqnarray}

To compare with $T^{(N)}_{\mu_1\ldots\mu_M}$ we need to put $\nu_1=\nu_2=\ldots =\nu_N=1$, $\lambda=l$, $\kappa_2=n_1$, $\kappa_3=n_2$, $\ldots$,
$\kappa_N=n_{N-1}$,
\begin{eqnarray}
\label{tensor_J_J_3a_N}
&& \hspace*{-5mm}
J^{(N)}_{\mu_1\ldots\mu_M}(D;1,\ldots,1)\Bigr|_{p_1=0,\; p_2=k_1,\ldots , p_N=k_1+\ldots+k_{N-1}} 
\nonumber \\
&& =
\mathop{\sum_{l, n_1, \ldots, n_{N-1}}}_{2l+n_1+\ldots +n_{N-1}=M} 
\left(-\frac{1}{2}\right)^l\;
\left(\prod\limits_{i=1}^{N-1} {n_i}!\right)
\left\{ [g]^l [k_1]^{n_1} [k_1\!+\!k_2]^{n_2}\ldots 
[k_1\!+\ldots +\!k_{N-1}]^{n_{N-1}}\right\}_{\mu_1\ldots\mu_M}\;
\nonumber \\ 
%&& \hspace*{5mm}%\times 
%\pi^{-l-n_1-\ldots -n_{N-1}}\; 
%\nonumber 
%\\ 
&& \hspace*{5mm}\times 
\pi^{-l-n_1-\ldots -n_{N-1}} J^{(N)}\left(D\!+\!2l\!+\!2n_1+\ldots +\!2n_{N-1}; 
1, 1\!+\!n_1,\ldots, 1\!+\!n_{N-1}\right)\Bigr|_{p_1=0,\; p_2=k_1,\ldots , p_N=k_1+\ldots+k_{N-1}} . 
\hspace*{-20mm}
\nonumber \\
&& {\hspace*{5mm}}
\end{eqnarray}

In this way, we arrive at
\begin{eqnarray}
\label{Z_to_J_mapping_2}
&& Z_{\underset{2l}{\underbrace{\scriptstyle{0...0}}}\;
\underset{n_1}{\underbrace{\scriptstyle{1...1}}}\;
\underset{n_2}{\underbrace{\scriptstyle{2...2}}}\;
\ldots
\underset{n_{N-1}}{\underbrace{\scriptstyle{(N-1)...(N-1)}}}}
\nonumber \\ && \hspace*{5mm}
= \frac{\mu^{4-D} e^{\gamma_E (4-D)/2}}
{{\rm i}\pi^{l+n_1+\ldots +n_{N-1}+D/2}}
\left(\prod\limits_{i=1}^{N-1} {n_i}! \right)
\left(-\frac{1}{2}\right)^l
\nonumber \\ && \hspace*{10mm} \times
J^{(N)}\left(D\!+\!2l\!+\!2n_1+\ldots +\!2n_{N-1}; 
1, 1\!+\!n_1,\ldots, 1\!+\!n_{N-1}\right)\Bigr|_{p_1=0,\; p_2=k_1,\ldots , p_N=k_1+\ldots+k_{N-1}} . 
\hspace*{5mm}
\end{eqnarray}
To be consistent with the standard Passarino-Veltman notations, for $N=2,3,4,5,\ldots$ the notation $Z$ should be replaced by
$B$, $C$, $D$, $E$, etc. 

%============================

For the two-point case ($N=2$) we get
%In this way, we arrive at
\begin{equation}
\label{B_to_J_mapping_2}
B_{\stackrel{\underbrace{\scriptstyle{0\ldots 0}}}{2l}\; \stackrel{\underbrace{\scriptstyle{1\ldots 1}}}{n}}
= \frac{\mu^{4-D} e^{\gamma_E (4-D)/2} n!}
{{\rm i}\pi^{l+n+D/2}}
\left(-\frac{1}{2}\right)^l
J^{(2)}\left(D+2l+2n; 1, 1+n\right)\Bigr|_{p_1=0,\; p_2=k_1} \; . 
\end{equation}
If we have only one external momentum we will usually suppress its
index, $k_1=k$. 

%============================

For the three-point case ($N=3$) we get
%In this way, we arrive at
\begin{eqnarray}
\label{C_to_J_mapping_2}
%C_{\stackrel{\underbrace{\scriptstyle{0\ldots 0}}}{2l}\; 
%\stackrel{\underbrace{\scriptstyle{1\ldots 1}}}{n_1}\;
%\stackrel{\underbrace{\scriptstyle{2\ldots 2}}}{n_2}}
C_{\underset{2l}{\underbrace{\scriptstyle{0...0}}}\;
\underset{n_1}{\underbrace{\scriptstyle{1...1}}}\;
\underset{n_2}{\underbrace{\scriptstyle{2...2}}}}
&=& \frac{\mu^{4-D} e^{\gamma_E (4-D)/2} n_1! n_2!}
{{\rm i}\pi^{l+n_1+n_2+D/2}}
\left(-\frac{1}{2}\right)^l
\nonumber \\ && \times
J^{(3)}\left(D+2l+2n_1+2n_2; 1, 1+n_1, 1+n_2\right)\Bigr|_{p_1=0,\; p_2=k_1,\; p_3=k_1+k_2} \; ,
\end{eqnarray}
where the scalar integral $J^{(3)}$ on the rhs depends on the following momentum invariants: $(p_1-p_2)^2=k_1^2$ (for the incoming momentum opposite to the line with $m_3$), $(p_2-p_3)^2=k_2^2$ (for the incoming momentum opposite to the line with $m_1$), and $(p_3-p_1)^2=(k_1+k_2)^2$ (for the incoming momentum opposite to the line with $m_2$).

In the occurring three-point integral, we can combine any pair of denominators by using the Feynman parametrization trick, e.g.,
\begin{eqnarray}
\label{Fp3}
&& \hspace*{-5mm}
J^{(3)}\left(D+2l+2n_1+2n_2; 1, 1+n_1, 1+n_2\right)\Bigr|_{p_1=0,\; p_2=k_1,\; p_3=k_1+k_2}
\nonumber \\
&& = \int \frac{{\rm d}^{D+2l+2n_1+2n_2}q}
{\left[q^2-m_1^2\right] \left[(k_1+q)^2-m_2^2\right]^{1+n_1} 
\left[(k_1+k_2+q)^2-m_3^2\right]^{1+n_2}}
\nonumber \\
&& = \frac{(n_1+n_2+1)!}{n_1! n_2!}
\int\limits_0^1 {\rm d}x \; x^{n_1} \bar{x}^{n_2}
\int \frac{{\rm d}^{D+2l+2n_1+2n_2}q}
{\left[q^2-m_1^2\right] 
\left[(k_1+\bar{x}k_2+q)^2-x m_2^2-\bar{x}m_3^2+x\bar{x}k_2^2\right]^{2+n_1+n_2}}
\nonumber \\
&& = \frac{(n_1\!+\!n_2\!+\!1)!}{n_1! n_2!}
\int\limits_0^1 {\rm d}x \; x^{n_1} \bar{x}^{n_2}
J^{(2)}(D+2l+2n_1+2n_2; 1, 2+n_1+n_2)\Bigr|_{\footnotesize{\begin{array}{l} p_1=0,\; p_2=k_1+\bar{x}k_2, \\[-3mm] m_2^2 \leftrightarrow x m_2^2\!+\!\bar{x}m_3^2\!-\!x\bar{x}k_2^2\end{array}}} \; , 
%J^{(2)}(D+2l+2n_1+2n_2; 1, 2+n_1+n_2)\Bigr|_{p_1=0,\; p_2=k_1+\bar{x}%k_2, \; m_2^2 \leftrightarrow x m_2^2+\bar{x}m_3^2-x\bar{x}k_2^2} \; , 
\end{eqnarray}
where $\bar{x}=1-x$.

%============================

For the four-point case ($N=4$), we get
\begin{eqnarray}
\label{D_to_J_mapping_2}
D_{\underset{2l}{\underbrace{\scriptstyle{0...0}}}\;
\underset{n_1}{\underbrace{\scriptstyle{1...1}}}\;
\underset{n_2}{\underbrace{\scriptstyle{2...2}}}\;
\underset{n_3}{\underbrace{\scriptstyle{3...3}}}}
&=& \frac{\mu^{4-D} e^{\gamma_E (4-D)/2} n_1! n_2! n_3!}
{{\rm i}\pi^{l+n_1+n_2+n_3+D/2}}
\left(-\frac{1}{2}\right)^l
\nonumber \\ && \hspace*{-40mm} \times
J^{(4)}\left(D\!+\!2l\!+\!2n_1\!+\!2n_2\!+\!2n_3; 
1, 1\!+\!n_1, 1\!+\!n_2, 1\!+\!n_3\right)\Bigr|_{p_1=0,\; p_2=k_1,\; p_3=k_1+k_2,\; p_4=k_1+k_2+k_3} \; . \hspace*{5mm}
\end{eqnarray}

In the occurring four-point integral we can combine any triple of denominators by using the Feynman parametrization trick, e.g.,
\begin{eqnarray}
\label{Fp4}
&& \hspace*{-5mm}
J^{(4)}\left(D+2l+2n_1+2n_2+2n_3; 1, 1+n_1, 1+n_2, 1+n_3\right)\Bigr|_{p_1=0,\; p_2=k_1,\; p_3=k_1+k_2,\; p_4=k_1+k_2+k_3}
\nonumber \\
&& = \!\int\! \frac{{\rm d}^{D+2l+2n_1+2n_2+2n_3}q}
{\left[q^2-m_1^2\right] \left[(k_1+q)^2-m_2^2\right]^{1+n_1} 
\left[(k_1+k_2+q)^2-m_3^2\right]^{1+n_2}
\left[(k_1+k_2+k_3+q)^2-m_4^2\right]^{1+n_3}}
\nonumber \\
&& = \frac{(n_1+n_2+n_3+2)!}{n_1! n_2! n_3!}
\int\limits_0^1 \int\limits_0^1
{\rm d}x \; {\rm d}y \; x^{n_1} \bar{x}^{n_2+n_3+1}
y^{n_2} \bar{y}^{n_3}
\nonumber \\
&& \hspace*{3mm} \times
\!\int\! \frac{{\rm d}^{D+2l+2n_1+2n_2+2n_3}q}
{\left[q^2\!-\!m_1^2\right] 
\left[(k_1\!+\!\bar{x}k_2\!+\!\bar{x}\bar{y}k_3\!+\!q)^2
\!-\!x m_2^2\!-\!\bar{x}y m_3^2\!-\!\bar{x}\bar{y} m_4^2
\!+\!x\bar{x}(k_2\!+\!\bar{y}k_3)^2\!+\!\bar{x}y\bar{y}k_3^2
\right]^{3+n_1+n_2+n_3}}
\nonumber \\
&& = \frac{(n_1+n_2+n_3+2)!}{n_1! n_2! n_3!}
\int\limits_0^1 \int\limits_0^1
{\rm d}x \; {\rm d}y \; x^{n_1} \bar{x}^{n_2+n_3+1}
y^{n_2} \bar{y}^{n_3}
\nonumber \\
&& \hspace*{3mm} \times
J^{(2)}(D\!+\!2l\!+\!2n_1\!+\!2n_2\!+\!2n_3; 
1, 3\!+\!n_1\!+\!n_2\!+\!n_3)\Bigr|_{\footnotesize{\begin{array}{l} 
p_1=0,\; p_2=k_1+\bar{x}k_2+ \bar{x}\bar{y}k_3,\\[-3mm] 
m_2^2 \leftrightarrow x m_2^2\!+\!\bar{x}y m_3^2
\!+\!\bar{x}\bar{y}m_4^2
\!-\!x\bar{x}(k_2+\bar{y}k_3)^2
\!-\!\bar{x}y\bar{y}k_3^2\end{array}}}\; , 
\nonumber \\
&& \qquad
\end{eqnarray}
where $\bar{x}=1-x$ and $\bar{y}=1-y$.
%For the two-point integral we can use the same recurrence relations 
%as before. 
%============================

\subsection{Recurrence relations and the momentum expansion}
%============================

In the two-point case, according to Eq.~(\ref{B_to_J_mapping_2}), we need to calculate the integrals $J^{(2)}\left(D+2l+2n; 1, 1+n\right)$ with $l\geq 0$ and $n\geq 0$. 
Using Feynman parameters we can express higher functions in terms of
the integrals $J^{(2)}$. In the three- and four-point cases, according to Eqs.~(\ref{Fp3}) and (\ref{Fp4}), we need to calculate 
the integrals  
$J^{(2)}(D+2l+2n_1+2n_2; 1, 2+n_1+n_2)$ and
$J^{(2)}(D+2l+2n_1+2n_2+2n_3; 1, 3+n_1+n_2+n_3)$, respectively. 
In general, for an $N$-point function, we would need to deal with
the integrals 
\begin{equation}
\label{before_rec}
J^{(2)}(D+2l+2n; 1, N-1+n),
\quad {\rm with} \quad n=n_1+\ldots n_{N-1} \; .
\end{equation}

To decrease the index (power of propagator) $N-1+n$, 
we can use the recurrence relation~(\ref{tar98b}) for the case 
$\nu_1=1$,
\begin{eqnarray}
\label{tar98b_1}
J^{(2)}(D+2;1,\nu_2+1) &=&
-\frac{\pi}{2 \nu_2 k^2}
\left[
(k^2+m_1^2-m_2^2) J^{(2)}(D;1,\nu_2)
\right.
\nonumber \\ && \hspace*{16mm}
\left.
- J^{(2)}(D;1,\nu_2-1)
+ J^{(2)}(D;0,\nu_2)
\right].
\end{eqnarray}
Whenever one of the indices on the rhs becomes zero (like, e.g., in $J^{(2)}(D;0,\nu_2)$), this is a tadpole integral which can be expressed in terms of $J^{(2)}(D;0,1)$ or $J^{(2)}(D;0,1)$ using Eqs.~(\ref{tadpole3}) and (\ref{tadpole4}). After that, for the remaining integrals $J^{(2)}\left(D+2l; 1, 1\right)$ we can use the recurrence relation~(\ref{d_red}) for the case $\nu_1=\nu_2=1$,
\begin{eqnarray}
\label{d_red_11}
J^{(2)}(D+2;1,1) &\!\!\!=\!\!\!& -\frac{\pi}{2k^2 (D-1)}
\left[
\Delta J^{(2)}(D;1,1)
\right.
\nonumber \\ &&
\left.
+(k^2+m_1^2-m_2^2) J^{(2)}(D;1,0)
+(k^2-m_1^2+m_2^2) J^{(2)}(D;0,1)
\right]  \; ,
\hspace*{5mm}
\end{eqnarray}
with
\begin{equation}
\label{Delta}
\Delta\equiv
\Delta(m_1^2, m_2^2, k^2) 
= -\lambda(m_1^2, m_2^2, k^2)
= 4 m_1^2 m_2^2 - (k^2-m_1^2-m_2^2)^2 \; ,
\end{equation}
where $\lambda(m_1^2, m_2^2, k^2)$ is the standard notation 
for the K\"all\'en function (other representations of $\Delta$ are collected in Eq.~(\ref{Delta2})).

Another way to deal with the integrals~(\ref{before_rec}) is to use recurrence relations (\ref{ibp1}) and (\ref{ibp2}). In this way, we can  bring them to the integrals $J^{(2)}(D+2l+2n; 1, 1)$ (with the same value of the space-time dimension as the original ones) plus tadpoles, and then apply Eq.~(\ref{d_red_11}) as many times as needed. However, this option would involve more steps, and it would produce very cumbersome intermediate expressions because of the presence of $\Delta$ in the denominators of (\ref{ibp1}) and (\ref{ibp2}). We found the way based on Eqs.~(\ref{tar98b_1}) and (\ref{d_red_11}) to be more efficient.  

Let us first consider the two- and three-point cases (we will discuss the higher cases later). In the two-point case ($N=2$), using $n$ times Eq.~(\ref{tar98b_1}) and $l$ times Eq.~(\ref{d_red_11}), we reduce $J^{(2)}\left(D+2l+2n; 1, 1+n\right)$ to $J^{(2)}(D;1,1)$ plus tadpoles. In the three-point case ($N=3$), using $n+1$ times Eq.~(\ref{tar98b_1}) and $l$ times Eq.~(\ref{d_red_11}), we reduce $J^{(2)}\left(D+2l+2n; 1, 2+n\right)$ to $J^{(2)}(D-2;1,1)$ plus tadpoles. If we want to use the same basis for the two-point integrals, we need to apply Eq.~(\ref{d_red_11}) one more time (shifting $D\to D-2$).

In this way, in the two- and three-point cases, starting from the integrals~(\ref{before_rec}) with $D=4-2\varepsilon$ we bring them to the basis of $(2-2\varepsilon)$-dimensional integrals. Namely, by using recurrence relations with respect to the powers of propagators $\nu_i$ and the space-time dimension $D$, we can express all the relevant integrals $J^{(2)}\left(4-2\varepsilon+2l+2n; 1, N-1+n\right)$ (with $N=2$ and $N=3$) in terms of the master integral $J^{(2)}(2-2\varepsilon;1,1)$, as well as the tadpoles 
\begin{equation}
\label{2-2ep_tadpole_integrals}
J^{(2)}(2-2\varepsilon;1,0) = -{\rm i} \pi^{1-\varepsilon}\;
\Gamma(\varepsilon)\; (m_1^2)^{-\varepsilon}
\quad {\rm and} \quad
J^{(2)}(2-2\varepsilon;0,1) = -{\rm i} \pi^{1-\varepsilon}\;
\Gamma(\varepsilon)\; (m_2^2)^{-\varepsilon}
\end{equation}
(see Appendix~A for more details). Analytical results for the integral $J^{(2)}(2-2\varepsilon;1,1)$ (including the relevant terms of the $\varepsilon$-expansion) are collected in Appendix~B. This procedure provides analytical results for all required Passarino-Veltman 
functions~(\ref{B_to_J_mapping_2}).
% occurring on the r.h.s. of Eqs.~(\ref{tensor_T_2}) and 
% (\ref{tensor_T_2_3}).

When using the recurrence relations~(\ref{tar98b_1}) and (\ref{d_red_11}) we are getting powers of $k^2$ in the denominator. In particular, when reducing $J^{(2)}\left(4-2\varepsilon+2l+2n; 1, N-1+n\right)$ (with $N=2$ or $N=3$) to $J^{(2)}(2-2\varepsilon;1,1)$ and tadpoles~(\ref{2-2ep_tadpole_integrals}) the maximal power is $(k^2)^{l+n+1}$, i.e.,
\begin{eqnarray}
\label{2-2*ep_basis_1a}
J^{(2)}\left(4\!-\!2\varepsilon\!+\!2l\!+\!2n; 1, N\!-\!1\!+\!n\right)
&=& 
\frac{\pi^{l+n+1}}{(k^2)^{l+n+1}}
\Bigl[
R_{N,l,n}^{(1,1)}(m_1, m_2, k^2, \varepsilon) 
J^{(2)}\left(2-2\varepsilon; 1, 1\right)
\nonumber \\ && \hspace*{17mm}
+ R_{N,l,n}^{(1,0)}(m_1, m_2, k^2, \varepsilon) J^{(2)}\left(2-2\varepsilon; 1, 0\right)
\nonumber \\ && \hspace*{17mm}
+ R_{N,l,n}^{(0,1)}(m_1, m_2, k^2, \varepsilon) 
J^{(2)}\left(2-2\varepsilon; 0, 1\right)
\Bigr]
\; , \hspace*{8mm}
\end{eqnarray}
where $R_{N,n,l}^{(1,1)}$, $R_{N,n,l}^{(1,0)}$ and $R_{N,n,l}^{(0,1)}$ are algebraic coefficients which are polynomial in $k^2$.

To make sure that the resulting expression~(\ref{2-2*ep_basis_1a}) is not singular as $k^2\to 0$, let us employ the small momentum expansion of the integral $J^{(2)}(2-2\varepsilon;1,1)$. According to Eq.~(\ref{J2_F4dd}), the terms of the small-$k^2$ expansion of $J^{(2)}(2-2\varepsilon;1,1)$ up to $(k^2)^{j_0}$ can be presented as
\begin{eqnarray}
\label{J2_F4dd_}
J_{[j_0]}^{(2)}(2-2\varepsilon;1,1) &\!\!=\!\!& 
\sum\limits_{j=0}^{j_0}
(k^2)^j\; \frac{(1+\varepsilon)_j}{(m_2^2-m_1^2)^{1+2j}}
\nonumber \\ &&
\times \Biggl\{
J^{(2)}(2-2\varepsilon;0,1)\;
\sum\limits_{l=0}^j
\frac{j!}{l!(j-l)!}\;
\frac{(m_2^2)^l (m_1^2)^{j-l}}
{(1-\varepsilon)_l (1+\varepsilon)_{j-l}}\;
\nonumber \\ && \hspace*{5mm}
-J^{(2)}(2-2\varepsilon;1,0)\;
\sum\limits_{l=0}^j
\frac{j!}{l!(j-l)!}\;
\frac{(m_1^2)^l (m_2^2)^{j-l}}
{(1-\varepsilon)_l (1+\varepsilon)_{j-l}}\;
\Biggr\} \; , \hspace*{8mm}
\end{eqnarray}
so that
\begin{equation}
J_{[\infty]}^{(2)}(2-2\varepsilon;1,1)
= J^{(2)}(2-2\varepsilon;1,1) \; .
\end{equation}
If we subtract the expansion (\ref{J2_F4dd_}) from $J^{(2)}(2-2\varepsilon;1,1)$, the difference will be of the order $(k^2)^{j_0+1}$, and it can be presented as
\begin{equation}
\label{barJ}
J^{(2)}(2-2\varepsilon;1,1)
-J_{[j_0]}^{(2)}(2-2\varepsilon;1,1)
=(k^2)^{j_0+1} 
\bar{J}_{j_0+1}^{(2)}(2-2\varepsilon;1,1) \; .
\end{equation}

In our case, we need to put $j_0=n+l$. Using Eq.~(\ref{barJ}), we get
\begin{equation}
\label{barJ_}
J^{(2)}(2-2\varepsilon;1,1)
= J_{[n+l]}^{(2)}(2-2\varepsilon;1,1)
+ (k^2)^{n+l+1} 
\bar{J}_{n+l+1}^{(2)}(2-2\varepsilon;1,1) \; .
\end{equation}
Combining Eqs.~(\ref{2-2*ep_basis_1a}), (\ref{J2_F4dd_}) and (\ref{barJ_}) we get
\begin{eqnarray}
\label{2-2*ep_basis_1b}
J^{(2)}\left(4\!-\!2\varepsilon\!+\!2l\!+\!2n; 1, N\!-\!1\!+\!n\right)
&\!=\!& 
\frac{\pi^{l+n+1}}{(k^2)^{l+n+1}}
\Bigl[
R_{N,l,n}^{(1,1)}(m_1, m_2, k^2, \varepsilon) 
(k^2)^{n+l+1}
\bar{J}_{n+l+1}^{(2)}\left(2-2\varepsilon; 1, 1\right)
\hspace*{-8mm}
\nonumber \\ && \hspace*{17mm}
+ \widetilde{R}_{N,l,n}^{(1,0)}(m_1, m_2, k^2, \varepsilon) J^{(2)}\left(2-2\varepsilon; 1, 0\right)
\nonumber \\ && \hspace*{17mm}
+ \widetilde{R}_{N,l,n}^{(0,1)}(m_1, m_2, k^2, \varepsilon) 
J^{(2)}\left(2-2\varepsilon; 0, 1\right)
\Bigr]
\; , \hspace*{8mm}
\end{eqnarray}
where $\widetilde{R}_{N,l,n}^{(1,0)}$ and $\widetilde{R}_{N,l,n,}^{(0,1)}$ include $R_{N,l,n}^{(1,0)}$ and $R_{N,l,n}^{(0,1)}$ plus the polynomial (in $k^2$) contributions coming from $R_{N,l,n}^{(1,1)} J_{[n+l]}^{(2)}(2-2\varepsilon;1,1)$ (see Eq.~(\ref{J2_F4dd_})). The absence of singularities in $k^2$ means that in $\widetilde{R}_{N,l,n}^{(1,0)}$ and $\widetilde{R}_{N,l,n}^{(0,1)}$ all the powers of $k^2$ less than $n+l+1$ should cancel, so that
\begin{eqnarray}
\widetilde{R}_{N,l,n}^{(1,0)}(m_1, m_2, k^2, \varepsilon)
&=& (k^2)^{l+n+1}\bar{R}_{N,l,n}^{(1,0)}(m_1, m_2, k^2, \varepsilon),
\nonumber \\
\widetilde{R}_{N,l,n}^{(0,1)}(m_1, m_2, k^2, \varepsilon)
&=& (k^2)^{l+n+1}\bar{R}_{N,l,n}^{(0,1)}(m_1, m_2, k^2, \varepsilon),
\end{eqnarray}
where $\bar{R}_{N,l,n}^{(1,0)}$ and $\bar{R}_{N,l,n}^{(0,1)}$ are also polynomial in $k^2$. In this way, we arrive at
\begin{eqnarray}
\label{2-2*ep_basis_1c}
J^{(2)}\left(4-2\varepsilon+2l+2n; 1, N-1+n\right)
&=& 
\pi^{l+n+1}
\Bigl[
R_{N,l,n}^{(1,1)}(m_1, m_2, k^2, \varepsilon) 
\bar{J}_{l+n+1}^{(2)}\left(2-2\varepsilon; 1, 1\right)
\nonumber \\ && \hspace*{15mm}
+ \bar{R}_{N,l,n}^{(1,0)}(m_1, m_2, k^2, \varepsilon) J^{(2)}\left(2-2\varepsilon; 1, 0\right)
\nonumber \\ && \hspace*{15mm}
+ \bar{R}_{N,l,n}^{(0,1)}(m_1, m_2, k^2, \varepsilon) 
J^{(2)}\left(2-2\varepsilon; 0, 1\right)
\Bigr]
\; . \hspace*{8mm}
\end{eqnarray}

In particular, this yields the following result for the function (\ref{B_to_J_mapping_2}):
\begin{eqnarray}
\label{2-2*ep_basis_2}
B_{\stackrel{\underbrace{\scriptstyle{0\ldots 0}}}{2l}\; \stackrel{\underbrace{\scriptstyle{1\ldots 1}}}{n}}
&=& \frac{\mu^{2\varepsilon} e^{\gamma_E \varepsilon} n!}
{{\rm i}\pi^{1-\varepsilon}}
\left(-\frac{1}{2}\right)^l
\Bigl[
R_{2,l,n}^{(1,1)}(m_1, m_2, k^2, \varepsilon) 
\bar{J}_{l+n+1}^{(2)}\left(2-2\varepsilon; 1, 1\right)
\nonumber \\ && \hspace*{30mm}
+ \bar{R}_{2,l,n}^{(1,0)}(m_1, m_2, k^2, \varepsilon) J^{(2)}\left(2-2\varepsilon; 1, 0\right)
\nonumber \\ && \hspace*{30mm}
+ \bar{R}_{2,l,n}^{(0,1)}(m_1, m_2, k^2, \varepsilon) 
J^{(2)}\left(2-2\varepsilon; 0, 1\right)
\Bigr] \; .
\end{eqnarray}
Note that the coefficient functions $R_{N,l,n}^{(1,1)}$, $\bar{R}_{N,l,n}^{(1,0)}$ and $\bar{R}_{N,l,n}^{(0,1)}$ do not have poles in $\varepsilon$ because in the recurrence relations~(\ref{tar98b_1}) and (\ref{d_red_11}) the only $D$-dependent factor in the denominator is $(D-1)$ which would never produce $\varepsilon$ for even dimensions. 

We can split the function (\ref{2-2*ep_basis_2}) into two parts, the first one containing the 1-point (tadpole-like) integrals $J^{(2)}\left(2-2\varepsilon; 1, 0\right)$ and $J^{(2)}\left(2-2\varepsilon; 0, 1\right)$, and the second one involving the genuine (subtracted) 2-point integral $\bar{J}_{n+l+1}^{(2)}\left(2-2\varepsilon; 1, 1\right)$:
\begin{equation}
B_{\stackrel{\underbrace{\scriptstyle{0\ldots 0}}}{2l}\; \stackrel{\underbrace{\scriptstyle{1\ldots 1}}}{n}}
\equiv B_{\{2l, n\}}(k^2, m_1^2, m_2^2)
= B_{\{2l, n\}}^{{\rm 1-point}}(k^2, m_1^2, m_2^2)
+ B_{\{2l, n\}}^{{\rm 2-point}}(k^2, m_1^2, m_2^2) ,
\end{equation}
with
\begin{eqnarray}
\label{1pt_part}
B_{\{2l, n\}}^{{\rm 1-point}}(k^2, m_1^2, m_2^2)
&=& \frac{\mu^{2\varepsilon} e^{\gamma_E \varepsilon} n!}
{{\rm i}\pi^{1-\varepsilon}}
\left(-\frac{1}{2}\right)^l
\Bigl[
\bar{R}_{2,l,n}^{(1,0)}(m_1, m_2, k^2, \varepsilon) J^{(2)}\left(2-2\varepsilon; 1, 0\right)
\nonumber \\ && \hspace*{33mm}
+ \bar{R}_{2,l,n}^{(0,1)}(m_1, m_2, k^2, \varepsilon) 
J^{(2)}\left(2-2\varepsilon; 0, 1\right)
\Bigr] , \hspace*{-5mm}
\\
B_{\{2l, n\}}^{{\rm 2-point}}(k^2, m_1^2, m_2^2)
&=& \frac{\mu^{2\varepsilon} e^{\gamma_E \varepsilon} n!}
{{\rm i}\pi^{1-\varepsilon}}
\left(-\frac{1}{2}\right)^l
R_{2,l,n}^{(1,1)}(m_1, m_2, k^2, \varepsilon) 
\bar{J}_{n+l+1}^{(2)}\left(2-2\varepsilon; 1, 1\right) .
\end{eqnarray}
Note that all UV-singularities are in $B_{\{2l, n\}}^{{\rm 1-point}}(k^2, m_1^2, m_2^2)$, namely in the tadpole integrals (\ref{2-2ep_tadpole_integrals}), whereas
the term $B_{\{2l, n\}}^{{\rm 2-point}}(k^2, m_1^2, m_2^2)$ is UV-finite. 

For the four-point function, the situation is a bit more complicated.
Let us start from the integral~(\ref{before_rec}), 
$J^{(2)}(D+2l+2n; 1, 3+n)$, and use recurrence relations~(\ref{tar98b_1}) and (\ref{d_red_11}). If we stop the recurrence procedure 
when the space-time dimension becomes $D-2$ (i.e., $2-2\varepsilon$),
then among the remaining integrals we may have not only
$J^{(2)}(2-2\varepsilon;1,1)$, 
$J^{(2)}(2-2\varepsilon;1,0)$ and 
$J^{(2)}(2-2\varepsilon;0,1)$, but also
$J^{(2)}(2-2\varepsilon;1,2)$ (this happens at $l=0$). 
The integral $J^{(2)}(2-2\varepsilon;1,2)$ is not
independent: using the relation~(\ref{ibp2}) it can be expressed as
\begin{eqnarray}
\label{ibp2a}
J^{(2)}(D;1,2) &=& \frac{1}{\Delta}
\Bigl[ (D-3)(k^2+m_1^2-m_2^2) J^{(2)}(D;1,1)
\nonumber \\ &&
- (D-2) J^{(2)}(D;1,0) 
- \frac{(D-2)(k^2-m_1^2-m_2^2)}{2m_1^2} J^{(2)}(D;0,1)
\Bigr] \; .
\end{eqnarray}
If we use Eq.~(\ref{ibp2a}) for $J^{(2)}(2-2\varepsilon;1,2)$ we
would get for $N=4$ a representation similar 
to (\ref{2-2*ep_basis_1a}), but the occurring coefficient functions
$R_{4,l,n}^{(1,1)}$, etc., will not be polynomial in $k^2$, because of the presence of $\Delta$ (see Eq.~(\ref{Delta})) in their denominators.
In this way, we would get rather cumbersome expressions for the
higher-order Passarino-Veltman functions.

Another way is to keep the $J^{(2)}(2-2\varepsilon;1,2)$ contributions
as an extra term 
\begin{equation}
R_{N,l,n}^{(1,2)}(m_1,m_2,k^2,\varepsilon) J^{(2)}(2-2\varepsilon;1,2)
\end{equation}
in Eq.~(\ref{2-2*ep_basis_1a}), as well as in Eqs.~(\ref{2-2*ep_basis_1b}) and (\ref{2-2*ep_basis_1c}). For the small-$k^2$ expansion we can
use the derivative of Eq.~(\ref{J2_F4dd_}) with respect to $m^2$,
\begin{equation}
J_{[j_0]}^{(2)}(2-2\varepsilon;1,2) 
= \frac{\partial}{\partial m_2^2}
J_{[j_0]}^{(2)}(2-2\varepsilon;1,1) \; , 
\end{equation} 
which can be calculated automatically. In this way, for the four-point case we get the following decomposition:
\begin{eqnarray}
\label{2-2*ep_basis_1c_4}
J^{(2)}\left(4-2\varepsilon+2l+2n; 1, 3+n\right)
&=& 
\pi^{l+n+1}
\Bigl[
R_{4,l,n}^{(1,1)}(m_1, m_2, k^2, \varepsilon) 
\bar{J}_{l+n+1}^{(2)}\left(2-2\varepsilon; 1, 1\right)
\nonumber \\ && \hspace*{15mm}
+ R_{4,l,n}^{(1,2)}(m_1, m_2, k^2, \varepsilon) 
\bar{J}_{l+n+1}^{(2)}\left(2-2\varepsilon; 1, 2\right)
\nonumber \\ && \hspace*{15mm}
+ \bar{R}_{4,l,n}^{(1,0)}(m_1, m_2, k^2, \varepsilon) J^{(2)}\left(2-2\varepsilon; 1, 0\right)
\nonumber \\ && \hspace*{15mm}
+ \bar{R}_{4,l,n}^{(0,1)}(m_1, m_2, k^2, \varepsilon) 
J^{(2)}\left(2-2\varepsilon; 0, 1\right)
\Bigr]
\; . \hspace*{8mm}
\end{eqnarray}

In the same way, for the five-point function we would get 
in Eq.~(\ref{2-2*ep_basis_1a}) an extra
term involving $J^{(2)}(2-2\varepsilon;1,3)$, etc.

%============================

\subsection{The dispersion approach}
%============================

The subtracted integral $\bar{J}_{n+l+1}^{(2)}\left(2-2\varepsilon; 1, 1\right)$ is represented using the dispersive approach.  We chose this approach because it allows us to replace the one-loop insertion in the two-loop topology with an effective propagator $1/(s-k^2-\rm{i}0)$ integrated over an $s$ parameter.  Subtracting dispersively-represented series expansion of $J_{[n+l]}^{(2)}\left(2-2\varepsilon; 1, 1\right) $ around $k^2\rightarrow0$ from $ J^{(2)}\left(2-2\varepsilon; 1, 1\right) $ renders the dispersive one-loop insertion multiply subtracted and convergent.  Following Eq.~(\ref{barJ_}), we can write dispersively

\begin{eqnarray*}
(k^2)^{n+l+1}\bar{J}_{n+l+1}^{(2)}\left(2-2\varepsilon; 1, 1\right)
&=&
\frac{{\rm i}}{\pi}\;
\!\!\int\limits_{(m_1+m_2)^2}^{\infty}\!\!\!\! {\rm d}s\;
{\rm{Im}}\left[{\rm i}^{-1}J^{(2)}\left(2-2\varepsilon; 1, 1\right)\right]_s
\left[
\frac{1}{s\!-\!k^2\!-\!{\rm i}0}-\!\sum\limits_{j=0}^{n+l}\frac{(k^2)^j}{s^{j+1}}
\right]
\\
&=&
\frac{{\rm i}}{\pi}\; (k^2)^{n+l+1}
\int\limits_{(m_1+m_2)^2}^{\infty} {\rm d}s\;
\frac{{\rm{Im}}\left[{\rm i}^{-1}J^{(2)}\left(2-2\varepsilon; 1, 1\right)\right]_s}{s^{n+l+1}\; (s-k^2-{\rm i}0)}\; .
\end{eqnarray*}

As a result, for the subtracted $\bar{J}_{n+l+1}^{(2)}\left(2-2\varepsilon; 1, 1\right)$ we get
\begin{equation}
\label{disp11}
\bar{J}_{n+l+1}^{(2)}\left(2-2\varepsilon; 1, 1\right)
= \frac{{\rm i}}{\pi}\;
\int\limits_{(m_1+m_2)^2}^{\infty} {\rm d}s\;
\frac{{\rm{Im}}\left[{\rm i}^{-1}J^{(2)}\left(2-2\varepsilon; 1, 1\right)\right]_s}{s^{n+l+1}\; (s-k^2-{\rm i}0)} \; ,
\end{equation}
where (see Eq.~(\ref{ImJ2_2m2ep_exact}))
\begin{equation}
\label{ImJ2_2m2ep_exact_}
{\rm Im}\left[{\rm i}^{-1} J^{(2)}(2-2\varepsilon;1,1)\right]_s
= 2\pi^{1-\varepsilon}\frac{\Gamma(1-\varepsilon)}
{\Gamma(1-2\varepsilon)}\; 
\frac{\pi}{\sqrt{-\Delta_s}}
\left(\frac{s}{-\Delta_s}\right)^{\varepsilon} \;
\end{equation}
(the subscript $s$ means that we substitute $k^2\to s$).
The first two terms ($\varepsilon^0$ and $\varepsilon^1$) of the $\varepsilon$-expansion of 
${\rm{Im}}\left[{\rm i}^{-1}J^{(2)}\left(2-2\varepsilon; 1, 1\right)\right]$ are given in Eq.~(\ref{ImJ2_2m2ep_exp}). Note that the 
appearance of the factor $1/s^{n+l}$ in the integrand of 
Eq.~(\ref{disp11}) provides better convergence of the dispersive
integral. This is another advantage of subtracting the first terms
of the Taylor expansion in $k^2$.

To deal with the four-point function (see Eq.~(\ref{2-2*ep_basis_1c_4})), we also need the dispersive integral representation for $\bar{J}_{n+l+1}^{(2)}\left(2-2\varepsilon; 1, 2\right)$. 
To derive it,  let us differentiate Eq.~(\ref{disp11}) with respect to $m_2^2$,
\begin{equation}
\bar{J}_{n+l+1}^{(2)}\left(2-2\varepsilon; 1, 2\right)
= \frac{\partial}{\partial m_2^2}
\bar{J}_{n+l+1}^{(2)}\left(2-2\varepsilon; 1, 1\right) \; .
\end{equation}
For the function in the integrand we get
\begin{eqnarray}
\label{deriv_mm2}
\frac{\partial}{\partial m_2^2}
{\rm{Im}}\left[{\rm i}^{-1}J^{(2)}\left(2-2\varepsilon; 1, 1\right)\right]_s
&=&
{\rm Im}\left[{\rm i}^{-1} J^{(2)}(2-2\varepsilon;1,2)\right]_s
\nonumber \\
&=& -\frac{(1+2\varepsilon)(s+m_1^2-m_2^2)}{\Delta_s}
{\rm Im}\left[{\rm i}^{-1} J^{(2)}(2-2\varepsilon;1,1)\right]_s \; .
\hspace*{8mm} 
\end{eqnarray}
The same result can be obtained by using Eq.~(\ref{ibp2a}) and taking into account that the tadpoles do not contribute to the imaginary part.
Note that the limit of integration in (\ref{disp11}) also depends on $m_2$.

Taking into account that separate terms may have singularities as $s\to (m_1+m_2)^2$, let us shift the lower limit by a small positive $\delta$,
\begin{equation}
s_{\delta} = (m_1+m_2)^2 + \delta, 
\end{equation}
and at the end consider the limit $\delta\to 0$. Differentiating Eq.~(\ref{disp11}) with respect to $m_2^2$ we get
\begin{eqnarray}
\label{jtemp1}
\bar{J}_{n+l+1}^{(2)}\left(2-2\varepsilon; 1, 2\right)
&=& \frac{{\rm i}}{\pi}\;
\int\limits_{s_{\delta}}^{\infty} 
\frac{{\rm d}s}{s^{n+l+1}\; (s-k^2-{\rm i}0)}\;
\frac{\partial}{\partial m_2^2}
{\rm{Im}}\left[{\rm i}^{-1}J^{(2)}\left(2-2\varepsilon; 1, 1\right)\right]_s
\nonumber \\ &&
- \frac{{\rm i}}{\pi}\; \frac{m_1+m_2}{m_2}\;
\frac{1}{s_{\delta}^{n+l+1}\; (s_{\delta}-k^2-{\rm i}0)}\;
{\rm{Im}}\left[{\rm i}^{-1}J^{(2)}\left(2-2\varepsilon; 1, 1\right)\right]_{s_{\delta}} \; .
\hspace*{8mm} 
\end{eqnarray}

To proceed, let us use the analytic result given in Eq.~(\ref{ImJ2_2m2ep_exact_}) to calculate the derivatives
\[
\frac{\partial}{\partial s}
\left\{ s^{-\varepsilon}
{\rm{Im}}\left[{\rm i}^{-1}J^{(2)}\left(2-2\varepsilon; 1, 1\right)\right]_s
\right\}
= 2\pi^{1-\varepsilon}\frac{\Gamma(1-\varepsilon)}
{\Gamma(1-2\varepsilon)}\; 
\pi\; 
\left(\frac{\partial \Delta_s}{\partial s}\right)
\left(\frac{\partial}
{\partial \Delta_s}(-\Delta_s)^{-1/2-\varepsilon}\right) \; ,
\]
\[
\frac{\partial}{\partial m_2^2}
{\rm{Im}}\left[{\rm i}^{-1}J^{(2)}\left(2-2\varepsilon; 1, 1\right)\right]_s
= 2\pi^{1-\varepsilon}\frac{\Gamma(1-\varepsilon)}
{\Gamma(1-2\varepsilon)}\; 
\pi\; s^{\varepsilon}
\left(\frac{\partial \Delta_s}{\partial m_2^2}\right)
\left(\frac{\partial}
{\partial \Delta_s}(-\Delta_s)^{-1/2-\varepsilon}\right) \; .
\]
Combining these equations we get
\begin{equation}
\frac{\partial}{\partial m_2^2}
{\rm{Im}}\left[{\rm i}^{-1}J^{(2)}\left(2\!-\!2\varepsilon; 1, 1\right)\right]_s 
= s^{\varepsilon} 
\frac{\left(\partial \Delta_s/\partial m_2^2\right)}
{\left(\partial \Delta_s/\partial s\right)} 
\frac{\partial}{\partial s} \left\{ s^{-\varepsilon}
{\rm{Im}}\left[{\rm i}^{-1}J^{(2)}\left(2\!-\!2\varepsilon; 1, 1\right)\right]_s\right\} \; .
\end{equation}
Taking into account that
\[
\frac{\partial \Delta_s}{\partial m_2^2} = 2 (s+m_1^2-m_2^2),
\qquad {\rm and} \qquad
\frac{\partial \Delta_s}{\partial s} = -2 (s-m_1^2-m_2^2)
\]
we arrive at
\begin{equation}
\frac{\partial}{\partial m_2^2}
{\rm{Im}}\left[{\rm i}^{-1}J^{(2)}\left(2\!-\!2\varepsilon; 1, 1\right)\right]_s 
= -s^{\varepsilon} \frac{s+m_1^2-m_2^2}{s-m_1^2-m_2^2} \ \
\frac{\partial}{\partial s} \left\{ s^{-\varepsilon}
{\rm{Im}}\left[{\rm i}^{-1}J^{(2)}\left(2\!-\!2\varepsilon; 1, 1\right)\right]_s\right\}\; .
\end{equation}

After transforming the derivative with respect to $m_2^2$ into the derivative with respect to $s$, we can apply integration by parts to the integral on the rhs of Eq.~(\ref{jtemp1}),  
\begin{eqnarray}
\label{jtemp2}
&&\frac{{\rm i}}{\pi}\;
\int\limits_{s_{\delta}}^{\infty} 
\frac{{\rm d}s}{s^{n+l+1}\; (s-k^2-{\rm i}0)}\;
\frac{\partial}{\partial m_2^2}
{\rm{Im}}\left[{\rm i}^{-1}J^{(2)}\left(2-2\varepsilon; 1, 1\right)\right]_s
\nonumber \\ && 
= - \frac{{\rm i}}{\pi}\;
\int\limits_{s_{\delta}}^{\infty} 
\frac{s^{\varepsilon}\;{\rm d}s}{s^{n+l+1}\; (s-k^2-{\rm i}0)}\;
\frac{s+m_1^2-m_2^2}{s-m_1^2-m_2^2} 
\frac{\partial}{\partial s} \left\{ s^{-\varepsilon}
{\rm{Im}}\left[{\rm i}^{-1}J^{(2)}\left(2\!-\!2\varepsilon; 1, 1\right)\right]_s\right\}
\nonumber \\ && 
= \frac{{\rm i}}{\pi}\;
\int\limits_{s_{\delta}}^{\infty} {\rm d}s\;
s^{-\varepsilon}\;
{\rm{Im}}\left[{\rm i}^{-1}J^{(2)}\left(2-2\varepsilon; 1, 1\right)\right]_s 
\frac{\partial}{\partial s} \left[
\frac{s^{\varepsilon}}{s^{n+l+1}\; (s-k^2-{\rm i}0)}
\frac{s+m_1^2-m_2^2}{s-m_1^2-m_2^2}\right]
\nonumber \\ && \hspace*{5mm}
+ \frac{{\rm i}}{\pi} 
\frac{s_{\delta}+m_1^2-m_2^2}{s_{\delta}-m_1^2-m_2^2} \ 
\frac{1}{s_{\delta}^{n+l+1}\; (s_{\delta}-k^2-{\rm i}0)}\;
{\rm{Im}}\left[{\rm i}^{-1}J^{(2)}\left(2-2\varepsilon; 1, 1\right)\right]_{s_{\delta}}
\hspace*{8mm} 
\end{eqnarray}

Recalling that $s_{\delta}=(m_1+m_2)^2+\delta$, we can see that
in the limit $\delta\to 0$ the last term on the rhs of 
Eq.~(\ref{jtemp2}) exactly cancels the non-integral term 
in Eq.~(\ref{jtemp1}). Since the the first (integral) term 
on the rhs of
Eq.~(\ref{jtemp2}) is finite as $\delta\to 0$, we can put $\delta=0$. 
In this way, we get 
%\begin{equation}
%\label{disp12}
%\bar{J}_{n+l+1}^{(2)}\left(2-2\varepsilon; 1, 2\right)
%= \frac{{\rm i}}{\pi}\;
%\int\limits_{(m_1+m_2)^2}^{\infty} {\rm d}s\;
%\frac{{\rm{Im}}\left[{\rm i}^{-1}J^{(2)}\left(2-2\varepsilon; 1, 2\right)\right]}{s^{n+l+1}\; (s-k^2-{\rm i}0)} \; .
%\end{equation}
\begin{eqnarray}
\label{disp12a}
\bar{J}_{n+l+1}^{(2)}\left(2-2\varepsilon; 1, 2\right)
&=& \frac{{\rm i}}{\pi}\;
\int\limits_{(m_1+m_2)^2}^{\infty} {\rm d}s\;
s^{-\varepsilon}\;
{\rm{Im}}\left[{\rm i}^{-1}J^{(2)}\left(2-2\varepsilon; 1, 1\right)\right]_s
\nonumber \\ &&
\times
\frac{\partial}{\partial s} \left[
\frac{s^{\varepsilon}}{s^{n+l+1}\; (s-k^2-{\rm i}0)} \
\frac{s+m_1^2-m_2^2}{s-m_1^2-m_2^2}\right] \; .
\end{eqnarray}

Using Eq.~(\ref{disp12a}) we can also get another representation,
\begin{eqnarray}
\label{disp12b}
\bar{J}_{n+l+1}^{(2)}\left(2\!-\!2\varepsilon; 1, 2\right)
&=& \frac{{\rm i}}{\pi}\!
\int\limits_{(m_1+m_2)^2}^{\infty} \!\!\!\!\!{\rm d}s
\Bigl\{
s^{-\varepsilon}\;
{\rm{Im}}\left[{\rm i}^{-1}J^{(2)}\left(2\!-\!2\varepsilon; 1, 1\right)\right]_s
\!- (k^2)^{-\varepsilon}\;
{\rm{Im}}\left[{\rm i}^{-1}J^{(2)}\left(2\!-\!2\varepsilon; 1, 1\right)\right]_{k^2}
\Bigr\}
\nonumber \\ &&
\qquad \qquad \times
\frac{\partial}{\partial s} \left[
\frac{s^{\varepsilon}}{s^{n+l+1}\; (s-k^2-{\rm i}0)} \
\frac{s+m_1^2-m_2^2}{s-m_1^2-m_2^2}\right]
\nonumber \\ &&
- \frac{{\rm i}}{\pi}\; \frac{m_1+m_2}{m_2} \
\frac{\left[(m_1+m_2)^2\right]^{\varepsilon-n-l-1}}{(m_1+m_2)^2-k^2-{\rm i}0}
(k^2)^{-\varepsilon}\;
{\rm{Im}}\left[{\rm i}^{-1}J^{(2)}\left(2\!-\!2\varepsilon; 1, 1\right)\right]_{k^2}
\; ,
\end{eqnarray}
where $k^2\leftrightarrow k^2+{\rm i}0$, in the same way as in the prescription
$1/(s-k^2-{\rm i}0)$.

Using Eq.~(\ref{disp12b}) may lead to numerical issues (slow convergence, etc.),  because after taking the derivative with respect to $s$ in the integrand we obtain the term   $1/(s-k^2-{\rm i}0)^2$.
To get an alternative representation,  we start from Eq.~(\ref{jtemp1}) and substitute the result of Eq.~(\ref{deriv_mm2}) for the derivative w.r.t. $m_2^2$,
\begin{eqnarray}
\label{disp12c1}
\bar{J}_{n+l+1}^{(2)}\left(2\!-\!2\varepsilon; 1, 2\right)
&=& -\frac{{\rm i}}{\pi} (1+2\varepsilon)
\int\limits_{s_{\delta}}^{\infty} {\rm d}s\;
\frac{s+m_1^2-m_2^2}{s^{n+l+1}(s-k^2-{\rm i}0)\Delta_s}
{\rm{Im}}\left[{\rm i}^{-1}J^{(2)}\left(2\!-\!2\varepsilon; 1, 1\right)\right]_s
\nonumber \\ &&
-\frac{{\rm i}}{\pi}  \frac{m_1+m_2}{m_2} \
\frac{1}{s_{\delta}^{n+l+1}(s_{\delta}-k^2-{\rm i}0)}
{\rm{Im}}\left[{\rm i}^{-1}J^{(2)}\left(2\!-\!2\varepsilon; 1, 1\right)\right]_{s_{\delta}}
\; . \hspace*{8mm}
\end{eqnarray}
Note that the integral in Eq.~(\ref{disp12c1}) is singular as $\delta\to 0$, because
\begin{equation}
\Delta_s = -\left[ s - (m_1+m_2)^2\right]
\left[ s - (m_1-m_2)^2\right]
= - (s-s_0)(s-s_1) \, ,
\end{equation}
with $s_0\equiv (m_1+m_2)^2$, $s_1\equiv (m_1-m_2)^2$. To separate the finite and the divergent contributions, let us employ the identity
\begin{equation}
\frac{1}{s-k^2-{\rm i}0}
= - \frac{s-s_{\delta}}{(s-k^2-{\rm i}0)(s_{\delta}-k^2-{\rm i}0)}
+ \frac{1}{s_{\delta}-k^2-{\rm i}0} \, .
\end{equation}
In this way, we get
\begin{eqnarray}
\label{disp12c2}
\bar{J}_{n+l+1}^{(2)}\left(2\!-\!2\varepsilon; 1, 2\right)
&=& \frac{{\rm i}}{\pi}  
\frac{1+2\varepsilon}{s_{\delta}-k^2-{\rm i}0}
\int\limits_{s_{\delta}}^{\infty} {\rm d}s\;
\frac{(s+m_1^2-m_2^2)(s-s_{\delta})}{s^{n+l+1}(s-k^2-{\rm i}0)\Delta_s}
{\rm{Im}}\left[{\rm i}^{-1}J^{(2)}\left(2\!-\!2\varepsilon; 1, 1\right)\right]_s
\nonumber \\ &&
-\frac{{\rm i}}{\pi}  
\frac{1+2\varepsilon}{s_{\delta}-k^2-{\rm i}0}
\int\limits_{s_{\delta}}^{\infty} {\rm d}s\;
\frac{s+m_1^2-m_2^2}{s^{n+l+1}\Delta_s}\;
{\rm{Im}}\left[{\rm i}^{-1}J^{(2)}\left(2\!-\!2\varepsilon; 1, 1\right)\right]_s
\nonumber \\ &&
-\frac{{\rm i}}{\pi}  \frac{m_1+m_2}{m_2} \
\frac{1}{s_{\delta}^{n+l+1}(s_{\delta}-k^2-{\rm i}0)} 
{\rm{Im}}\left[{\rm i}^{-1}J^{(2)}\left(2\!-\!2\varepsilon; 1, 1\right)\right]_{s_{\delta}}
\; . \hspace*{8mm}
\end{eqnarray}
The first integral on the rhs of Eq.~(\ref{disp12c2}) is finite as $\delta\to 0$, so that we can put $\delta=0$ and substitute $(s-s_{\delta})/\Delta_s=(s-s_0)/\Delta_s=-1/(s-s_1)$. To deal with the second integral let us employ the analytic result given in Eq.~(\ref{ImJ2_2m2ep_exact_}) to calculate the derivative
\[
\frac{\partial}{\partial s}
\left\{ s^{-\varepsilon}
{\rm{Im}}\left[{\rm i}^{-1}J^{(2)}\left(2-2\varepsilon; 1, 1\right)\right]_s
\right\}
= \frac{(1+2\varepsilon)(s-m_1^2-m_2^2)}{\Delta_s}\;
s^{-\varepsilon} 
{\rm{Im}}\left[{\rm i}^{-1}J^{(2)}\left(2-2\varepsilon; 1, 1\right)\right]_s \, .
\]
Therefore,
\[
\frac{1+2\varepsilon}{\Delta_s}\; 
{\rm{Im}}\left[{\rm i}^{-1}J^{(2)}\left(2-2\varepsilon; 1, 1\right)\right]_s
=
\frac{s^{\varepsilon}}{s-m_1^2-m_2^2}\;
\frac{\partial}{\partial s}
\left\{ s^{-\varepsilon}
{\rm{Im}}\left[{\rm i}^{-1}J^{(2)}\left(2-2\varepsilon; 1, 1\right)\right]_s
\right\} \, .
\]
Integrating by parts, we can transform the second integral in Eq.~(\ref{disp12c2}) as
\begin{eqnarray}
\label{disp12c3}
&& \hspace*{-10mm}
-\frac{{\rm i}}{\pi}  
\frac{1+2\varepsilon}{s_{\delta}-k^2-{\rm i}0}
\int\limits_{s_{\delta}}^{\infty} {\rm d}s\;
\frac{s+m_1^2-m_2^2}{s^{n+l+1}\Delta_s}\;
{\rm{Im}}\left[{\rm i}^{-1}J^{(2)}\left(2\!-\!2\varepsilon; 1, 1\right)\right]_s
\nonumber \\ 
&=& -\frac{{\rm i}}{\pi}  
\frac{1}{s_{\delta}-k^2-{\rm i}0}
\int\limits_{s_{\delta}}^{\infty} {\rm d}s\;
\frac{s^{\varepsilon}(s+m_1^2-m_2^2)}{s^{n+l+1}(s-m_1^2-m_2^2)}\;
\frac{\partial}{\partial s}
\left\{ s^{-\varepsilon}
{\rm{Im}}\left[{\rm i}^{-1}J^{(2)}\left(2-2\varepsilon; 1, 1\right)\right]_s
\right\}
\nonumber \\ 
&=&
\frac{{\rm i}}{\pi}  
\frac{1}{s_{\delta}-k^2-{\rm i}0} \
\frac{s_{\delta}+m_1^2-m_2^2}{s_{\delta}^{n+l+1}(s_{\delta}-m_1^2-m_2^2)}\;
{\rm{Im}}\left[{\rm i}^{-1}J^{(2)}\left(2-2\varepsilon; 1, 1\right)\right]_{s_{\delta}}
\nonumber \\ &&
+ \frac{{\rm i}}{\pi} 
\frac{1}{s_{\delta}-k^2-{\rm i}0}
\int\limits_{s_{\delta}}^{\infty} {\rm d}s\;
{\rm{Im}}\left[{\rm i}^{-1}J^{(2)}\left(2\!-\!2\varepsilon; 1, 1\right)\right]_s \ s^{-\varepsilon}
\frac{\partial}{\partial s}
\left[ 
\frac{s^{\varepsilon}(s+m_1^2-m_2^2)}{s^{n+l+1}(s-m_1^2-m_2^2)}
\right] \; . \hspace*{8mm}
\end{eqnarray}
We can see that in the limit $\delta\to 0$ the first (non-integral) term on the rhs of Eq.~(\ref{disp12c3}) exactly cancels the non-integral term in Eq.~(\ref{disp12c2}). The remaining integrals contributing to $\bar{J}_{n+l+1}^{(2)}\left(2\!-\!2\varepsilon; 1, 2\right)$ are finite as $\delta\to 0$. In this way, putting $\delta=0$ we arrive at the following alternative representation:  
\begin{eqnarray}
\label{disp12c}
\bar{J}_{n+l+1}^{(2)}\left(2\!-\!2\varepsilon; 1, 2\right)
&=& -\frac{{\rm i}}{\pi} \frac{1+2\varepsilon}{s_0\!-\!k^2\!-\!{\rm i}0}
\int\limits_{s_0}^{\infty} {\rm d}s\;
\frac{s+m_1^2-m_2^2}{s^{n+l+1}(s-s_1)(s-k^2-{\rm i}0)}
{\rm{Im}}\left[{\rm i}^{-1}J^{(2)}\left(2\!-\!2\varepsilon; 1, 1\right)\right]_s
\nonumber \\ &&
+\frac{{\rm i}}{\pi}\frac{1}{s_0\!-\!k^2\!-\!{\rm i}0}
\int\limits_{s_0}^{\infty} {\rm d}s\;
{\rm{Im}}\left[{\rm i}^{-1}J^{(2)}\left(2\!-\!2\varepsilon; 1, 1\right)\right]_s
s^{-\varepsilon}
\frac{\partial}{\partial s} \left[
\frac{s^{\varepsilon}\;(s+m_1^2-m_2^2)}{s^{n+l+1}\;(s\!-\!m_1^2\!-\!m_2^2)}\right] .
\hspace*{-10mm}
\nonumber \\ &&
\hspace*{10mm}
\end{eqnarray}
Using partial fractioning in the denominator of the first integral 
in Eq.~(\ref{disp12c}) and identifying some contributions as specific cases of Eq.~(\ref{disp11}), we can get the following representation:
\begin{eqnarray}
\label{disp12d}
\bar{J}_{n+l+1}^{(2)}\left(2\!-\!2\varepsilon; 1, 2\right)
&=& 
-(1+2\varepsilon)\frac{k^2+m_1^2-m_2^2}{\Delta}\;
\bar{J}_{n+l+1}^{(2)}\left(2\!-\!2\varepsilon; 1, 1\right)
\nonumber \\ &&
+(1+2\varepsilon)\frac{2m_1(m_1-m_2)}{\Delta}\;
\bar{J}_{n+l+1}^{(2)}\left(2\!-\!2\varepsilon; 1, 1\right)\Bigr|_{k^2=s_1}
\nonumber \\ &&
+\frac{{\rm i}}{\pi}\frac{1}{s_0\!-\!k^2\!-\!{\rm i}0}
\int\limits_{s_0}^{\infty} {\rm d}s\;
{\rm{Im}}\left[{\rm i}^{-1}J^{(2)}\left(2\!-\!2\varepsilon; 1, 1\right)\right]_s
s^{-\varepsilon}
\frac{\partial}{\partial s} \left[
\frac{s^{\varepsilon}\;(s+m_1^2-m_2^2)}{s^{n+l+1}\;(s\!-\!m_1^2\!-\!m_2^2)}\right] .
\hspace*{-10mm}
\nonumber \\ &&
\hspace*{10mm}
\end{eqnarray}
Furthermore, using integration by parts we can evaluate the remaining integral 
in Eq.~(\ref{disp12d}) as
\begin{eqnarray}
&& \hspace{-30mm}
\frac{{\rm i}}{\pi}
\int\limits_{s_0}^{\infty} {\rm d}s\;
{\rm{Im}}\left[{\rm i}^{-1}J^{(2)}\left(2\!-\!2\varepsilon; 1, 1\right)\right]_s
s^{-\varepsilon}
\frac{\partial}{\partial s} \left[
\frac{s^{\varepsilon}\;(s+m_1^2-m_2^2)}{s^{n+l+1}\;(s\!-\!m_1^2\!-\!m_2^2)}\right]
\nonumber \\ &&
= \frac{{\rm i}}{\pi}
\frac{\partial}{\partial m_2^2}
\int\limits_{s_0}^{\infty} \frac{{\rm d}s}{s^{n+l+1}}\;
{\rm{Im}}\left[{\rm i}^{-1}J^{(2)}\left(2\!-\!2\varepsilon; 1, 1\right)
\right]_s
\nonumber \\ &&
= \frac{\partial}{\partial m_2^2}
\bar{J}_{n+l}^{(2)}\left(2\!-\!2\varepsilon; 1, 1\right)\Bigr|_{k^2=0}\; ,
\end{eqnarray}
where $\bar{J}_{n+l}^{(2)}\left(2\!-\!2\varepsilon; 1, 1\right)\Bigr|_{k^2=0}$
is the $(n+l)$-th coefficient of the small-$k^2$ expansion of 
Eq.~(\ref{J2_F4dd_}), namely, 
\begin{eqnarray}
\label{J2_F4dd_npl}
\bar{J}_{n+l}^{(2)}\left(2\!-\!2\varepsilon; 1, 1\right)\Bigr|_{k^2=0}
&\!\!=\!\!& 
\frac{(1+\varepsilon)_{n+l}}{(m_2^2-m_1^2)^{1+2n+2l}}
\nonumber \\ &&
\times \Biggl\{
J^{(2)}(2-2\varepsilon;0,1)\;
\sum\limits_{r=0}^{n+l}
\frac{(n+l)!}{r!(n+l-r)!}\;
\frac{(m_2^2)^r (m_1^2)^{n+l-r}}
{(1-\varepsilon)_r (1+\varepsilon)_{n+l-r}}\;
\nonumber \\ && \hspace*{5mm}
-J^{(2)}(2-2\varepsilon;1,0)\;
\sum\limits_{r=0}^j
\frac{(n+l)!}{r!(n+l-r)!}\;
\frac{(m_1^2)^r (m_2^2)^{n+l-r}}
{(1-\varepsilon)_r (1+\varepsilon)_{n+l-r}}\;
\Biggr\} \; . \hspace*{8mm}
\end{eqnarray}
In this way, we get
\begin{eqnarray}
\label{disp12e}
\bar{J}_{n+l+1}^{(2)}\left(2\!-\!2\varepsilon; 1, 2\right)
&=& 
-(1+2\varepsilon)\; \frac{k^2+m_1^2-m_2^2}{\Delta}\;
\bar{J}_{n+l+1}^{(2)}\left(2\!-\!2\varepsilon; 1, 1\right)
\nonumber \\ &&
+(1+2\varepsilon)\; \frac{2m_1(m_1-m_2)}{\Delta}\;
\bar{J}_{n+l+1}^{(2)}\left(2\!-\!2\varepsilon; 1, 1\right)\Bigr|_{k^2=s_1}
\nonumber \\ &&
+\frac{1}{s_0\!-\!k^2\!-\!{\rm i}0}\;
\frac{\partial}{\partial m_2^2}
\bar{J}_{n+l}^{(2)}\left(2\!-\!2\varepsilon; 1, 1\right)\Bigr|_{k^2=0}\; .
\end{eqnarray}
or 
\begin{eqnarray}
\label{disp12f}
\bar{J}_{n+l+1}^{(2)}\left(2\!-\!2\varepsilon; 1, 2\right)
&=& 
\frac{1}{s_0\!-\!k^2\!-\!{\rm i}0}
\Biggl\{
-(1+2\varepsilon)\; 
\bar{J}_{n+l+1}^{(2)}\left(2\!-\!2\varepsilon; 1, 1\right)
\nonumber \\ &&
-(1\!+\!2\varepsilon) \frac{2m_1(m_1\!-\!m_2)}{k^2-s_1}\;
\Bigl[
\bar{J}_{n+l+1}^{(2)}\left(2\!-\!2\varepsilon; 1, 1\right)
-\bar{J}_{n+l+1}^{(2)}\left(2\!-\!2\varepsilon; 1, 1\right)\Bigr|_{k^2=s_1}
\Bigr]
\nonumber \\ &&
+ \frac{\partial}{\partial m_2^2}
\bar{J}_{n+l}^{(2)}\left(2\!-\!2\varepsilon; 1, 1\right)\Bigr|_{k^2=0}
\Biggr\}
\; .
\end{eqnarray}
The dispersion integral representation for $\bar{J}_{n+l+1}^{(2)}\left(2\!-\!2\varepsilon; 1, 1\right)$ is given in Eq.~(\ref{disp11}), and the combination of integrals in the second line of Eq.~(\ref{disp12f}) can be presented as
\begin{equation}
\nonumber
\frac{1}{k^2\!-\!s_1}\;
\Bigl[
\bar{J}_{n+l+1}^{(2)}\left(2\!-\!2\varepsilon; 1, 1\right)
-\bar{J}_{n+l+1}^{(2)}\left(2\!-\!2\varepsilon; 1, 1\right)\Bigr|_{k^2=s_1}
\Bigr]
= \frac{{\rm i}}{\pi}\;
\int\limits_{s_0}^{\infty} {\rm d}s\;
\frac{{\rm{Im}}\left[{\rm i}^{-1}J^{(2)}\left(2-2\varepsilon; 1, 1\right)\right]_s}{s^{n+l+1} (s\!-\!k^2\!-\!{\rm i}0)(s\!-\!s_1\!-\!{\rm i}0)} \; . 
\end{equation}

%============================
%\eject

\section{Numerical Examples}

In this section, we provide a numerical comparison of three- and four-point
functions calculated using techniques outlined in the last section 
and Collier \citep{COL1}-\citep{COL4} numerical library. 

\begin{figure}
\begin{centering}
\includegraphics[scale=0.3]{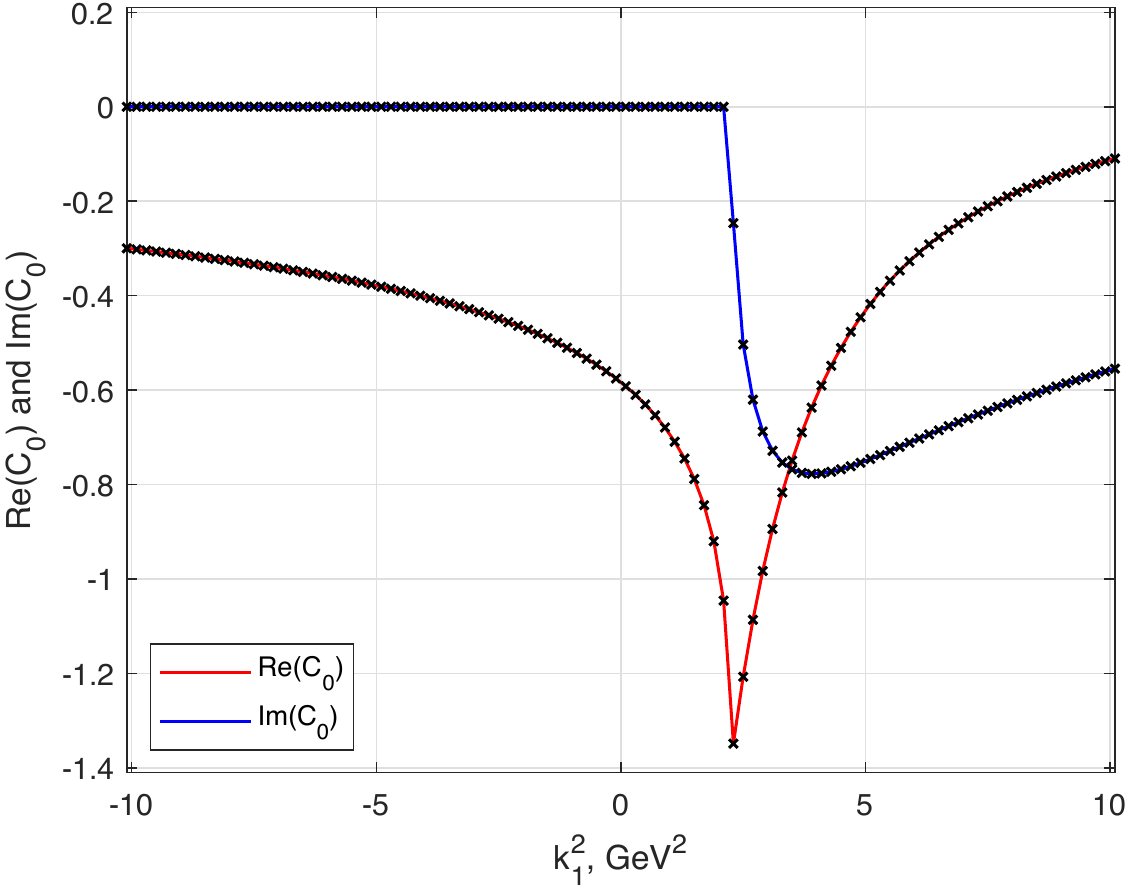}\includegraphics[scale=0.3]{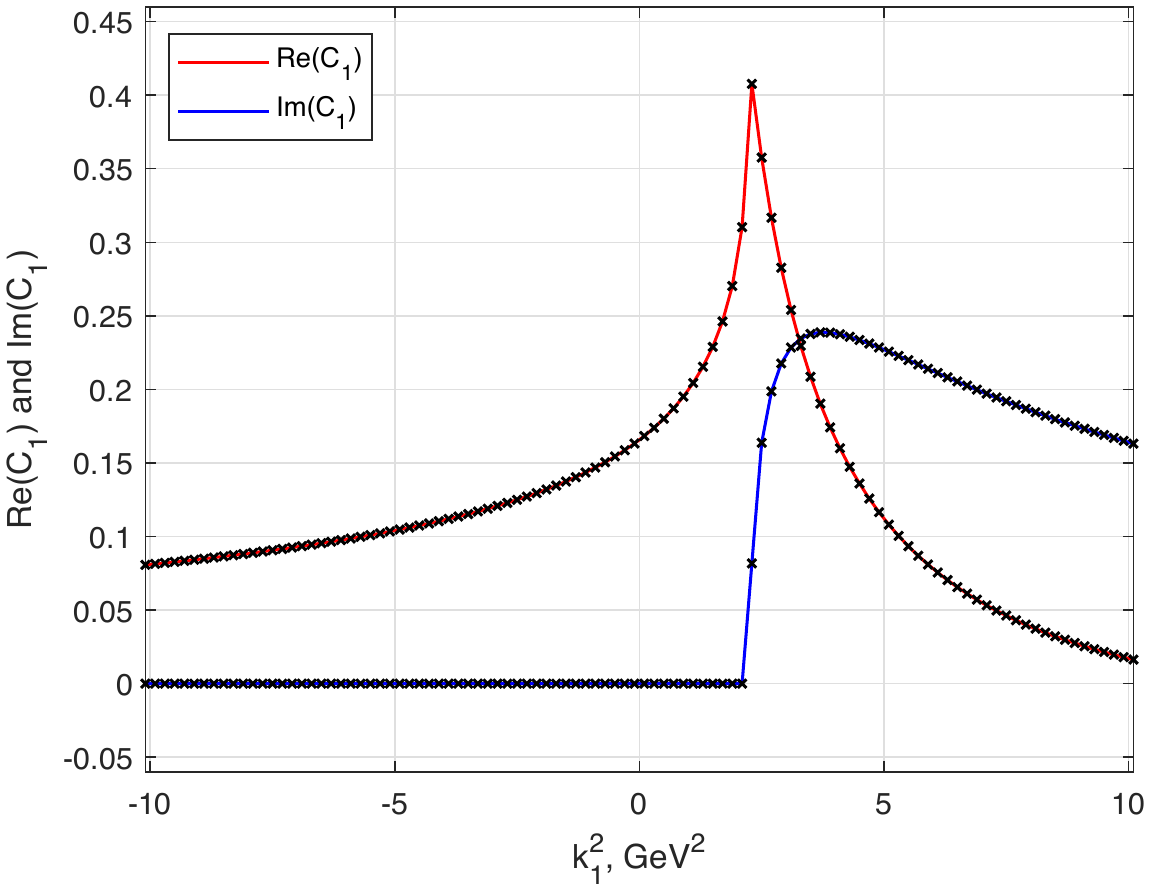}\includegraphics[scale=0.3]{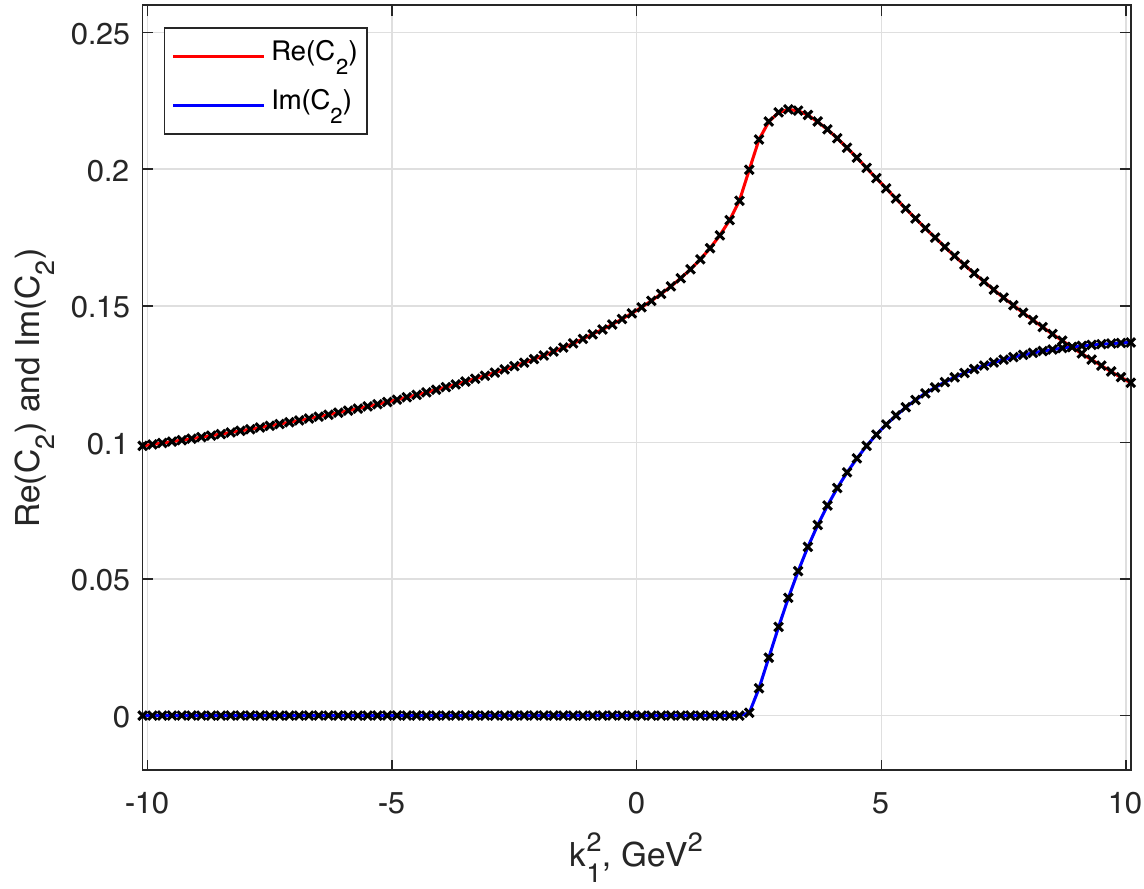}

\includegraphics[scale=0.3]{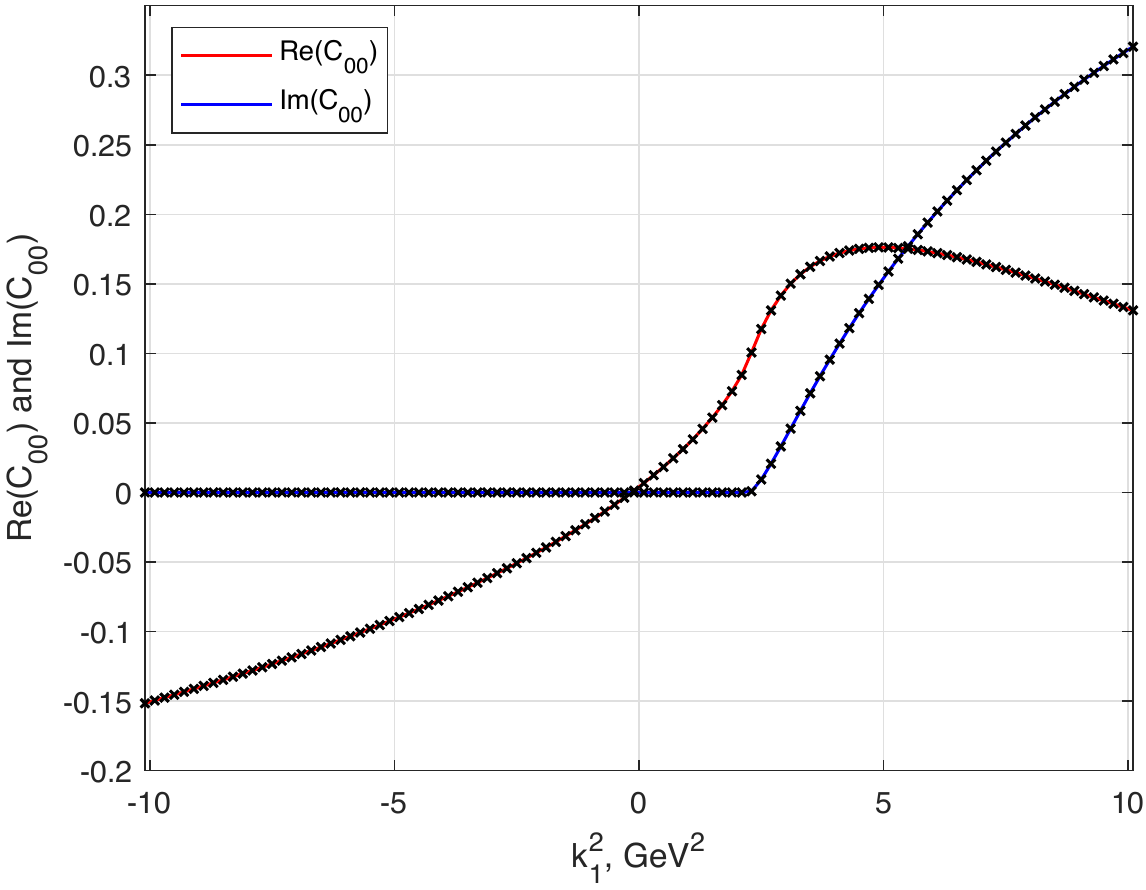}\includegraphics[scale=0.3]{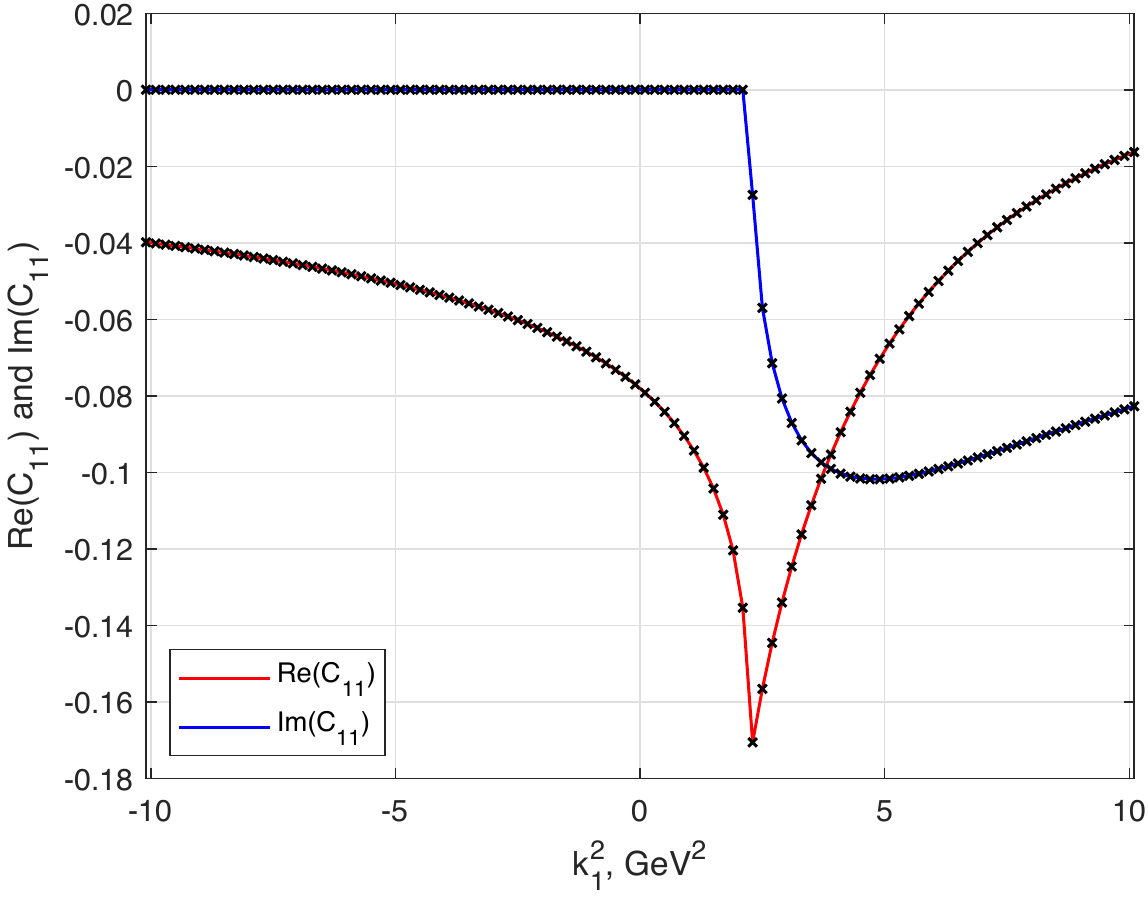}\includegraphics[scale=0.3]{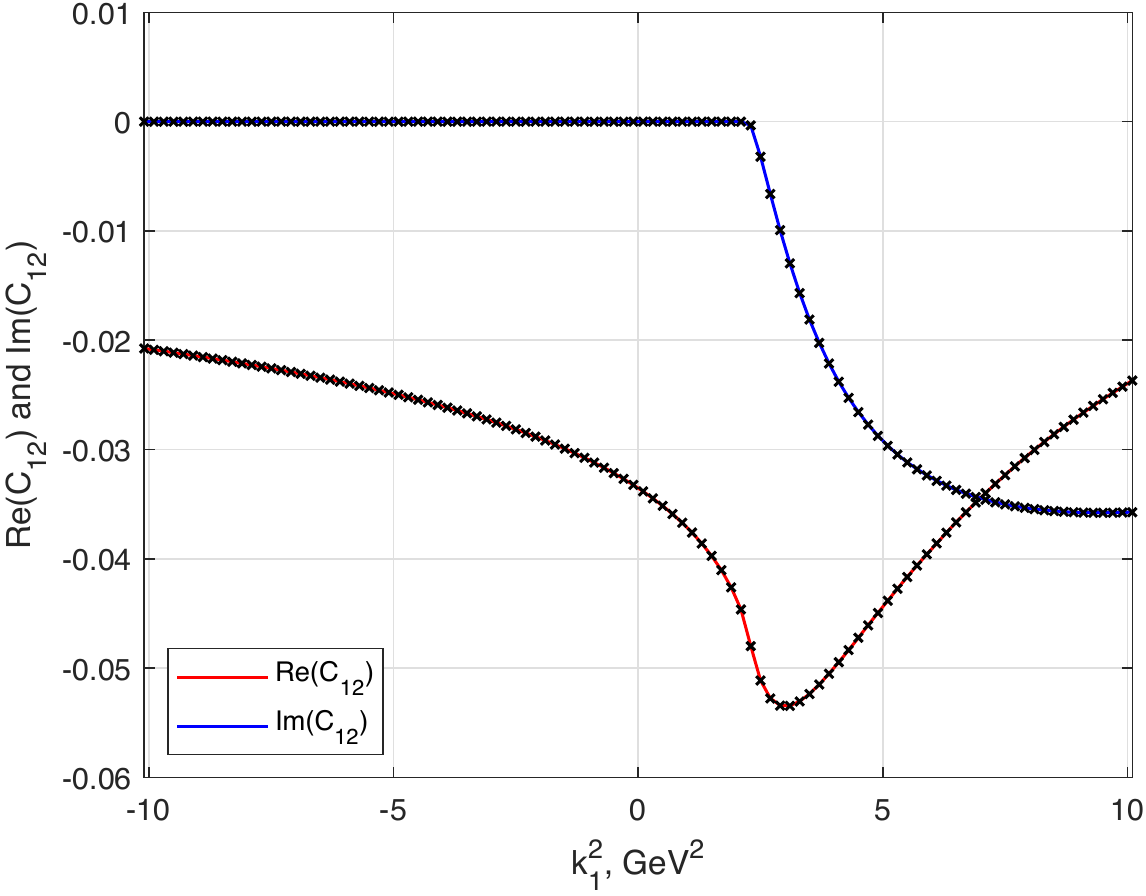}
\par\end{centering}
%\caption{Numerical results for the three-point functions $C_{2l,n_{1},n_{2}}\equiv C_{2l,n_{1},n_{2}}\left(k_{1}^{2},k_{2}^{2},\left(k_{1}+k_{2}\right)^{2},m_{1},m_{2},m_{3}\right)$:
%$C_{0}$, $C_{1}$, $C_{2}$, $C_{00}$, $C_{11}$ and $C_{12}$ ($k_{2}^{2}=-1.5~{\rm GeV}^{2}$,
%$(k_{1}+k_{2})^{2}=m_{3}^{2}$, $m_{1}=0.5~{\rm GeV}$, $m_{2}=1.0~{\rm GeV}$,
%$m_{3}=1.5~{\rm GeV}$). Crossed dots are results based on this work
%and solid lines are produced from Collier library.}
\caption{Numerical results for the three-point functions $C_{0}$, $C_{1}$, $C_{2}$, $C_{00}$, $C_{11}$ and $C_{12}$ ($k_{2}^{2}=-1.5~{\rm GeV}^{2}$,
$(k_{1}+k_{2})^{2}=m_{3}^{2}$, $m_{1}=0.5~{\rm GeV}$, $m_{2}=1.0~{\rm GeV}$,
$m_{3}=1.5~{\rm GeV}$).  The functions $C_{2l,n_{1},n_{2}}$ are defined in Eq.~(\ref{C_to_J_mapping_2}). Crossed dots are results based on this work
and solid lines are produced from the Collier library.}

\label{fig-exmpl-C}
\end{figure}

\begin{figure}
\begin{centering}
\includegraphics[scale=0.4]{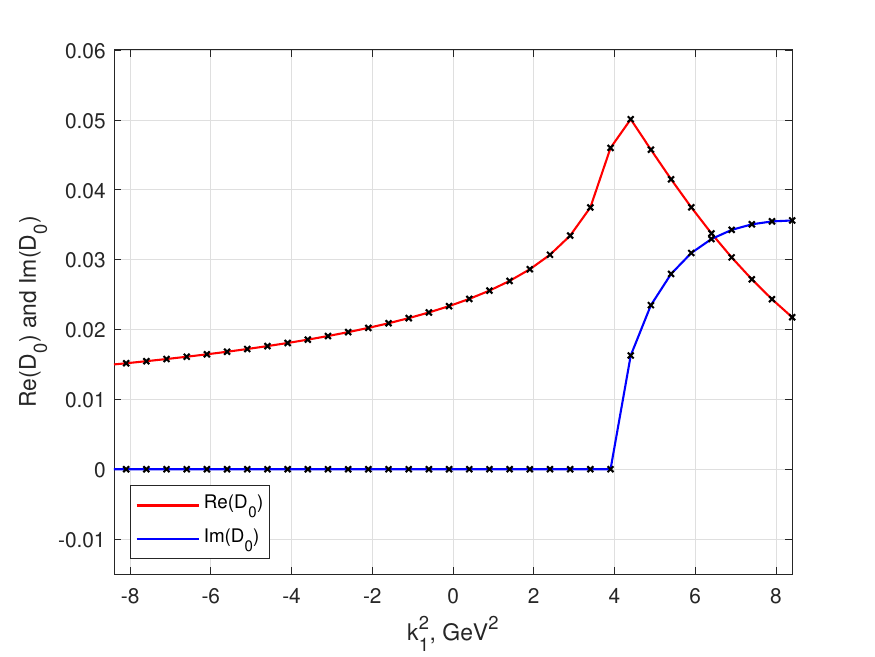}\includegraphics[scale=0.4]{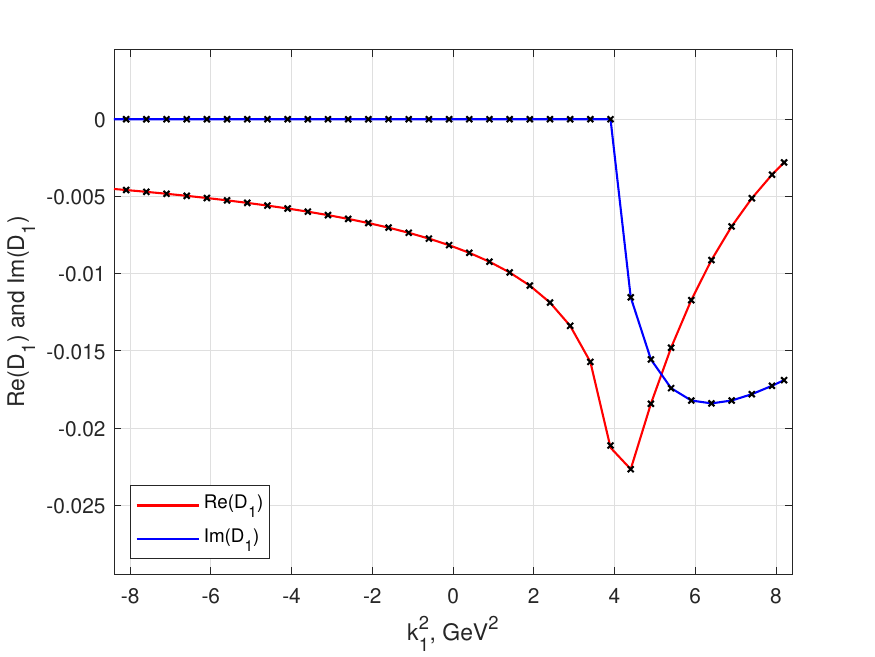}\includegraphics[scale=0.4]{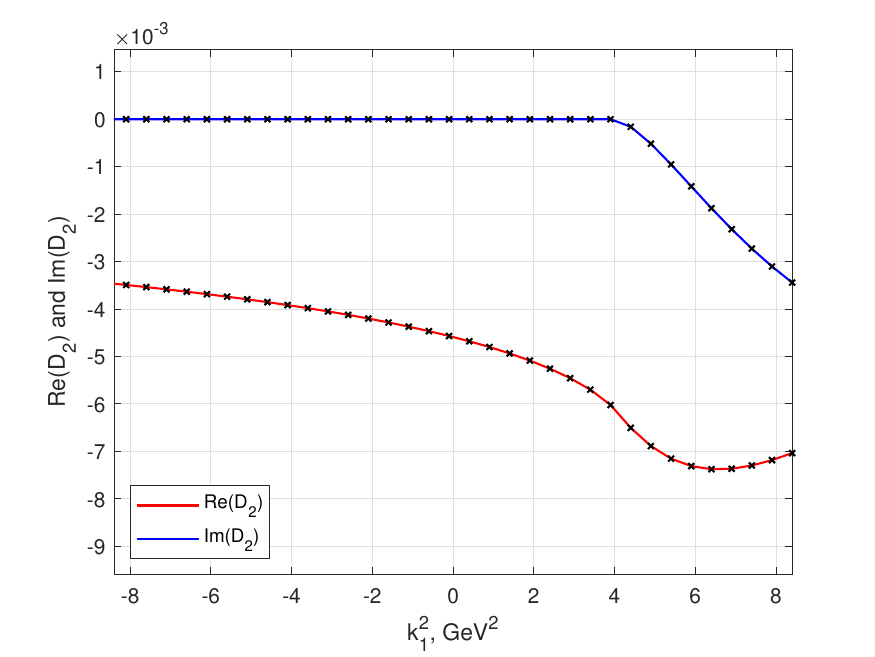} \\
\includegraphics[scale=0.4]{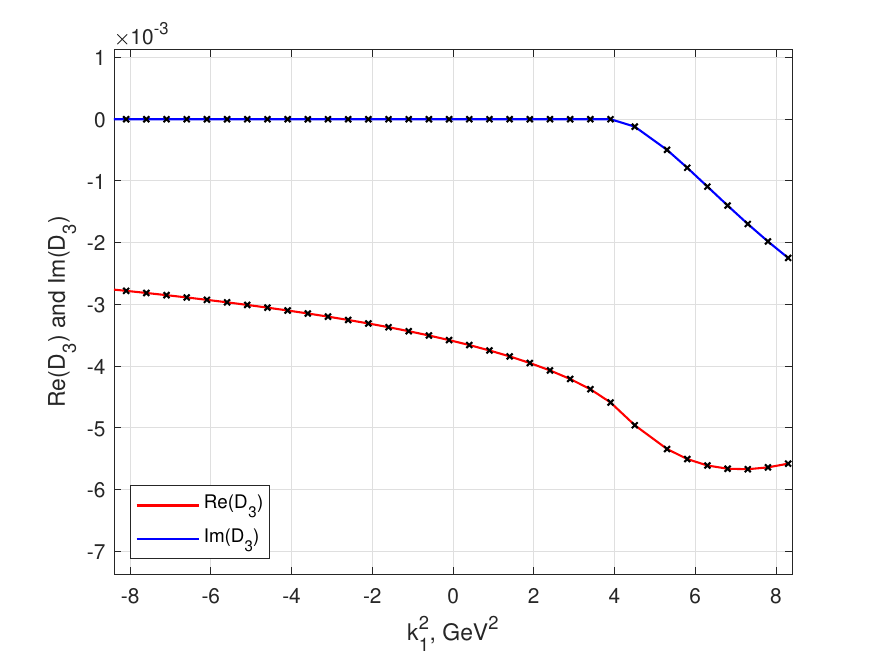}\includegraphics[scale=0.4]{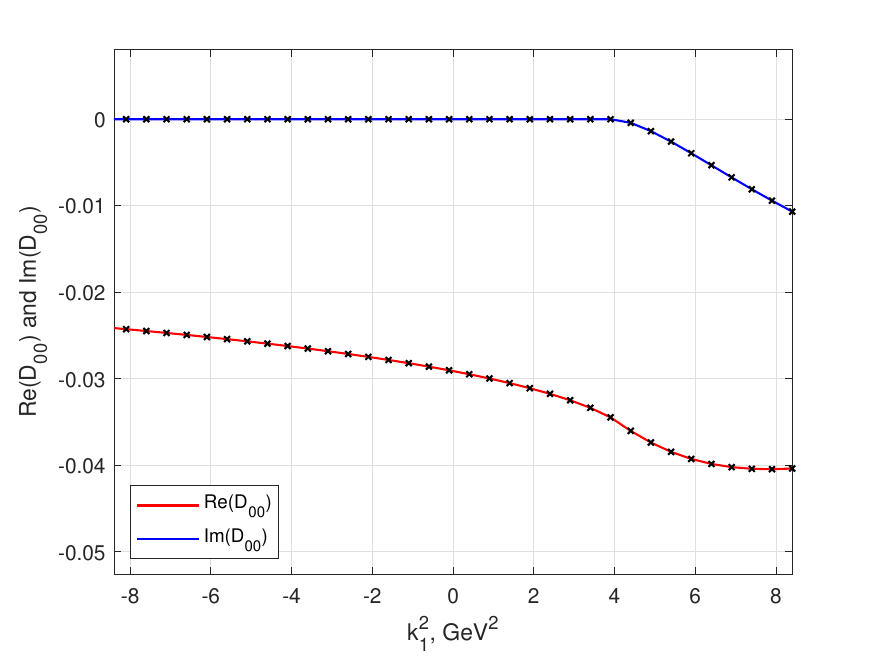}
\par\end{centering}
%\caption{Numerical results for the four-point functions $D_{2l,n_{1},n_{2},n_{3}}\equiv D_{2l,n_{1},n_{2},n_{3}}\left(k_{1}^{2},k_{2}^{2},k_{3}^{2},k_{4}^{2},\left(k_{1}+k_{2}\right)^{2},\left(k_{2}+k_{3} 
% \right)^{2},m_{1},m_{2},m_{3},m_{4}\right)$:
%$D_{0}$ ($k_{2}^{2}=-1.5~{\rm GeV}^{2}$,$k_{3}^{2}=-2.5~{\rm GeV}^{2}$,$k_{4}^{2}=m_{4}^{2}~{\rm GeV}^{2}$,
%$(k_{1}\cdot k_{3})=4.0~{\rm GeV}^{2}$, $(k_{2}\cdot k_{3})=-1.0~{\rm GeV}^{2}$
%$m_{1}=1.5~{\rm GeV}$, $m_{2}=0.5~{\rm GeV}$, $m_{3}=2.0~{\rm GeV}$,
%$m_{4}=2.5~{\rm GeV}$). Crossed dots are results based on this work
%and solid line is produced from Collier library.}
\caption{Numerical results for the four-point functions
$D_{2l,n_1,n_2,n_3}$ ($k_{2}^{2}=-1.5~{\rm GeV}^{2}$,  $k_{3}^{2}=-2.5~{\rm GeV}^{2}$,  $k_{4}^{2}=m_{4}^{2}$,
$(k_{1}\cdot k_{3})=4.0~{\rm GeV}^{2}$, $(k_{2}\cdot k_{3})=-1.0~{\rm GeV}^{2}$
$m_{1}=1.5~{\rm GeV}$, $m_{2}=0.5~{\rm GeV}$, $m_{3}=2.0~{\rm GeV}$,
$m_{4}=2.5~{\rm GeV}$). The functions $D_{2l,n_{1},n_{2},n_{3}}$ are defined in Eq.~(\ref{D_to_J_mapping_2}).  Crossed dots are results based on this work and solid lines are produced from the Collier library.}

\label{fig-exmpl-D}
\end{figure}

For numerical integration over the dispersive and Feynman parameters, 
we have used the \textit{Mathematica}, GlobalAdaptive method. As it can be seen
from Fig.~\ref{fig-exmpl-C}, results are in excellent agreement
with Collier. Numerical comparison for four-point functions is given
in Fig.~\ref{fig-exmpl-D}.
As in the case of three-point functions, the four-point example shows
that we have rather good consistency with Collier.  
At this point, we are ready to apply derived many-point functions in dispersive representation
to the evaluation of two-loop diagrams.

\section{Roadmap to two-Loop calculations}

\begin{figure}
\begin{centering}
\includegraphics[scale=0.9]{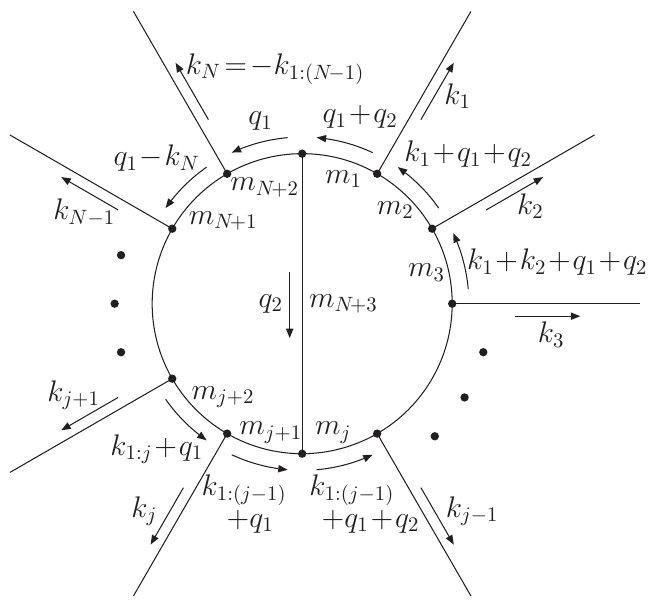}
\par\end{centering}
\caption{The two-loop planar $N$-point diagram ($k_{1:j}\equiv k_{1}+\ldots+k_{j}$). }

\label{Fig-two-loop-gen}
\end{figure}

As we can see from the previous sections we have successfully represented
one-loop (up to multiplicity four) integrals with an arbitrary tensor
rank using recurrence and dispersive methods.  In addition,  we were
able to reduce higher multiplicity PV functions to tthe wo-point result.
Finally,  we have adopted the dispersive technique introduced in \citep{AA1}-\citep{AB19}  to 
subtracted two-point functions.  Now we have analytical results for
PV functions with polynomial terms in $k^{2}$ and the dispersive term
carrying a propagator-like structure $\propto\frac{1}{(s-k^{2}-i0)}$.
This particular representation is most valuable for applications in
two-loop calculations for any possible particle physics models. It is worth mentioning that, in this section, we will consider only planar topology for the two-loop graphs.  In the case of non-planar graphs,  we can re-arrange momenta in the loop integral and reduce the result to the  dispersive representation considered in \citep{AACR}. First,
if we consider a $\left(j+2\right)$-point Feynman graph as an
insertion (index $\left(j+2\right)$ means that we have $j$ number
of external and two internal legs in the insertion) into the two-loop
planar topology (see Fig.~\ref{Fig-two-loop-gen}), then
polynomial terms in external or second-loop momenta will be a part
of the numerator algebra and the dispersive contribution will be treated
as an additional propagator in the second-loop integral. 
Let us define the two-loop integral in a similar way as it was done
in Eq.~(\ref{tensor_T_1_N}),
\begin{align}
U_{\mu_{1}\ldots\mu_{M+L}}^{(N)} & =\left(\frac{\mu^{4-D}e^{\gamma_{E}(4-D)/2}}{{\rm i}\pi^{D/2}}\right)^{2}\int\int {\rm d}^{D}q_{1}\;{\rm d}^{D}q_{2}\nonumber \\
\nonumber \\
 & \times\frac{{q_1}_{\mu_1}\ldots{q_1}_{\mu_{M}}\;{q_2}_{\mu_{M+1}}\ldots {q_2}_{\mu_{M+L}} }{\left[q_{1}^{2}-m_{N+2}^{2}\right]\left[q_{2}^{2}-m_{N+3}^{2}\right]\left[\left(q_{1}+q_{2}\right)^{2}-m_{1}^{2}\right]\left[\left(k_{1}+q_{1}+q_{2}\right)^{2}-m_{2}^{2}\right]}\nonumber \\
\nonumber \\
 & \times\frac{1}{\left[\left(k_{1}+k_{2}+q_{1}+q_{2}\right)^{2}-m_{3}^{2}\right]...\left[\left(k_{1}+...+k_{j-1}+q_{1}+q_{2}\right)^{2}-m_{j}^{2}\right]}\nonumber \\
\nonumber \\
 & \times\frac{1}{\left[\left(k_{1}+...+k_{j-1}+q_{1}\right)^{2}-m_{j+1}^{2}\right]...\left[\left(k_{1}+...+k_{N-1}+q_{1}\right)^{2}-m_{N+1}^{2}\right]}.\label{eq:1r}
\end{align}
We start our evaluation with the integration over one of the loop
momentum. The next step is to apply tensor decomposition, and reduce one-loop
insertion of Eq.~(\ref{eq:1r}) to $J^{(j+2)}\left(D+2l+2n_{1}+\ldots+2n_{j+1};1,1+n_{1},\ldots,1+n_{j+1}\right)$.
After that, we can apply Feynman trick to reduce the number of the propagators
in the first-loop integration to two, which will result in the two-point
function with one of the propagators in the power of $\left(1+j+n_{1}+\ldots+n_{j+1}\right)$:
$J^{(2)}\left(D+2l+2n_{1}+\ldots+2n_{j+1};1,1+j+n_{1}+\ldots+n_{j+1}\right)$.
Using the recurrence approach, we can reduce the number of the dimensions to
$D=2-2\varepsilon$ and the propagator's power to one, resulting in the insertion
expressed as a subtracted UV-finite $\bar{J}_{l+n_{1}+\ldots+n_{j+1}}^{(2)}\left(2-2\varepsilon;1,1\right)$
two-point function and two UV-divergent tadpoles, $J^{(2)}\left(2-2\varepsilon;1,0\right)$
and $J^{(2)}\left(2-2\varepsilon;0,1\right)$, multiplied by the polynomials
in external and second-loop momenta. Subtracted $\bar{J}_{l+n_{1}+\ldots+n_{j+1}}^{(2)}\left(2-2\varepsilon;1,1\right)$
can be expressed dispersively (see Eq.~(\ref{disp11})), and in the second-loop
integral, we will receive an additional propagator and terms in the numerator
expressed as polynomials in the momenta. Second-loop integration is now
reduced to one-loop integral where we can apply well-tested packages,
such as X \citep{HP1, HP2}, FeynCalc \cite{FeyC1, FeyC2},
FormCalc \citep{Hahn}, and Form \citep{Form} to complete two-loop evaluation.

\begin{figure}
\begin{centering}
\includegraphics[scale=0.9]{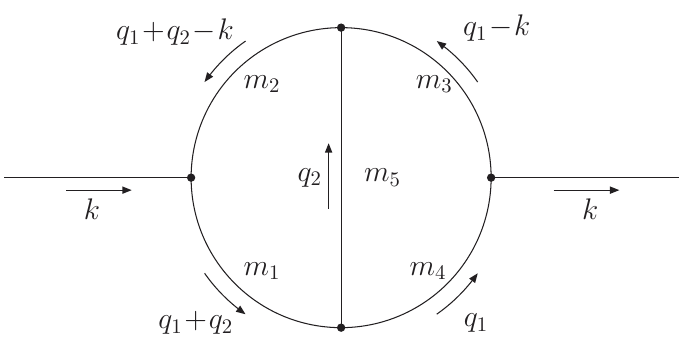}
\par\end{centering}
\caption{Two-loop scalar example.}

\label{fig-two-loop-exmpl}
\end{figure}
In order to demonstrate how the outlined roadmap can be applied to two-loop
calculations, we chose to consider the well-known example originally introduced
in~\citep{Bohm}. We start our example (see Fig.~\ref{fig-two-loop-exmpl})
with the two-loop integral
\begin{align}
U^{(2)}&=\left(\frac{\mu^{4-D}e^{\gamma_{E}(4-D)/2}}{{\rm i}\pi^{D/2}}\right)^{2}\nonumber \\ 
&\times \int  \int\frac{{\rm d}^{D}q_{1}\;{\rm d}^{D}q_{2}}{\left[q_{1}^{2}-m_{4}^{2}\right]\left[q_{2}^{2}-m_{5}^{2}\right]\left[\left(q_{1}+q_{2}\right)^{2}-m_{1}^{2}\right]\left[\left(q_{1}+q_{2}-k\right)^{2}-m_{2}^{2}\right]\left[\left(q_{1}-k\right)^{2}-m_{3}^{2}\right]},\label{eq:2r}
\end{align}
and first perform integration over the loop momentum $q_{2}$. That
results in a three-point integral
\begin{align}
J^{(3)}\left(D;1,1,1\right)
\Bigr|_{\footnotesize{\begin{array}{l}
p_{1}=0, p_{2}=q_{1}, p_{3}=q_1-k\\[-3mm]
m_{1}^{2}\leftrightarrow m_{5}^{2}, m_{2}^{2}\leftrightarrow m_{1}^{2},
m_{3}^2\leftrightarrow m_2^2
\end{array}}}
& \!\!\!\! =  \int\frac{{\rm d}^{D}q_{2}}{\left[q_{2}^{2}\!-\!m_{5}^{2}\right]\left[\left(q_{1}+q_{2}\right)^{2}\!-\!m_{1}^{2}\right]\left[\left(q_{1}+q_{2}-k\right)^{2}\!-\!m_{2}^{2}\right]}.\label{eq:3r}
\end{align}
Using Eq.~(\ref{Fp3}) (with $\bar{x}=1-x$), we get
\begin{align}
J^{(3)}\left(D;1,1,1\right)= & \int\limits_{0}^{1}{\rm d}x\;J^{(2)}\left(D;1,2\right)
\Bigr|_{\footnotesize{\begin{array}{l}
p_{1}=0,p_{2}=q_{1}-\bar{x}k\\[-3mm]
m_{1}^{2}\leftrightarrow m_{5}^{2},m_{2}^{2}\leftrightarrow xm_{1}^{2}+\bar{x}m_{2}^{2}-x\bar{x}k^{2}
\end{array}}}\nonumber \\
= & \int\limits_{0}^{1}{\rm d}x\int\frac{{\rm d}^{D}q_{2}}{\left[q_{2}^{2}-m_{5}^{2}\right]\left[\left(q_{1}-\bar{x}k+q_{2}\right)^{2}-xm_{1}^{2}-\bar{x}m_{2}^{2}+x\bar{x}k^{2}\right]^{2}}, 
\label{eq:4r}
\end{align}
which effectively gives us a reduction of the three-point integral to a two-point one. Here it is assumed that $D=4-2\varepsilon$.
Applying the recursive approach outlined in Sec.~II, we can
lower the second power of the last propagator in Eq.(\ref{eq:4r}) and
arrive at the subtracted $\bar{J}_1^{(2)}\left(2-2\varepsilon;1,1\right)$
plus tadpole terms containing $J^{(2)}\left(2-2\varepsilon;1,0\right)$ and
$J^{(2)}\left(2-2\varepsilon;0,1\right)$ (see Eq.~(\ref{2-2*ep_basis_1c})). In this way, Eq.~(\ref{eq:3r})
can be written in the following form:
\begin{align}
J^{(3)}\left(4-2\varepsilon;1,1,1\right) 
=\frac{1}{2}\int\limits_{0}^{1}{\rm d}x
\Biggl\{ &
{\rm i} \pi^{2-\varepsilon} \Gamma(\varepsilon)\;
\frac{(m_5^2)^{-\varepsilon}-(m_{12x}^2)^{-\varepsilon}}{m_{5}^{2}-m_{12x}^{2}}
\nonumber \\ &
-\left[\left(q_{1}-\bar{x}k\right)^{2}+m_{5}^{2}-m_{12x}^{2}\right]\pi \bar{J}_1^{(2)}\left(2-2\varepsilon;1,1\right)\Biggr\},
%\nonumber
\label{eq:4.5r}
\end{align}
with $m_{12x}^{2}=xm_{1}^{2}+\bar{x}m_{2}^{2}-x\bar{x}k^{2}$.
For the subtracted
$\bar{J}_1^{(2)}\left(2-2\varepsilon;1,1\right)$, we can apply dispersive
representation stemming from Eqs.~(\ref{disp11}) and (\ref{ImJ2_2m2ep_exact_}),
\begin{align*}
\bar{J}_1^{(2)}\left(2-2\varepsilon;1,1\right)=\frac{{\rm i}}{\pi}\;
\int\limits_{(m_{5}+m_{12x})^{2}}^{\infty} {\rm d}s\; \frac{{\rm{Im}}\left[{\rm i}^{-1}J^{(2)}\left(2-2\varepsilon; 1, 1\right)\right]_s}{s\left[s-\left(q_{1}-\bar{x}k\right)^{2}-{\rm i}0\right]} \; .
\end{align*}
Since the insertion $J^{(3)}\left(D;1,1,1\right)$, and the entire two-loop
integral are UV-finite, we can consider the limit $\varepsilon\to 0$,
\begin{align}
J^{(3)}\left(4;1,1,1\right) & =\frac{1}{2}\int\limits_{0}^{1}{\rm d}x
\left\{ \frac{{\rm i} \pi^{2}}{m_{5}^{2}-m_{12x}^{2}}\ln\frac{m_{12x}^{2}}{m_{5}^{2}}-\left[\left(q_{1}-\bar{x}k\right)^{2}+m_{5}^{2}-m_{12x}^{2}\right]\pi \bar{J}_1^{(2)}\left(2;1,1\right)\right\}.
\label{eq:5r}
\end{align}
Note that at
$\varepsilon=0$ the imaginary part of ${\rm i}^{-1}J^{(2)}(2-2\varepsilon;1,1)$ has a simple structure:
\begin{align*}
{\rm Im}\left[{\rm i}^{-1}J^{(2)}\left(2; 1, 1\right)\right]_s
=\frac{2\pi^2}{\sqrt{\left(s-m_{12x}^{2}-m_{5}^{2}\right)^{2}-4m_{12x}^{2}m_{5}^{2}}},
\end{align*}

At this point, we are ready to complete integration
over second-loop momentum $q_{1}$,
\begin{align}
U^{(2)} =& \frac{1}{2}\int\limits_{0}^{1}{\rm d}x
\Biggl\{\frac{1}{m_{5}^{2}-m_{12x}^{2}}\ln\frac{m_{12x}^{2}}{m_{5}^{2}}\;B_{0} \nonumber \\
 & +\frac{1}{\pi^2}\int\limits_{(m_{5}+m_{12x})^{2}}^{\infty}{\rm d}s\; 
\frac{{\rm{Im}}\left[{\rm i}^{-1}J^{(2)}\left(2; 1, 1\right)\right]_s}{s}\left[\left(s+m_{5}^{2}-m_{12x}^{2}\right)C_{0}+B_{0}\right]\Biggr\}, 
\label{eq:6r}
\end{align}
where in the second-loop integration we have used the usual PV functions
without dispersive representation. In Eq.~(\ref{eq:6r}), the three-point
function has the following arguments: $C_{0}\equiv C_{0}\left(k^{2},x^{2}k^{2},\bar{x}^{2}k^{2},m_{4}^{2},m_{3}^{2},s\right)$
(here we have used the following mapping of arguments for $C_{0}\equiv C_{0}\left(k_{1}^{2},k_{2}^{2},(k_{1}+k_{2})^{2},m_{1}^{2},m_{2}^{2},m_{3}^{2}\right)$
), and the two-point function $B_{0}\equiv B_{0}\left(k^{2},m_{4}^{2},m_{3}^{2}\right)$.
Since $B_{0}\left(k^{2},m_{4}^{2},m_{3}^{2}\right)$ does not depend
on either dispersive or Feynman parameters, we can evaluate the dispersive
integral multiplied by $B_{0}\left(k^{2},m_{4}^{2},m_{3}^{2}\right)$
analytically. As a result, the first term in Eq.~(\ref{eq:6r}) cancels
out with dispersive integration times $B_{0}\left(k^{2},m_{4}^{2},m_{3}^{2}\right)$.
The final two-loop result has a rather simple form, 
\begin{equation}
U^{(2)} =\frac{1}{2\pi^2}\int\limits_{0}^{1}{\rm d}x\int\limits_{(m_{5}+m_{12x})^{2}}^{\infty}\frac{{\rm d}s}{s}\left(s+m_{5}^{2}-m_{12x}^{2}\right)\;{\rm{Im}}\left[{\rm i}^{-1}J^{(2)}\left(2; 1, 1\right)\right]_s C_{0}.
\label{eq:7r}
\end{equation}
The three-point function can also be written analytically,
\begin{align*}
C_{0}= & \frac{1}{s-m_{43x}^{2}}\left[ x\mathfrak{\ Disc}\left(x^{2}k^{2},m_{3}^{2},s\right)+\bar{x}\ \mathfrak{Disc}\left(\bar{x}^{2}k^{2},m_{3}^{2},s\right)-\mathfrak{Disc}\left(k^{2},m_{3}^{2},m_{4}^{2}\right)\right]
\end{align*}
\begin{align}
+ & \frac{1}{2x\bar{x}k^{2}}\left(\ln\frac{m_{3}^{2}}{s}-x\ln\frac{m_{3}^{2}}{m_{4}^{2}}\right).\label{eq:8r}
\end{align}
Here, $m_{43x}^{2}=xm_{4}^{2}+\bar{x}m_{3}^{2}-x\bar{x}k^{2}$ and
$\mathfrak{Disc}\left(k^{2},m_{1}^{2},m_{2}^{2}\right)$ is a discontinuity
of the two-point function, which contains a branch cut from $(m_{1}+m_{2})^{2}$
to infinity and has the following structure:
\begin{align}
\mathfrak{Disc}\left(k^{2},m_{1}^{2},m_{2}^{2}\right) & =\frac{\sqrt{-\Delta\left(k^{2},m_{1}^{2},m_{2}^{2}\right)}}{k^{2}}\ln\left(\frac{m_{1}^{2}+m_{2}^{2}-k^{2}+\sqrt{-\Delta\left(k^{2},m_{1}^{2},m_{2}^{2}\right)}}{2m_{1}m_{2}}\right).\label{eq:9r}
\end{align}
At this point, using Eqs.~(\ref{eq:7r})--(\ref{eq:9r}), we can reproduce
numerical results for the two-loop graph $U^{(2)}$ in Fig.~\ref{fig-two-loop-exmpl}.
To make a comparison to the earlier works \citep{Bohm}
and \citep{AA1} we will use $m_{1}=2.0$~GeV, $m_{2}=1.0$~GeV, $m_{3}=4.0$~GeV, $m_{4}=5.0$~GeV and $m_{5}=3.0$~GeV.
\begin{table}
\begin{centering}
\begin{tabular}{|c|c|c|c|c|c|c|}
\hline 
$k^{2}$ (GeV$^{2}$) & This work & $\Delta t$ (sec) & \citep{AA1} & $\Delta t$ (sec) & \citep{Bohm} (Table~1) & $\Delta t$ (sec)\tabularnewline
\hline 
\hline 
$-50.0$ & $-0.08295$ & $1.0$ & $-0.08296$ & $75.0$ & - & -\tabularnewline
\hline 
$-10.0$ & $-0.18399$ & $0.7$ & $-0.18399$ & $22.0$ & - & -\tabularnewline
\hline 
$-5.0$ & $-0.22180$ & $0.7$ & $-0.22178$ & $17.0$ & - & -\tabularnewline
\hline 
$-1.0$ & $-0.26923$ & $0.7$ & $-0.26919$ & $8.0$ & - & -\tabularnewline
\hline 
$-0.5$ & $-0.27704$ & $0.7$ & $-0.27712$ & $9.0$ & - & -\tabularnewline
\hline 
$-0.1$ & $-0.28372$ & $0.7$ & $-0.28360$ & $9.0$ & - & -\tabularnewline
\hline 
$0.1$ & $-0.28723$ & $0.7$ & $-0.28714$ & $9.0$ & $-0.28724$ & $0.6$\tabularnewline
\hline 
$0.5$ & $-0.29458$ & $0.7$ & $-0.29443$ & $9.0$ & $-0.29459$ & $0.7$\tabularnewline
\hline 
$1.0$ & $-0.30451$ & $0.7$ & $-0.30449$ & $10.0$ & $-0.30452$ & $0.7$\tabularnewline
\hline 
$5.0$ & $-0.45250$ & $0.8$ & $-0.45230$ & $14.0$ & $-0.45252$ & $0.7$\tabularnewline
\hline 
$10.0$ & $-0.48807-0.35309{\rm i}$ & $4.0$ & $-0.48810-0.35318{\rm i}$ & $30.0$ & $-0.48815-0.35322{\rm i}$ & $0.7$\tabularnewline
\hline 
$50.0$ & $0.17390-0.11804{\rm i}$ & $76.0$ & $0.17335-0.11781{\rm i}$ & $1120.0$ & $0.17390-0.11808{\rm i}$ & $1.4$\tabularnewline
\hline 
\end{tabular}
\par\end{centering}
\caption{Numerical results for $U^{(2)}$ (Fig.~\ref{fig-two-loop-exmpl})
calculated from Eq.~(\ref{eq:7r}) and compared to \citep{AA1}
and \citep{Bohm}. The masses are $m_{1}=2.0$ GeV, $m_{2}=1.0$
GeV, $m_{3}=4.0$ GeV, $m_{4}=5.0$ GeV and $m_{5}=3.0$ GeV. }

\label{Tbl-r-1}
\end{table}
For the numerical integration, we shifted $k^{2}$ and all the masses
by ${\rm i}\times10^{-16}$ to remove singular behavior at the poles in Eq.~(\ref{eq:7r}).
Numerical results in Table~\ref{Tbl-r-1} are in very good agreement
if compared to previously obtained values in \citep{AA1}
 and \citep{Bohm}. It is worth noting that in this work, as well as in Ref~\citep{AA1},
\textit{Mathematica} was used to complete numerical integration using the GlobalAdaptive
method. In Ref.~\citep{Bohm}, the QUADPACK routine was applied
for numerical integration.

\eject
\section{Conclusion}

In this paper, we continue the development of the dispersive approach for the calculation of multi-loop Feynman diagrams. This study builds upon our previous work, where we introduced a general framework based on representing multi-point Passarino-Veltman functions in a two-point function basis, thereby allowing the replacement of sub-loop diagrams by effective propagators. In the present work, we extend this framework by employing shifted space-time dimensions in the tensor decomposition of the sub-loops, together with recurrence relations that systematically lower both the dimensionality and the powers of propagators. These relations algebraically minimize the number of basic dispersive integrals required for numerical evaluation.
Furthermore, the complexity of the resulting expressions can be reduced by subtracting a finite number of terms from the small-momentum expansion, which significantly improves the convergence of the dispersion integrals. Compared to the differentiation-based approach with respect to internal masses used in our earlier study, this algebraic reduction scheme proves substantially more efficient numerically, reducing computation time and enabling the treatment of more complex topologies.
Our method complements recent advances in two-loop electroweak calculations by providing a semi-analytical pathway that combines the dispersive representation of sub-loops, dimension-recurrence identities, and the two-point-basis decomposition. Since obtaining fully analytic results for general two-loop, multi-leg electroweak amplitudes remains exceptionally challenging, our approach offers a scalable and robust alternative. Instead of pursuing closed-form solutions for each diagram, we transform the problem into a compact set of well-behaved dispersive integrals amenable to stable numerical evaluation, enabling precision predictions directly applicable to current and upcoming experiments such as MOLLER, P2, and Belle~II.
Looking ahead, this framework establishes a solid foundation for the automation of multi-loop calculations in a dispersive representation. The next steps will involve implementing the reduction and integration algorithms into a numerical library optimized for precision electroweak observables and extending the method to full two-loop amplitudes with multiple mass scales. Such developments will enable comprehensive,  \textit{ab-initio} predictions for a broad class of processes relevant to the ongoing and future precision programs at JLab, MESA, and KEK, bridging the gap between analytical theory and phenomenological applications.
Ultimately, the goal is not merely the refinement of individual calculations but the construction of a predictive,  \textit{ab-initio} framework capable of interpreting deviations in upcoming experiments as definitive signals of new physics. In this broader context, the ongoing development of the dispersive approach, along with canonical two-loop methods and numerical integration tools, represents an essential component of the global precision-physics enterprise.

\begin{acknowledgments}
This work was supported in part by the Natural Sciences and Engineering Research Council of Canada (NSERC).
\end{acknowledgments}

\newpage 

\section*{Appendix A: Recurrence relations}
According to the general notation~(\ref{tensor_T_1_N}) and (\ref{JN_scalar_def}), we define the scalar two-point integrals as
\begin{equation}
\label{J2_scalar_def}
J^{(2)}(D;\nu_1,\nu_2) = 
\int \frac{{\rm d}^D q}
{\left[q^2-m_1^2\right]^{\nu_1} 
\left[(k+q)^2-m_2^2\right]^{\nu_2}} \;.
\end{equation}
Using the integration by parts technique~\cite{Tkachov,ChT} we get the following recurrence relations for these integrals~\cite{BDS}:
\begin{eqnarray}
\label{ibp1}
{\mathbf 1}^{+} J^{(2)}(D;\nu_1, \nu_2) &\!\!\!\!=\!\!\!\!& 
\frac{1}{\nu_1 \Delta}
\left\{ \left[ (k^2-m_1^2) (D-\nu_1-2\nu_2)+m_2^2 (D-3\nu_1) \right] 
\right.
%\hspace{20mm}
\nonumber \\ && \hspace{10mm}
\left.
-2 \nu_2 m_2^2 {\mathbf 1}^{-}{\mathbf 2}^{+} 
- \nu_1 (k^2-m_1^2-m_2^2) {\mathbf 1}^{+}{\mathbf 2}^{-}
\right\} J^{(2)}(D;\nu_1, \nu_2),
% \hspace*{7mm}
% \end{eqnarray}
% % and an analogous result for $J^{(2)}(\nu_1, \nu_2+1)$,
% \begin{eqnarray}
% \label{ibp2}
\\ 
\label{ibp2}
{\mathbf 2}^{+} J^{(2)}(D;\nu_1, \nu_2) &\!\!\!\!=\!\!\!\!&
\frac{1}{\nu_2 \Delta}
\left\{ \left[ (k^2-m_2^2) (D-2\nu_1-\nu_2)+m_1^2 (D-3\nu_2) \right] 
\right.
%\hspace{20mm}
\nonumber \\ && \hspace*{10mm}
\left.
-2 \nu_1 m_1^2 {\mathbf 1}^{+} {\mathbf 2}^{-}  
- \nu_2 (k^2-m_1^2-m_2^2) {\mathbf 1}^{-} {\mathbf 2}^{+} \right\}
J^{(2)}(D;\nu_1, \nu_2) ,
\hspace*{7mm}
\end{eqnarray}
with
${\mathbf 1}^{\pm}J^{(2)}(D;\nu_1, \nu_2)=J^{(2)}(D;\nu_1\pm 1, \nu_2)$, 
${\mathbf 2}^{\pm}J^{(2)}(D;\nu_1, \nu_2)=J^{(2)}(D;\nu_1, \nu_2\pm 1)$, 
and
\begin{eqnarray}
\label{Delta2}
\Delta\equiv
\Delta(m_1^2, m_2^2, k^2) 
&=& 2 k^2 m_1^2 + 2 k^2 m_2^2 + 2 m_1^2 m_2^2 - (k^2)^2 - m_1^4 - m_2^4
\nonumber \\
&=& 4 m_1^2 m_2^2 - (k^2-m_1^2-m_2^2)^2 
\nonumber \\
&=& - \left[ k^2-(m_1+m_2)^2\right]\; \left[ k^2-(m_1-m_2)^2\right]
\nonumber \\
&=& -\lambda(m_1^2, m_2^2, k^2) \; ,
\end{eqnarray}
where $\lambda(m_1^2, m_2^2, k^2)$ is the standard notation 
for the K\"all\'en function.
Note that the sum of the indices $\nu_1$ and $\nu_2$ on the rhs of Eqs.~(\ref{ibp1})--(\ref{ibp2}) is less by one than
their sum on the lhs. Therefore, by using these relations,  all the integrals with higher integer 
$\nu$'s can be expressed in terms of the integral 
$J^{(2)}(D;1,1)$ and the massive tadpoles
\begin{eqnarray}
\label{tadpole1}
J^{(2)}(D;\nu_1,0) &\!\!=\!\!& 
{\rm i}^{1-2\nu_1} \pi^{D/2} \frac{\Gamma(\nu_1\!-\!D/2)}{\Gamma(\nu_1)}
(m_1^2)^{D/2-\nu_1}
= (-m_1^2)^{1-\nu_1} 
\frac{\Gamma(\nu_1-D/2)}{\Gamma(\nu_1) \Gamma(1\!-\!D/2)} J^{(2)}(D;1,0),
\nonumber \\
&& \; \\
\label{tadpole2}
J^{(2)}(D;0,\nu_2) &\!\!=\!\!& 
{\rm i}^{1-2\nu_2} \pi^{D/2} \frac{\Gamma(\nu_2\!-\!D/2)}{\Gamma(\nu_2)}
(m_2^2)^{D/2-\nu_2}
= (-m_2^2)^{1-\nu_2} 
\frac{\Gamma(\nu_2-D/2)}{\Gamma(\nu_2) \Gamma(1\!-\!D/2)} J^{(2)}(D;0,1).
\nonumber \\
&& \;
\end{eqnarray}

Furthermore, to bring the shifted values of the space-time dimension $D$ back to $4-2\varepsilon$ (or $2-2\varepsilon$) we can use the following relation, which can be obtained by using the geometrical approach~\cite{DD-JMP,D-NIM} or the functional relations~\cite{Tarasov97}:
\begin{eqnarray}
\label{d_red_gen}
J^{(2)}(D+2;\nu_1,\nu_2) &\!\!\!=\!\!\!& -\frac{\pi}{2k^2 (D-\nu_1-\nu_2+1)}
\left[
\Delta J^{(2)}(D;\nu_1,\nu_2)
\right.
\nonumber \\ && \hspace*{40mm}
+(k^2+m_1^2-m_2^2) J^{(2)}(D;\nu_1,\nu_2-1)
\nonumber \\ && \hspace*{40mm}
\left.
+(k^2-m_1^2+m_2^2) J^{(2)}(D;\nu_1-1,\nu_2)
\right]  \; .
\hspace*{5mm}
\end{eqnarray}
In particular, for $\nu_1=\nu_2=1$,  Eq.~(\ref{d_red_gen}) yields
\begin{eqnarray}
\label{d_red}
J^{(2)}(D+2;1,1) &\!\!\!=\!\!\!& -\frac{\pi}{2k^2 (D-1)}
\left[
\Delta J^{(2)}(D;1,1)
\right.
\nonumber \\ &&
\left.
+(k^2+m_1^2-m_2^2) J^{(2)}(D;1,0)
+(k^2-m_1^2+m_2^2) J^{(2)}(D;0,1)
\right]  \; .
\hspace*{5mm}
\end{eqnarray}

To deal with the occurring tadpole integrals, we can use the following formulae (which follow from Eqs.~(\ref{tadpole1}) and (\ref{tadpole2})):
\begin{eqnarray}
\label{tadpole3}
J^{(2)}(D+2j;1,0) &=& \pi^j (m_1^2)^j 
\frac{\Gamma(1-D/2-j)}{\Gamma(1-D/2)} J^{(2)}(D;1,0),
\\
\label{tadpole4}
J^{(2)}(D+2j;0,1) &=& \pi^j (m_2^2)^j 
\frac{\Gamma(1-D/2-j)}{\Gamma(1-D/2)} J^{(2)}(D;0,1).
\end{eqnarray}
Using these relations, we can express any integral $J^{(2)}(D+2j;\nu_1,\nu_2)$ (with non-negative integers $j$, $\nu_1$ and $\nu_2$) in terms of three integrals, $J^{(2)}(D; 1,1)$, $J^{(2)}(D; 1,0)$, and $J^{(2)}(D; 0,1)$. 

Usually the recurrence w.ith respect to the space-time dimension $D$ stops at $D=4-2\varepsilon$. However, as an option, we can also use Eq.~(\ref{d_red}) one more time, 
to reduce $J^{(2)}(4-2\varepsilon; 1,1)$ to 
$J^{(2)}(2-2\varepsilon; 1,1)$ (which is UV-finite as $\varepsilon\to~0$),
\begin{eqnarray}
\label{d_red2}
J^{(2)}(4-2\varepsilon ;1,1) &\!\!\!=\!\!\!& 
-\frac{\pi}{2k^2 (1-2\varepsilon)}
\left[
\Delta J^{(2)}(2-2\varepsilon;1,1)
\right.
\nonumber \\ &&
\left.
+(k^2\!+\!m_1^2\!-\!m_2^2) J^{(2)}(2-2\varepsilon;1,0)
+(k^2\!-\!m_1^2\!+\!m_2^2) J^{(2)}(2-2\varepsilon;0,1)
\right]  \; ,
\hspace*{8mm}
\end{eqnarray}
and then use $J^{(2)}(2-2\varepsilon; 1,1)$, 
$J^{(2)}(2-2\varepsilon; 1,0)$, and $J^{(2)}(2-2\varepsilon; 0,1)$
as the master integrals.

Combining Eqs.~(\ref{ibp1}), (\ref{ibp2}), and (\ref{d_red_gen}), 
one can get another pair of useful relations~\cite{Tarasov97},
\begin{eqnarray}
\label{tar98a}
J^{(2)}(D+2;\nu_1+1,\nu_2) &=&
-\frac{\pi}{2 \nu_1 k^2}
\left[
(k^2-m_1^2+m_2^2) J^{(2)}(D;\nu_1,\nu_2)
\right.
\nonumber \\ && \hspace*{16mm}
\left.
+ J^{(2)}(D;\nu_1,\nu_2-1)
- J^{(2)}(D;\nu_1-1,\nu_2)
\right],
\\
\label{tar98b}
J^{(2)}(D+2;\nu_1,\nu_2+1) &=&
-\frac{\pi}{2 \nu_2 k^2}
\left[
(k^2+m_1^2-m_2^2) J^{(2)}(D;\nu_1,\nu_2)
\right.
\nonumber \\ && \hspace*{16mm}
\left.
- J^{(2)}(D;\nu_1,\nu_2-1)
+ J^{(2)}(D;\nu_1-1,\nu_2)
\right].
\end{eqnarray}
They can be used to simultaneously reduce one of the indices ($\nu_1$ 
or $\nu_2$) and the space-time dimension $D$. A nice property of Eqs.~(\ref{tar98a}) and (\ref{tar98b}) is the absence of $\Delta$ in the denominators.

%-------------------------------
\section*{Appendix B: $\varepsilon$ expansion of the master integral}
%-------------------------------

In general, the expansion of the master integral 
$J^{(2)}(4-2\varepsilon;1,1)$ is known to an arbitrary order in $\varepsilon$~\cite{Crete,D-ep,DK1}. Keeping terms up to the order $\varepsilon$,  we get
\begin{eqnarray}
\label{ep-exp11b}
J^{(2)}\left(4-2\varepsilon; 1, 1 \right)
&\!\!=\!\!&
\frac{\mbox{i}\pi^{2-\varepsilon}\Gamma(1+\varepsilon)}
{2(1-2\varepsilon)}
\Biggl\{
\frac{(m_1^2)^{-\varepsilon}+(m_2^2)^{-\varepsilon}}{\varepsilon}
+ \frac{m_1^2-m_2^2}{\varepsilon k^2}
\left[(m_1^2)^{-\varepsilon}-(m_2^2)^{-\varepsilon}\right]
\nonumber \\ && \hspace*{0mm}
+ \left[1 + \varepsilon \ln\left(\frac{k^2}{\Delta}\right)\right] F_1 
+ 2 \varepsilon F_2
+ {\cal{O}}(\varepsilon^2)
\Biggr\} \; , \hspace*{8mm}
\end{eqnarray}
where $\Delta\equiv \Delta(m_1^2,m_2^2,k^2)$ is defined in Eq.~(\ref{Delta2}).
For the integral in $2-2\varepsilon$ dimensions, we get (e.g., using Eq.~(\ref{d_red2}))
\begin{equation}
\label{ep-exp11b2}
J^{(2)}\left(2-2\varepsilon; 1, 1 \right)
=
-\mbox{i}\pi^{1-\varepsilon}\Gamma(1+\varepsilon)\;
\frac{k^2}{\Delta}
\left\{
\left[1 + \varepsilon \ln\left(\frac{k^2}{\Delta}\right)\right] F_1 
+ 2 \varepsilon F_2
+ {\cal{O}}(\varepsilon^2)
\right\} \; . \hspace*{8mm}
\end{equation}

Between the pseudothreshold and the threshold, when 
$(m_1-m_2)^2\leq k^2 \leq (m_1+m_2)^2$ and $\Delta\geq 0$, the functions $F_i$ can be presented as
\begin{equation}
F_i =
\frac{\sqrt{\Delta}}{k^2}
\sum\limits_{i=1}^2
\left[ \Ls{i}{\pi} - \Ls{i}{2\tau'_{0i}}\right] \; ,
\end{equation} 
where
\begin{equation}
\label{tau_prime}
%\tau'_{0i}=\frac{\pi}{2}-\tau_{0i}, \qquad
\cos\tau'_{01}=\frac{m_1^2-m_2^2+k^2}{2m_1\sqrt{k^2}}, \qquad
\cos\tau'_{02}=\frac{m_2^2-m_1^2+k^2}{2m_2\sqrt{k^2}},
\end{equation}
and the log-sine integrals are defined as
\begin{equation}
\Ls{j}{\theta}\equiv -\int_0^{\theta} \mbox{d}\theta'\; 
\ln^{j-1}\left|2\sin\frac{\theta}{2}\right| \;.
\end{equation}
In particular, $\Ls{1}{\theta}=-\theta$, and 
$\Ls{2}{\theta}=\Cl{2}{\theta}$, where 
\begin{equation}
\Cl{2}{\theta}
=\frac{1}{2\mbox{i}}\left[ \Li{2}{e^{\rm{i}\theta}}
-\Li{2}{e^{-\rm{i}\theta}}\right]
\end{equation}
is the Clausen function.
Therefore,
\begin{eqnarray}
\label{Ls1a}
F_1
&=& \frac{\sqrt{\Delta}}{k^2}\;
\sum\limits_{i=1}^2 \left[ \Ls{1}{\pi}-\Ls{1}{2\tau'_{0i}}
\right]
= -2\frac{\sqrt{\Delta}}{k^2}\;
\arccos{\left(\frac{m_1^2+m_2^2-k^2}{2m_1 m_2}\right)}\; ,
\\ 
\label{Ls2a}
F_2
&=& \frac{\sqrt{\Delta}}{k^2}
\sum\limits_{i=1}^2 \left[ \Ls{2}{\pi}-\Ls{2}{2\tau'_{0i}}\right]
= -\frac{\sqrt{\Delta}}{k^2}
\left[\Cl{2}{2\tau'_{01}}+\Cl{2}{2\tau'_{02}}\right] \; .
\end{eqnarray}

In other regions (where $\Delta<0$),  one can use analytic continuation. This process was described in Ref.~\cite{DK1}, at any order in $\varepsilon$. Introducing variables 
$z_i=e^{{\rm i}\sigma\theta_i}$, such that $\theta_i=2\tau'_{0i}$,
$\sigma=\pm1$, we get
\begin{eqnarray}
\label{Ls1b}
{\rm i}\sigma\left[ \Ls{1}{\pi}-\Ls{1}{\theta_i}
\right]
&=& \ln(-z_i) \; ,
\\ 
\label{Ls2b}
{\rm i}\sigma\left[ \Ls{2}{\pi}-\Ls{2}{\theta_i}\right]
&=& -\sfrac{1}{2} 
\left[\Li{2}{z_i} - \Li{2}{1/z_i}\right] \; .
\end{eqnarray}
In our case, the variables $z_1$ and $z_2$ can be presented as 
\begin{equation}
z_1 = \frac{k^2+m_1^2-m_2^2+\sqrt{-\Delta}}
{k^2+m_1^2-m_2^2-\sqrt{-\Delta}},
\qquad
z_2 = \frac{k^2-m_1^2+m_2^2+\sqrt{-\Delta}}
{k^2-m_1^2+m_2^2-\sqrt{-\Delta}}.
\end{equation}
In this way, we get
\begin{eqnarray}
F_1 &=& -\frac{\sqrt{-\Delta}}{k^2}
\left[
\ln\left(-z_1\right) + \ln\left(-z_2\right)
%\ln\left(z_1 z_2\right)
\right]
\; , 
\\
F_2 &=& -\frac{\sqrt{-\Delta}}{2k^2}
\left[
\Li{2}{\frac{1}{z_1}}
-\Li{2}{z_1}
+\Li{2}{\frac{1}{z_2}}
-\Li{2}{z_2}
\right] \; .
\end{eqnarray}
In particular, above the threshold (for $k^2>(m_1+m_2)^2$, where $z_1>1$ and $z_2>1$), 
we can explicitly separate the real and imaginary parts, 
\begin{eqnarray}
F_1 &=& -\frac{\sqrt{-\Delta}}{k^2}
\left[ \ln\left(z_1 z_2\right) - 2\rm{i}\pi \right]
\; , 
\\
F_2 &=& -\frac{\sqrt{-\Delta}}{2k^2}
\left[
2 \Li{2}{\frac{1}{z_1}}
+2\Li{2}{\frac{1}{z_2}}
-\sfrac{2}{3}\pi^2
\!+\!\sfrac{1}{2} \ln^2{z_1}
\!+\!\sfrac{1}{2} \ln^2{z_2}
-{\rm i}\pi\ln\left(z_1 z_2\right)
\right] . \hspace*{7mm}
\end{eqnarray}
Note that in Eq.~(\ref{ep-exp11b}) we also need to take care of the term
\begin{equation}
\ln\left(\frac{k^2}{\Delta}\right)F_1
\Rightarrow
\left[ 
\ln\left(\frac{k^2}{-\Delta}\right) + {\rm i}\pi
\right] F_1 \; .
\end{equation}
In this way, we get the following result for the imaginary part of 
${\rm i}^{-1} J^{(2)}\left(4-2\varepsilon; 1, 1 \right)$ above the threshold:
\begin{eqnarray}
\label{ImJ2_4m2ep_exp}
{\rm Im}\left[{\rm i}^{-1} J^{(2)}(4-2\varepsilon;1,1)\right]
&=& \frac{\pi^{2-\varepsilon}\Gamma(1+\varepsilon)}
{(1-2\varepsilon)}
\frac{\pi \sqrt{-\Delta}}{k^2}
\Biggl\{
1 + \varepsilon \ln\left(\frac{k^2}{-\Delta}\right) 
+ {\cal{O}}(\varepsilon^2)
\Biggr\}
\nonumber \\
&=&
\pi^{2-\varepsilon} e^{-\gamma\varepsilon}\;
\frac{\pi \sqrt{-\Delta}}{k^2}
\Biggl\{
1 + 2\varepsilon + \varepsilon \ln\left(\frac{k^2}{-\Delta}\right) 
+ {\cal{O}}(\varepsilon^2)
\Biggr\} \; . \hspace*{10mm}
\end{eqnarray}
For $J^{(2)}(2-2\varepsilon;1,1)$,  we get (e.g., using Eq.~(\ref{d_red2})):
\begin{eqnarray}
\label{ImJ2_2m2ep_exp}
{\rm Im}\left[{\rm i}^{-1} J^{(2)}(2-2\varepsilon;1,1)\right]
&=& 2\pi^{1-\varepsilon}\Gamma(1+\varepsilon)\;
\frac{\pi}{\sqrt{-\Delta}}
\Biggl\{
1 + \varepsilon \ln\left(\frac{k^2}{-\Delta}\right) 
+ {\cal{O}}(\varepsilon^2)
\Biggr\}
\nonumber \\
&=&
2\pi^{1-\varepsilon} e^{-\gamma\varepsilon}\;
\frac{\pi}{\sqrt{-\Delta}}
\Biggl\{
1 + \varepsilon \ln\left(\frac{k^2}{-\Delta}\right) 
+ {\cal{O}}(\varepsilon^2)
\Biggr\} \; . \hspace*{10mm}
\end{eqnarray}
One can also obtain the result for an arbitrary $\varepsilon$ (see ~\cite{DR,DD-JPA}), 
\begin{equation}
\label{ImJ2_2m2ep_exact}
{\rm Im}\left[{\rm i}^{-1} J^{(2)}(2-2\varepsilon;1,1)\right]
= 2\pi^{1-\varepsilon}\frac{\Gamma(1-\varepsilon)}
{\Gamma(1-2\varepsilon)}\; 
\frac{\pi}{\sqrt{-\Delta}}
\left(\frac{k^2}{-\Delta}\right)^{\varepsilon} \; .
\end{equation}

%-------------------------------
\section*{Appendix C: Small-$k^2$ expansion of the two-point function}
%-------------------------------

For small $k^2$ and arbitrary $m_1$ and $m_2$, using Eq.~(20) of Ref.~\cite{BD-TMF},  we get (for unit powers of propagators)
\begin{eqnarray}
\label{J2_F4a}
J^{(2)}(D;1,1) &\!\!=\!\!& -{\rm i} \pi^{D/2} (m_2^2)^{D/2-2}
\Gamma(1 - D/2)
\nonumber \\ &&
\times \Biggl\{
F_4\left( 1, 2-D/2; D/2, 2-D/2 \Bigl| 
\frac{k^2}{m_2^2}, \frac{m_1^2}{m_2^2} \right)
\nonumber \\ && \hspace*{3mm}
-\left(\frac{m_1^2}{m_2^2}\right)^{D/2-1} 
F_4\left( 1, 2-D/2; D/2, 2-D/2 \Bigl| 
\frac{k^2}{m_2^2}, \frac{m_1^2}{m_2^2} \right) \; ,
\hspace*{8mm}
\end{eqnarray}
where
\begin{equation}
\label{def_F4}
F_4(a,b;c,d|x,y)
=\sum\limits_{j_1=0}^{\infty} \sum\limits_{j_2=0}^{\infty}
\frac{(a)_{j_1+j_2}(b)_{j_1+j_2}}{(c)_{j_1}(d)_{j_2}}
\frac{x^{j_1}y^{j_2}}{j_1! j_2!}
\end{equation}
is Appell's hypergeometric function of two variables. The sum over $j_2$ produces the Gauss hypergeometric function,
\begin{equation}
\label{def_F4_2}
F_4(a,b;c,d|x,y)
=\sum\limits_{j=0}^{\infty}
\frac{x^j}{j!}
\frac{(a)_j (b)_j}{(c)_j}\;
{}_2F_1\left(\left.
\begin{array}{c}a+j, b+j\\d \end{array} \right| y \right) \; .
\end{equation}
Therefore, we can express the integral~(\ref{J2_F4a}) as
\begin{eqnarray}
\label{J2_F4b}
J^{(2)}(D;1,1) &\!\!=\!\!& -\frac{{\rm i} \pi^{D/2}}{m_2^2} 
\sum\limits_{j=0}^{\infty}
\left(\frac{k^2}{m_2^2}\right)^j \frac{1}{(D/2)_j}
\nonumber \\ &&
\times \Biggl\{
(m_2^2)^{D/2-1} (2-D/2)_j\;
{}_2F_1\left(\left.
\begin{array}{c}1+j, 2-D/2+j\\2-D/2 \end{array} \right|
\frac{m_1^2}{m_2^2} \right)
\nonumber \\ && \hspace*{5mm}
-(m_1^2)^{D/2-1} (D/2)_j\;
{}_2F_1\left(\left.
\begin{array}{c}1+j, D/2+j\\D/2 \end{array} \right|
\frac{m_1^2}{m_2^2} \right)
\Biggr\} \; .
\end{eqnarray}

The occurring $_2F_1$ functions can be transformed into truncating $_2F_1$ functions (see, e.g., Eq.~(7.3.1.26) of Ref.~\cite{PBM3}),
\begin{eqnarray}
\label{truncating1}
{}_2F_1\left(\left.
\begin{array}{c}1+j, 1+\alpha+j\\ 1+\alpha \end{array} \right|z \right)
&=& (1-z)^{-1-2j}
{}_2F_1\left(\left.
\begin{array}{c} -j, \alpha-j\\ 1+\alpha \end{array} \right|z \right) \; ,
\end{eqnarray}
where $\alpha=1-D/2$ in the first case and $\alpha=D/2-1$ in the second case. The resulting finite sums can be written as
\begin{equation}
\label{truncating2}
{}_2F_1\left(\left.
\begin{array}{c} -j, \alpha-j\\ 1+\alpha \end{array} \right|z \right)
= \sum\limits_{l=0}^j 
\frac{(\alpha-j)_l}{(\alpha+1)_l}\; 
%\Biggl( \begin{array}{c} j\\l \end{array} \Biggr) 
\frac{j!}{l!(j-l)!}
(-z)^l ,
\end{equation}
where the ratio of factorials is nothing but the binomial coefficient.

Using Eq.~(\ref{truncating2}),  we get
\begin{eqnarray}
\label{J2_F4c}
J^{(2)}(D;1,1) &\!\!=\!\!& -{\rm i} \pi^{D/2} 
\Gamma\left(1-\frac{D}{2}\right) 
\sum\limits_{j=0}^{\infty}
\frac{(k^2)^j}{(D/2)_j}\; \frac{1}{(m_2^2-m_1^2)^{1+2j}}
\nonumber \\ &&
\times \Biggl\{
(m_2^2)^{D/2-1} (2-D/2)_j\;
\sum\limits_{l=0}^j
\frac{(1-D/2-j)_l}{(2-D/2)_l}\;
\frac{j!}{l!(j-l)!} (-m_1^2)^l (m_2^2)^{j-l}
\nonumber \\ && \hspace*{5mm}
-(m_1^2)^{D/2-1} (D/2)_j\;
\sum\limits_{l=0}^j
\frac{(D/2-1-j)_l}{(D/2)_l}\;
\frac{j!}{l!(j-l)!} (-m_1^2)^l (m_2^2)^{j-l}
\Biggr\} \; . \hspace*{8mm}
\end{eqnarray}
Applying well-known transformations of the Pochhammer symbols, we arrive at an explicitly symmetric result, 
\begin{eqnarray}
\label{J2_F4ccc}
J^{(2)}(D;1,1) &\!\!=\!\!& -{\rm i} \pi^{D/2} 
\Gamma\left(1-\frac{D}{2}\right) 
\sum\limits_{j=0}^{\infty}
(k^2)^j\; \frac{(2-D/2)_j}{(m_2^2-m_1^2)^{1+2j}}
\nonumber \\ &&
\times \Biggl\{
(m_2^2)^{D/2-1}
\sum\limits_{l=0}^j
\frac{j!}{l!(j-l)!}\;
\frac{(m_2^2)^l (m_1^2)^{j-l}}{(D/2)_l (2-D/2)_{j-l}}\;
\nonumber \\ && \hspace*{5mm}
-(m_1^2)^{D/2-1}
\sum\limits_{l=0}^j
\frac{j!}{l!(j-l)!}\;
\frac{(m_1^2)^l (m_2^2)^{j-l}}{(D/2)_l (2-D/2)_{j-l}}\;
\Biggr\} \; . \hspace*{8mm}
\end{eqnarray}
This result can be also presented in terms of the tadpole integrals $J^{(2)}(D;1,0)$ and $J^{(2)}(D;0,1)$, 
\begin{eqnarray}
\label{J2_F4dd}
J^{(2)}(D;1,1) &\!\!=\!\!& 
\sum\limits_{j=0}^{\infty}
(k^2)^j\; \frac{(2-D/2)_j}{(m_2^2-m_1^2)^{1+2j}}
\nonumber \\ &&
\times \Biggl\{
J^{(2)}(D;0,1)\;
\sum\limits_{l=0}^j
\frac{j!}{l!(j-l)!}\;
\frac{(m_2^2)^l (m_1^2)^{j-l}}{(D/2)_l (2-D/2)_{j-l}}\;
\nonumber \\ && \hspace*{5mm}
-J^{(2)}(D;1,0)\;
\sum\limits_{l=0}^j
\frac{j!}{l!(j-l)!}\;
\frac{(m_1^2)^l (m_2^2)^{j-l}}{(D/2)_l (2-D/2)_{j-l}}\;
\Biggr\} \; . \hspace*{8mm}
\end{eqnarray}

It is easy to check that the limit $m_1=m_2\equiv m$ is regular. In this case, Eq.~(17) of Ref.~\cite{BD-TMF} yields
\begin{eqnarray}
\label{J2_mm}
J^{(2)}(D;1,1)\Bigr|_{k^2\to 0,\; m_1=m_2\equiv m}
&=& {\rm i} \pi^{D/2}\; (m^2)^{D/2-2} 
\Gamma\left(2-\frac{D}{2}\right)\;
{}_2F_1\left(\left.
\begin{array}{c} 1, 2-D/2 \\ 3/2 \end{array} \right| 
\frac{k^2}{4m^2} \right)
\nonumber \\
&=& {\rm i} \pi^{D/2}\; (m^2)^{D/2-2} 
\Gamma\left(2-\frac{D}{2}\right)\;
\sum\limits_{j=0}^{\infty}
\frac{(2-D/2)_j}{(3/2)_j}\; \left( \frac{k^2}{4m^2} \right)^j .
\hspace*{8mm}
\end{eqnarray}
Using Eq.~(\ref{J2_F4c}), we have checked that the first 20 terms
of its $k^2$ expansion in the limit $m_2\to~m_1$ are the same
as in Eq.~(\ref{J2_mm}). 

Let us also consider the case $m_1=0$, $m_2\equiv m$. In this case, using Eq.~(10) of Ref.~\cite{BD-TMF}, we get
\begin{eqnarray}
\label{J2_0m}
J^{(2)}(D;1,1)\Bigr|_{k^2\to 0,\; m_1=0,\; m_2\equiv m}
&\!\!=& -{\rm i} \pi^{D/2}\; (m^2)^{D/2-2} 
\Gamma\left(1-\frac{D}{2}\right)\;
{}_2F_1\left(\left.
\begin{array}{c} 1, 2-D/2 \\ D/2 \end{array} \right| 
\frac{k^2}{m^2} \right)
\nonumber \\
&\!\!=& -{\rm i} \pi^{D/2}\; (m^2)^{D/2-2} 
\Gamma\left(1-\frac{D}{2}\right)\;
\sum\limits_{j=0}^{\infty}
\frac{(2\!-\!D/2)_j}{(D/2)_j}\; \left( \frac{k^2}{m^2} \right)^j .
\hspace*{9mm}
\end{eqnarray}
Let us now consider the limit $m_1=0$, $m_2\equiv m$ in Eq.~(\ref{J2_F4c}). The third line (containing $(m_1^2)^{D/2-1}$) should be omitted because it corresponds to a massless tadpole. 
In the sum on the second line,  we only need to keep the term
with $l=j$ because all the others vanish. As a result,  we get
\[
-{\rm i} \pi^{D/2}\; (m^2)^{D/2-2} 
\Gamma\left(1-\frac{D}{2}\right)\;
\sum\limits_{j=0}^{\infty}
\frac{(D/2-1-j)_j}{(D/2)_j} \left( -\frac{k^2}{m^2} \right)^j .
\]
Transforming the Pochhammer symbol as 
$(D/2-1-j)_j=(-1)^j (2-D/2)_j$,  we reproduce the same result as 
in Eq.~(\ref{J2_0m}). 

%-------------------------------
\section*{Appendix D: Special cases of the two-point function}
%-------------------------------

In the special case $m_1=0$, $m_2\equiv m$ we can use Eqs.~(2.24) and (2.25) of Ref.~\cite{DK1}. In particular, in Eq.~(2.25) an arbitrary term of the $\varepsilon$ expansion is presented in terms of Nielsen polylogarithms $S_{a,b}(u)$, with $u\equiv k^2/m^2$. Taking into account that $S_{a,1}(u)=\Li{a+1}{u}$,  we get
\begin{eqnarray}
J^{(2)}(4-2\varepsilon;1,1)\Bigr|_{m_1=0,\; m_2\equiv m}
&=& {\rm i}\pi^{2-\varepsilon} (m^2)^{-\varepsilon} 
\frac{\Gamma(1+\varepsilon)}{1-2\varepsilon}
\Biggl\{
\frac{1}{\varepsilon}
-\frac{1-u}{2u\varepsilon}\left[(1-u)^{-2\varepsilon}-1\right]
\nonumber \\ &&
-\frac{\varepsilon (1-u)^{-2\varepsilon}}{u}\Li{2}{u} 
+{\cal{O}}(\varepsilon^2)
\Biggr\} .
\end{eqnarray}
The threshold corresponds to the point $u=1$ ($k^2=m^2$). To go beyond the threshold,  we can use
\[
\Li{2}{u}=-\Li{2}{\frac{1}{u}}+\sfrac{1}{3}\pi^2 -\sfrac{1}{2}\ln^2{u}
+{\rm i}\pi\ln{u} .
\]
Here the sign of the imaginary part is fixed according to the causal prescription ($k^2\leftrightarrow k^2+{\rm i}0$). Using
\[
\Li{2}{u} = -\Li{2}{1-u} + \sfrac{1}{6}\pi^2 -\ln{u}\ln(1-u)
\]
we can get as many terms of the ``on-shell'' expansion in powers and logarithms of $1-u=(m^2-k^2)/m^2$ as we like. In particular, at $k^2=m^2$ ($u=1$) we get
\begin{equation}
\label{special5}
J^{(2)}(4-2\varepsilon;1,1)\Bigr|_{m_1=0,\; m_2\equiv m, \; k^2=m^2}
= {\rm i}\pi^{2-\varepsilon} (m^2)^{-\varepsilon} 
\frac{\Gamma(\varepsilon)}{1-2\varepsilon} \; .
\end{equation}

For $m_1=m_2=0$,  we can use the well-known result
\[
J^{(2)}(4-2\varepsilon;1,1)\Bigr|_{m_1=m_2=0}
={\rm i} \pi^{2-\varepsilon} (-k^2)^{-\varepsilon}\;
\frac{\Gamma^2(1-\varepsilon)\;\Gamma(\varepsilon)}
{\Gamma(2-2\varepsilon)} \; .
\]

For $k^2=0$ and arbitrary $m_1$ and $m_2$,  we get (see, e.g., in Ref.~\cite{BD-TMF}, after Eq.~(21))
\begin{equation}
\label{k^2=0}
J^{(2)}(4-2\varepsilon;1,1)\Bigr|_{k^2=0}
= -{\rm i}\pi^{2-\varepsilon}\Gamma(-1+\varepsilon)\;
\frac{(m_1^2)^{1-\varepsilon}-(m_2^2)^{1-\varepsilon}}{m_1^2-m_2^2} \;.
\end{equation}
For $m_1=m_2\equiv m$,  Eq.~(\ref{k^2=0}) yields
\[
J^{(2)}(4-2\varepsilon;1,1)\Bigr|_{k^2=0,\; m_1=m_2\equiv m}
= {\rm i}\pi^{2-\varepsilon}\Gamma(\varepsilon)\;(m^2)^{-\varepsilon}\;,
\]
and for $m_1=0$, $m_2\equiv m$ we get
\[
J^{(2)}(4-2\varepsilon;1,1)\Bigr|_{k^2=0,\; m_1=0, \; m_2\equiv m}
= -{\rm i}\pi^{2-\varepsilon}\Gamma(-1+\varepsilon)\;
(m^2)^{-\varepsilon}\;.
\]

Considering the case $m_1\equiv m$, $m_2=\lambda$, $k^2=m^2$ and using the general hypergeometric representation, after some transformations we get
\begin{eqnarray}
J^{(2)}(4\!-\!2\varepsilon;1,1)\Bigr|_{k^2=m^2,\; m_1\equiv m, \;m_2\equiv \lambda}
&\!\!\!=\!\!\!& {\rm i}\pi^{2-\varepsilon}
\Biggl\{
\frac{\Gamma(\varepsilon)(m^2)^{-\varepsilon}}{1-2\varepsilon} 
\left( 1\!-\!\frac{\lambda^2}{2m^2}\right)
{}_2F_1\Biggl( 
\begin{array}{c}
1, \; \varepsilon \\ 1/2 \!+\! \varepsilon
\end{array}
\Biggr|
\frac{\lambda^2 (4m^2\!-\!\lambda^2)}{4m^4}
\Biggr)
\nonumber \\
&& + \Gamma\left(\sfrac{3}{2}\right)
\Gamma\left(-\sfrac{1}{2}+\varepsilon\right)
(m^2)^{-1/2} (\lambda^2)^{1/2-\varepsilon}
\left( 1 - \frac{\lambda^2}{4m^2}\right)^{1/2-\varepsilon}
\nonumber \\
&& + \frac{\Gamma(\varepsilon)}{2m^2} (\lambda^2)^{1-\varepsilon}
{}_2F_1\Biggl( 
\begin{array}{c}
1, \; \varepsilon \\ 3/2
\end{array}
\Biggr|
\frac{\lambda^2}{4m^2}
\Biggr)
\Biggr\} \; .
\end{eqnarray} 
In the limit $\lambda\to 0$,  we reproduce Eq.~(\ref{special5}).

Analogous results for special cases of 
$J^{(2)}(2-2\varepsilon;1,1)$ can be obtained by using Eq.~(\ref{d_red2}).
 
\eject


\begin{thebibliography}{99}

\bibitem{MOLLER} MOLLER Collaboration (J.~Benesch (Jefferson Lab)
et al.), JLAB-PHY-14-1986 (2014) {[}arXiv:1411.4088{]}.

\bibitem{P2} D.~Becker (U. Mainz, PRISMA \& Mainz U., Inst. Kernphys.)
et al., DOI: 10.1140/epja/i2018-12611-6 (2018) {[}arXiv:1802.04759{]}.

%\cite{JeffersonLabSoLID:2022iod}
\bibitem{SOLID}
J.~Arrington \textit{et al.,} [Jefferson Lab SoLID],
%``The solenoidal large intensity device (SoLID) for JLab 12 GeV,''
J. Phys. \textbf{G50}, 110501 (2023)
%doi:10.1088/1361-6471/acda21
[arXiv:2209.13357 [nucl-ex]].
%47 citations counted in INSPIRE as of 26 Oct 2025

\bibitem{EIC}
V.~Burkert et al.,
Prog. Part. Nucl. Phys. \textbf{131}, 104032 (2023) 
{[}arXiv:2211.15746 [nucl-ex]{]}. 
%	title = {Precision studies of QCD in the low energy domain of the EIC},

\bibitem{PV1979}
G.~Passarino and M.~J.~G.~Veltman,
%``One Loop Corrections for e+ e- Annihilation Into mu+ mu- in the Weinberg Model,''
Nucl. Phys. \textbf{B160}, 151-207 (1979).

%\cite{Hollik:1988ii}
\bibitem{Hollik1990}
W.~F.~L.~Hollik,
%``Radiative Corrections in the Standard Model and their Role for Precision Tests of the Electroweak Theory,''
Fortsch. Phys. \textbf{38}, 165-260 (1990).
%doi:10.1002/prop.2190380302
%631 citations counted in INSPIRE as of 25 Oct 2025

%\cite{Denner:1991kt}
\bibitem{Denner1993}
A.~Denner,
%``Techniques for calculation of electroweak radiative corrections at the one loop level and results for W physics at LEP-200,''
Fortsch. Phys. \textbf{41}, 307-420 (1993)
%doi:10.1002/prop.2190410402
[arXiv:0709.1075 [hep-ph]].
%1225 citations counted in INSPIRE as of 25 Oct 2025

%\cite{Tarasov:1996br}
\bibitem{Tarasov96}
O.~V.~Tarasov,
%``Connection between Feynman integrals having different values of the space-time dimension,''
Phys. Rev. D \textbf{54}, 6479-6490 (1996)
%doi:10.1103/PhysRevD.54.6479
[hep-th/9606018].
%589 citations counted in INSPIRE as of 25 Oct 2025

\bibitem{Tarasov97}
O.~V.~Tarasov, Nucl. Phys. {\bf B502}, 455--482 (1996) {[}hep-ph/9703319{]}.

\bibitem{PLB91}
A.~I.~Davydychev, Phys. Lett. {\bf B263}, 107--111 (1991).

\bibitem{GG}
W.~T.~Giele and E.~W.~N.~Glover, JHEP {\bf 04}, 029 (2004) {[}hep-ph/0402152{]}.

\bibitem{BGHPS}
T.~Binoth, J.~Ph.~Guillet, G.~Heinrich, E.~Pilon and C.~Schubert,
JHEP {\bf 10}, 015 (2005) {[}hep-ph/0504267{]}.

\bibitem{EGZ}
R.~K.~Ellis, W.~T.~Giele and G.~Zanderighi, Phys. Rev. {\bf D73}, 014027  (2006) {[}hep-ph/0508308{]}.

%\cite{Lee:2014ioa}
\bibitem{Lee}
R.~N.~Lee,
%``Reducing differential equations for multiloop master integrals,''
JHEP {\bf 04}, 108 (2015)
%doi:10.1007/JHEP04(2015)108
{[}arXiv:1411.0911 [hep-ph]{]}.
%297 citations counted in INSPIRE as of 26 Oct 2025

\bibitem{SmrBook}
V.~A.~Smirnov,
Evaluating Feynman integrals, Springer Tracts in Modern Physics, {\bf 211}, 1--244 (2004). 

\bibitem{Kotikov} 
A.~V.~Kotikov,
% Differential equations and Feynman integrals,
in "Anti-differentiation and the calculation of Feynman amplitudes" (2021), p. 235--259, {[}arXiv:2102.07424{]}.

\bibitem{BSch}
J.~Bl\"umlein and C.~Schneider,
% The SAGEX review on scattering amplitudes Chapter 4: Multi-loop Feynman integrals
J. Phys. {\bf A55}, 443005 (2022), {[}arXiv:2203.13015{]}.

\bibitem{BSSch}
J.~Bl\"umlein, M.~Saragnese and C.~Schneider,
% Hypergeometric structures in Feynman integrals
Ann. Math. Artif. Intell. {\bf 91}, 591--649 (2023) {[}arXiv:2111.15501{]}.

%\cite{Wasser:2022kwg}
\bibitem{Wasser}
P.~Wasser,
%``Scattering Amplitudes and Logarithmic Differential Forms,''
doi:10.25358/openscience-6801 (2025).
%5 citations counted in INSPIRE as of 26 Oct 2025

%\cite{Henn:2020lye}
\bibitem{Henn}
J.~Henn, B.~Mistlberger, V.~A.~Smirnov and P.~Wasser,
%``Constructing d-log integrands and computing master integrals for three-loop four-particle scattering,''
JHEP \textbf{04}, 167 (2020)
%doi:10.1007/JHEP04(2020)167
[arXiv:2002.09492 [hep-ph]].
%133 citations counted in INSPIRE as of 26 Oct 2025

%\cite{Aleksejevs:2011de}
\bibitem{AA-1}
A.~Aleksejevs, S.~Barkanova, Y.~Kolomensky, E.~Kuraev and V.~Zykunov,
%``Quadratic electroweak corrections for polarized Moller scattering,''
Phys. Rev. \textbf{D85}, 013007 (2012)
%doi:10.1103/PhysRevD.85.013007
[arXiv:1110.1750 [hep-ph]].
%17 citations counted in INSPIRE as of 25 Oct 2025

%\cite{Aleksejevs:2012xua}
\bibitem{AA-2}
A.~G.~Aleksejevs, S.~G.~Barkanova, Y.~M.~Bystritskiy, A.~N.~Ilyichev, E.~A.~Kuraev and V.~A.~Zykunov,
%``Double-box contributions to Moeller scattering in the standard model,''
Eur. Phys. J. \textbf{C72}, 2249 (2012).
%doi:10.1140/epjc/s10052-012-2249-x
%11 citations counted in INSPIRE as of 25 Oct 2025

%\cite{Aleksejevs:2013gxa}
\bibitem{AA-3}
A.~G.~Aleksejevs, S.~G.~Barkanova, V.~A.~Zykunov and E.~A.~Kuraev,
%``Estimating two-loop radiative effects in the MOLLER experiment,''
Phys. Atom. Nucl. \textbf{76}, 888--900 (2013).
%doi:10.1134/S1063778813070028
%16 citations counted in INSPIRE as of 25 Oct 2025

%\cite{Aleksejevs:2015dba}
\bibitem{AA-4}
A.~G.~Aleksejevs, S.~G.~Barkanova, Y.~M.~Bystritskiy, E.~A.~Kuraev and V.~A.~Zykunov,
%``NNLO Electroweak corrections for polarized M{\o}ller scattering: One-loop insertions to boxes,''
Phys. Part. Nucl. Lett. \textbf{12}, no.5, 645--656 (2015)
%doi:10.1134/S1547477115050039
[arXiv:1504.03560 [hep-ph]].
%11 citations counted in INSPIRE as of 25 Oct 2025

%\cite{Aleksejevs:2015zya}
\bibitem{AA-5}
A.~G.~Aleksejevs, S.~G.~Barkanova, Y.~M.~Bystritskiy, E.~A.~Kuraev and V.~A.~Zykunov,
%``Two-loop electroweak vertex corrections for polarized M{\o}ller scattering,''
Phys. Part. Nucl. Lett. \textbf{13}, no.3, 310-317 (2016)
%doi:10.1134/S1547477116030031
[arXiv:1508.07853 [hep-ph]].
%9 citations counted in INSPIRE as of 25 Oct 2025

\bibitem{AA1} A.~Aleksejevs, Phys. Rev. {\bf D98}, 036021 (2018) {[}hep-th/1804.08914{]}.

%\cite{Aleksejevs:2018xvj}
\bibitem{AACR}
A.~Aleksejevs, 
%``Crossed Topology in Two-Loop Dispersive Approach,''
[arXiv:1809.05592 [hep-th]].
%4 citations counted in INSPIRE as of 17 Nov 2025

\bibitem{AB19}
%\cite{Aleksejevs:2019oml}
A.~Aleksejevs and S.~Barkanova, (2019)
%``Dimensional Regularization and Dispersive Two-Loop Calculations,''
[arXiv:1905.07936 [hep-th]].

%\cite{Aleksejevs:2019rdz}
\bibitem{AB19a}
A.~Aleksejevs and S.~Barkanova,
%``Dispersive Two-Loop Calculations: Methodology and Applications,''
PoS LeptonPhoton2019, 090 (2019)
%doi:10.22323/1.367.0090
[arXiv:1912.04762 [hep-th]].

\bibitem{ChYCh}
X.-L.~Chen, P.-F.~Yang and W.~Chen, 
%Discontinuities of banana integrals in dispersion relation representation
Chin. Phys. Lett. {\bf 41}, 111101 (2024)
{[}arXiv:2405.19868{]}.

%\cite{Song:2021vru}
\bibitem{FR1a}
Q.~Song and A.~Freitas,
%``On the evaluation of two-loop electroweak box diagrams for $e^+e^- \to HZ$ production,''
JHEP \textbf{04}, 179 (2021)
%doi:10.1007/JHEP04(2021)179
[arXiv:2101.00308 [hep-ph]].
%24 citations counted in INSPIRE as of 17 Nov 2025

%\cite{Freitas:2022hyp}
\bibitem{FR1b}
A.~Freitas and Q.~Song,
%``Two-Loop Electroweak Corrections with Fermion Loops to e+e-{\textrightarrow}ZH,''
Phys. Rev. Lett. \textbf{130}, 031801 (2023)
%doi:10.1103/PhysRevLett.130.031801
[arXiv:2209.07612 [hep-ph]].
%16 citations counted in INSPIRE as of 17 Nov 2025

%\cite{Dubovyk:2016aqv}
\bibitem{Dubovyk1}
I.~Dubovyk, A.~Freitas, J.~Gluza, T.~Riemann and J.~Usovitsch,
%``The two-loop electroweak bosonic corrections to $\sin^2\theta^\textrm{b}_\textrm{eff}$,''
Phys. Lett. \textbf{B 762}, 184-189 (2016)
%doi:10.1016/j.physletb.2016.09.012
[arXiv:1607.08375 [hep-ph]].
%59 citations counted in INSPIRE as of 25 Oct 2025

%\cite{Dubovyk:2022frj}
\bibitem{Freitas1a}
I.~Dubovyk, A.~Freitas, J.~Gluza, K.~Grzanka, M.~Hidding and J.~Usovitsch,
%``Evaluation of multiloop multiscale Feynman integrals for precision physics,''
Phys. Rev. \textbf{D106}, L111301 (2022)
%doi:10.1103/PhysRevD.106.L111301
[arXiv:2201.02576 [hep-ph]].
%32 citations counted in INSPIRE as of 26 Oct 2025

%\cite{Erler:2022ckm}
\bibitem{Erler}
J.~Erler, R.~Ferro-Hern{\'a}ndez and A.~Freitas,
%``Hadronic effects in M{\o}ller scattering at NNLO,''
JHEP \textbf{08}, 183 (2022)
%doi:10.1007/JHEP08(2022)183
[arXiv:2202.11976 [hep-ph]].
%6 citations counted in INSPIRE as of 26 Oct 2025

%\cite{Schwanemann:2024kbg}
\bibitem{Schw}
N.~Schwanemann and S.~Weinzierl,
%``Electroweak double-box integrals for M{\o}ller scattering,''
SciPost Phys. \textbf{18}, no.6, 172 (2025)
%doi:10.21468/SciPostPhys.18.6.172
[arXiv:2412.07522 [hep-ph]].
%4 citations counted in INSPIRE as of 26 Oct 2025


\bibitem{Freitas1} 
A.~Freitas, Prog. Part. Nucl. Phys. {\bf 90}, 201--240 (2016) {[}arXiv:1604.00406 [hep-ph]{]}.

\bibitem{Kreimer} D.~Kreimer, Phys. Lett. {\bf B273}, 277-281 (1991).

\bibitem{Czarnecki} A.~Czarnecki, U.~Kilian and D.~Kreimer. Nucl. Phys. {\bf B433}, 259--275 (1995) {[}hep-ph/9405423{]}.

\bibitem{Frink} A.~Frink, U.~Kilian and D.~Kreimer. Nucl. Phys. {\bf B488}, 426--440 (1997) {[}hep-ph/9610285{]}.

\bibitem{Adams1} L.~Adams, C.~Bogner and S.~Weinzierl, J. Math. Phys.
{\bf 54}, 052303 (2013) {[}arXiv:1302.7004 [hep-ph]{]}.

\bibitem{Adams2} L.~Adams, C.~Bogner and S.~Weinzierl, J. Math. Phys.
{\bf 56}, 072303 (2015) {[}arXiv:1504.03255 [hep-ph]{]}.

\bibitem{Adams3} L.~Adams, C.~Bogner and S.~Weinzierl, J. Math. Phys.
{\bf 57}, 032304 (2016) {[}arXiv:1512.05630 [hep-ph]{]}.

\bibitem{Remiddi1} E.~Remiddi and L.~Tancredi, Nucl. Phys. {\bf B907}, 400--444 (2016) {[}arXiv:1602.01481 [hep-ph]{]}.

\bibitem{Bloch1} S.~Bloch, M.~Kerr and P.~Vanhove, Compos. Math. {\bf 151},
2329--2375 (2015) {[}arXiv:1406.2664 [hep-th]{]}.

\bibitem{Bloch2} S.~Bloch, M.~Kerr and P.~Vanhove, Adv. Theor. Math.
Phys. {\bf 21}, 1373--1453 (2017) {[}arXiv:1601.08181 [hep-th]{]}.

\bibitem{Borowka1} S.~Borowka, J.~Carter and G.~Heinrich, J. Phys. Conf.
Ser. {\bf 368}, 012051 (2012) {[}arXiv:1206.4908 [hep-ph]{]}.

\bibitem{Borowka2} S.~Borowka, J.~Carter and G.~Heinrich, Comput. Phys.
Commun. {\bf 184}, 396--408 (2013) {[}arXiv:1204.4152 [hep-ph]{]}.

\bibitem{Hollik-1} W.~Hollik, U.~Meier and S.~Uccirati, Nucl. Phys.
{\bf B731}, 213--224 (2005) {[}hep-ph/0507158{]}.

\bibitem{Hollik-2} A.~Freitas, W.~Hollik, W.~Walter and G~ Weiglein,
Nucl. Phys. {\bf B632}, 189--218 (2002) {[}hep-ph/0202131{]}.

\bibitem{Gluza2005} M.~Czakon, J.~Gluza and T.~Riemann, Phys. Rev. {\bf D71},
073009 (2005) {[}hep-ph/0412164{]}.

\bibitem{Gluza2008} S.~Actis, M.~Czakon, J.~Gluza and T.~Riemann,
Phys. Rev. Lett. {\bf 100}, 131602 (2008) {[}arXiv:0711.3847 [hep-ph]{]}.

\bibitem{FeyC3}
R.~Mertig, M.~B\"ohm and A.~Denner,
%``FEYN CALC: Computer algebraic calculation of Feynman amplitudes,''
Comput. Phys. Commun. \textbf{64}, 345--359 (1991).

%\cite{Erler:2025mub}
\bibitem{Erler1}
J.~Erler, Proc. Sci., CORFU2024 (2025) 235
%``Implications of Recent Experimental {\&} Theoretical Results on Electroweak Precision Tests,''
[arXiv:2505.03457 [hep-ph]].
%0 citations counted in INSPIRE as of 26 Oct 2025

%\cite{ParticleDataGroup:2024cfk}
\bibitem{PDG2024}
S.~Navas \textit{et al.} [Particle Data Group],
%``Review of particle physics,''
Phys. Rev. \textbf{D110}, 030001 (2024).
%doi:10.1103/PhysRevD.110.030001
%2966 citations counted in INSPIRE as of 26 Oct 2025

\bibitem{Hahn} T.~Hahn and M.~Perez-Victoria, Comput. Phys.
Commun. {\bf 118}, 153--165 (1999) {[}hep-ph/9807565{]}.

\bibitem{COL1}  
A.~Denner, S.~Dittmaier and L.~Hofer, 
Comput. Phys. Commun. {\bf 212}, 220--238 (2017) 
{[}arXiv:1604.06792{]}.

\bibitem{COL2}   
A.~Denner and S.~Dittmaier,  Nucl. Phys. {\bf B658},  175--202 (2003) {[}hep-ph/0212259{]}.

\bibitem{COL3}  A.~Denner and S.~Dittmaier,  Nucl. Phys. {\bf B734}, 62--115 (2006) {[}hep-ph/0509141{]}.

\bibitem{COL4}  A.~Denner and S.~Dittmaier,  Nucl. Phys. {\bf B844}, 199--242 (2011) {[}arXiv:1005.2076 [hep-ph]{]}.

\bibitem{HP1} H.~H.~Patel, Comput. Phys. Commun. {\bf 197}, 276--290  (2015) {[}arXiv:1503.01469 [hep-ph]{]}.

\bibitem{HP2} H.~H.~Patel,  Comput. Phys. Commun. {\bf 218}, 66--70 (2017) {[}arXiv:1612.00009 [hep-ph]{]}.

\bibitem{FeyC1}
V.~Shtabovenko, R.~Mertig and F.~Orellana,
Comput. Phys. Commun. \textbf{256}, 107478 (2020)
%doi:10.1016/j.cpc.2020.107478
{[}arXiv:2001.04407 [hep-ph]{]}.

\bibitem{FeyC2}
V.~Shtabovenko, R.~Mertig and F.~Orellana,
%``New Developments in FeynCalc 9.0,''
Comput. Phys. Commun. \textbf{207}, 432--444 (2016)
%doi:10.1016/j.cpc.2016.06.008
{[}arXiv:1601.01167 [hep-ph]{]}.

\bibitem{Form} J.~A.~M.~Vermaseren, "New features of FORM", (2000) math-ph/0010025.

\bibitem{Bohm} S.~Bauberger, M.~B\"ohm, Nucl. Phys. {\bf B445}, 25--46 (1995) {[}hep-ph/9501201 [hep-ph]{]}.

%%%%%%%%%%%%%%%%%%%%%%%%%%%%%%%%%%%%%%%%%%%%%%%%%%%%%%%%5

\bibitem{Tkachov}
F.~V.~Tkachov, Phys. Lett. {\bf B100}, 65--68 (1981).

\bibitem{ChT}
K.~G.~Chetyrkin and F.~V.~Tkachov, Nucl. Phys. {\bf B192}, 159--204 (1981).

\bibitem{BDS} 
F.~A.~Berends, A.~I.~Davydychev and V.~A.~Smirnov,
Nucl. Phys. {\bf B478}, 59--89 (1996) {[}hep-ph/9602396{]}.

\bibitem{DD-JMP}
A.~I.~Davydychev and R.~Delbourgo, J. Math. Phys. {\bf 39} 4299--4334 (1998) {[}hep-th/9709216{]}.

\bibitem{D-NIM}
A.~I.~Davydychev, Nucl. Instr. Meth. {\bf A559}, 293--297 (2006) {[}hep-th/0509233{]}.



\bibitem{Crete} 
A.~I.~Davydychev, Proc. Workshop "AIHENP-99", 
Heraklion, Greece, April 1999 
(Parisianou S.A., Athens, 2000), 219--225 {[}hep-th/9908032{]}.

\bibitem{D-ep}
A.~I.~Davydychev, Phys. Rev. {\bf D61}, 087701 (2000) {[}hep-ph/9910224{]}.

\bibitem{DK1} 
A.~I.~Davydychev and M.~Yu.~Kalmykov, 
Nucl. Phys. {\bf B605}, 266--318 (2001) {[}hep-th/0012189{]}.

\bibitem{DR}
R.~Delbourgo and M.~L.~Roberts, J. Phys. {\bf A36}, 1719--1728 (2003) {[}hep-th/0301004{]}.

\bibitem{DD-JPA}
A.~I.~Davydychev and R.~Delbourgo, J. Phys. {\bf A37}, 4871--4876  (2004) {[}hep-th/0311075{]}.

\bibitem{BD-TMF}
E.~E.~Boos and A.~I.~Davydychev, Teor. Mat. Fiz. {\bf 89} 56, (1991) [Theor. Math. Phys. {\bf 89}, 1052 (1991)].

\bibitem{PBM3}  
A.~P.~Prudnikov, Yu.~A.~Brychkov  and O.~I.~Marichev, Integrals and series. Additional chapters (Nauka, Moscow, 1986).

\end{thebibliography}
\end{document}